\documentclass[a4paper,11pt]{article}
\pdfoutput=1 

\usepackage{jheppub} 

\usepackage[T1]{fontenc} 
\usepackage{scalerel}
\usepackage{mathtools}
\usepackage{dsfont}

\title{\boldmath Entanglement Renormalization for Quantum Field Theories with Discrete Wavelet Transforms}

\author[a]{Daniele S.\,M. Alves}

\affiliation[a]{Theoretical Division, Los Alamos National Laboratory, Los Alamos, NM 87544, USA}

\preprint{LA-UR-24-23540}

\emailAdd{spier@lanl.gov}

\abstract{We propose an adaptation of Entanglement Renormalization for quantum field theories that, through the use of discrete wavelet transforms, strongly parallels the tensor network architecture of the \emph{Multiscale Entanglement Renormalization Ansatz} (a.k.a. MERA). Our approach, called wMERA, has several advantages of over previous attempts to adapt MERA to continuum systems. In particular, (i) wMERA is formulated directly in position space, hence preserving the quasi-locality and sparsity of entanglers; and (ii) it enables a built-in RG flow in the implementation of real-time evolution and in computations of correlation functions, which is key for efficient numerical implementations. As examples, we describe in detail two concrete implementations of our wMERA algorithm for free scalar and fermionic theories in (1+1) spacetime dimensions. Possible avenues for constructing wMERAs for interacting field theories are also discussed.}

\begin{document} 
\maketitle
\flushbottom

\section{Background: Entanglement Renormalization}
\label{ERGintro}

Ongoing advances in quantum technologies have opened up the possibility that quantum field theories (QFTs) might become simulatable in quantum computers in the medium-to-long term future, offering new methods to address long-standing and intractable problems that we have no hope to solve with classical computers. With such prospects, much thought and research efforts have been recently directed towards developing tools that are deliberately bypassed in the traditional, Euclidean space treatments of QFTs. Such tools include the construction of explicit Hilbert space representations of ground and excited states of QFTs---including their entanglement patterns---and real-time Hamiltonian evolution.

In building and understanding these tools, it pays off to borrow some of the wisdom developed in condensed matter physics,  which has traditionally used the framework of wavefunctions, entanglement, and coarse graining to understand the properties and dynamics of condensed matter systems. In particular, a significant insight that emerged in the early 90s was that entanglement plays a significantly role in how fundamental degrees of freedom reorganize themselves as effective degrees of freedom under Renormalization Group (RG) evolution.

Until the early 90s, there were notorious difficulties in numerically implementing real-space block-spin renormalization of quantum spin chains. In block-spin renormalization, a coarse-graining step consists of mapping a block of sites of the fine-grained lattice into a single site of the coarse-grained lattice via a decimation procedure that reduces the dimensionality of the Hilbert space. In Wilson's numerical RG  \cite{Wilson:1974mb}, this is implemented via diagonalization of the Hamiltonian restricted to the block's Hilbert space and its subsequent truncation, keeping only the lowest energy eigenstates of the block. This procedure, however, generically fails to retain the whole support of the ground state, and provides rather unreliable results for a variety of lattice systems  \cite{PhysRevB.19.4876,XIANG1992861,PhysRevLett.42.1492}, with the exception of Hamiltonians which can be put into a special form. In 1992, White and Noack  \cite{White:1992zza} understood the fundamental difficulty with this approach: by neglecting interactions of the block with the rest of the lattice, Wilson's numerical RG fails to capture entanglement between neighboring blocks, and therefore cannot preserve ground state properties, regardless of any reasonable increase in the number of states kept.

The shortcomings of the standard numerical RG approach were overcome by White's Density Matrix Renormalization Group (DMRG)  \cite{White:1992zz}. Instead of the block's Hamiltonian spectrum, this approach considered the spectrum of the block's reduced density matrix, and truncated its Hilbert space by discarding eigenstates of the reduced density matrix with eigenvalues smaller than a given threshold, determined by the desired level of accuracy. In this way, the entanglement properties of the ground state could be retained to any desired precision and a proper RG flow could be generated.

Eventually, the DMRG was understood in terms of the quantum information theoretic concept of {\it tensor networks} (TNs), and recast as a variational optimization method over the class of TNs known as Matrix Product States (MPS)  \cite{Schollwock11,Verstraete:2008}. Essentially, TNs represent quantum many-body states in terms of networks of interconnected tensors, which in
turn capture the relevant entanglement properties of the wave function. TNs provide an unbiased approach to numerical quantum many-body physics; they have been shown to be highly efficient and accurate tools to compute physical observables, while also being free from the sign problem that plagues Monte Carlo sampling methods in Euclidean space (such as in, e.g., Lattice QFT), and also suitable to simulate real-time dynamics.
 
The success of TNs in describing properties of quantum many-body systems can be attributed to the efficiency of optimized TNs in identifying and keeping track of the relevant degrees of freedom. This, in turn, is directly related to the fact that the low energy states of realistic, gapped local Hamiltonians are not generic states in the Hilbert space: they are heavily constrained by locality, and their entanglement is limited by area laws.
%
%
In particular, a key insight of TNs is that, for systems described by local Hamiltonians, coarse graining must be preceded by the removal of short-range entanglement
by local unitary transformations, so that short-distance degrees of freedom can be safely traced out while preserving ground state entanglement properties \cite{White:1992zza,Vidal:2007hda}. A particular class of TNs in which this is self evident and intuitive is called MERA, an acronym for {\it Multiscale Entanglement Renormalization Ansatz} \cite{Evenbly:2007hxg}.

In MERA, short range entanglement across the boundaries of adjacent blocks is first removed by local unitary transformations, or {\it disentanglers}, and then a coarse-graining transformation can be applied to trace out short distance degrees of freedom while safely preserving ground state entanglement properties. An important consequence of the use of disentanglers is that the dimension of the effective sites no longer needs to grow with each coarse graining step, and therefore an efficient description of critical fixed points becomes feasible \cite{2011arXiv1109.5334E,Pfeifer09}.
MERA has been successfully applied to study strongly interacting systems in low dimensions, including the emergence of symmetry breaking order and topological order  \cite{Aguado08,PhysRevB.79.195123}, quantum critical points  \cite{2008PhRvL.101r0503G,Pfeifer09}, real time evolution, fermionic and anyonic systems (which are intractable by quantum Monte Carlo)  \cite{2010PhRvA..81a0303C,2010PhRvA..81e0303P}, as well as a lattice realization of the holographic principle  \cite{Swingle:2009bg,Swingle:2012wq}.
 
\begin{figure}
\centering\includegraphics[width=0.95\textwidth]{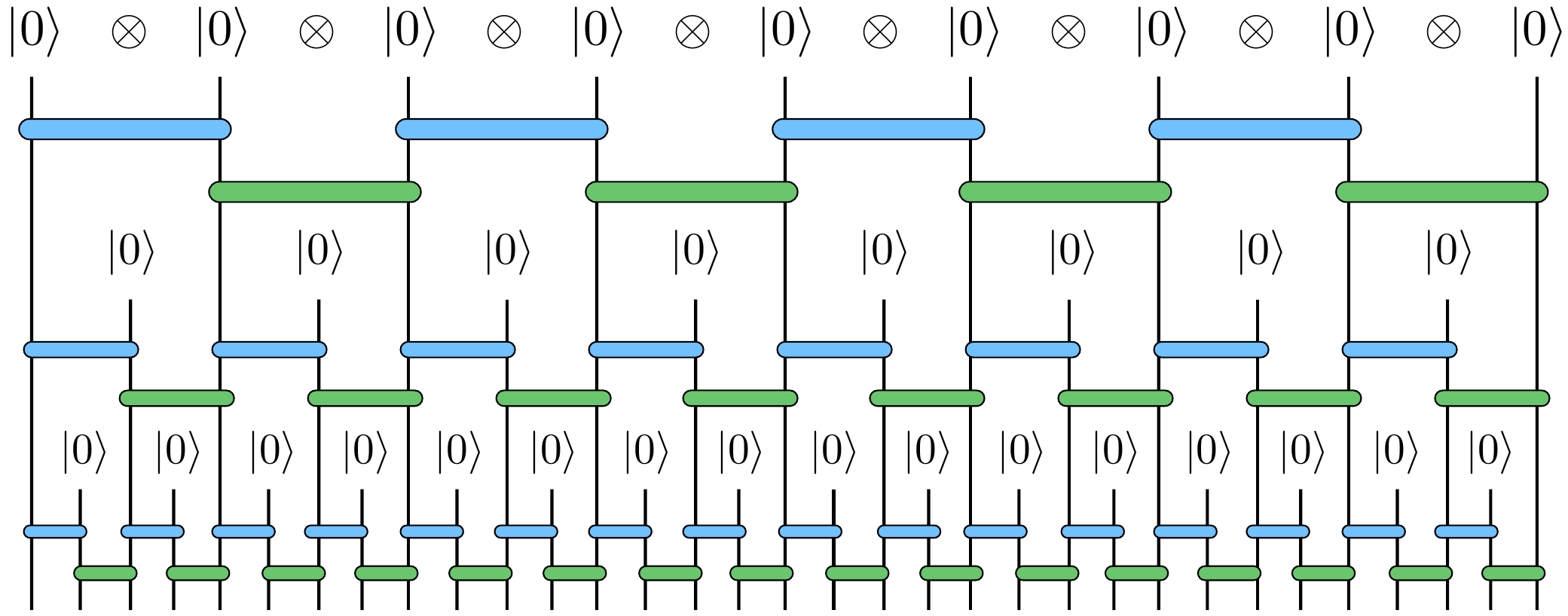}
\caption{A (non-technical) depiction of the MERA architecture. From bottom to top, it can be viewed as an RG flow, whereas from top to bottom, it can be viewed as the construction of the ground state ansatz described in the text.}
\label{MERAcartoon} 
\end{figure}

The construction of a MERA to represent the ground state wavefunction of a system essentially amounts to a reverse RG flow. Consider, as a matter of example, a lattice system governed by a local Hamiltonian. The starting point of the MERA is an unentangled state of $n$ lattice sites, e.g., the product state $|P\,\rangle=|0 \rangle^{\otimes n}$. The reverse RG flow starts by acting on $|P\,\rangle$ with quasi-local unitary transformations\footnote{Meaning transformations that act on neighboring sites separated by a maximum finite distance.} to create entanglement between neighboring sites. These quasi-local unitaries are called {\it entanglers}, and their collective action over the entire system is represented by an operator $U_0$ such that
\begin{equation}
|\Psi_0\rangle~\equiv~U_{0\,}|P\,\rangle~=~U_0\Big[|0 \rangle^{\otimes n}\Big].
\end{equation}
A \emph{``fine-graining''} mapping then follows by the introduction of $n$ new lattice sites intercalated with the original sites, effectively reducing the lattice spacing by half. These new sites are also initialized as a product state, so that the augmented, finer-lattice state is given by
\begin{equation}
|\Psi_0^\prime\rangle~\equiv~W_1|\Psi_0\rangle~\equiv~\Big[|0 \rangle^{\otimes n}\Big]\otimes|\Psi_0\rangle.
\end{equation}
The next step is to entangle the newly introduced sites with the rest of the lattice via new quasi-local entanglers represented by $U_1$, yielding
\begin{equation}
|\Psi_1\rangle~=~U_1|\Psi_0^\prime\rangle.\end{equation}
New layers of fine-graining and entangling operations can then be successively applied to flow deeper into the UV, until a ground state ansatz at the desired UV resolution (or until a fixed point) is reached, namely,
\begin{equation}
|\Psi_{\text{\tiny UV}}\rangle~=\,\left[\prod_{j=0}^{j_\text{\tiny UV}}U_{\!j}W_{\!j}\,\right]\!\!|P\,\rangle.
\end{equation}
The parameters of the entanglers $U_{\!j}$ are variationally optimized (by minimizing the expectation value of the Hamiltonian) to recover the ground state of the lattice system. A (non-technical) depiction of the MERA described above is shown in figure \ref{MERAcartoon}.

In the example described above, the entanglement introduced between adjacent sites at scale $j$ gives rise to entanglement between $2^{(j_{\text{\tiny UV}}-j)}$ sites in the final ansatz $|\Psi_{\text{\tiny UV}}\rangle$. Hence, this iterative approach of introducing quasi-local entanglement at each scale allows one to capture long distance correlations with a series of manageable, quasi-local unitary operations. In essence, MERA works because (i) entanglement is quasi-local in {\it position space} (this stems from interactions being local) and (ii) the entanglement between effective degrees of freedom at a given resolution is insensitive to shorter-range entanglement at much higher resolutions (this stems from the {\it decoupling principle}  \cite{Appelquist:1974tg}, i.e., the insensitivity of a low energy effective theory to its UV completion).

The general properties aforementioned are not exclusive to condensed matter systems described by lattices; indeed, locality of interactions and decoupling are also properties of continuum relativistic systems such as local QFTs. Hence, one would expect that the MERA principles could be generalized to build variational classes for QFTs. And indeed, in 2011, Haegeman {\it et al.}\,\cite{Haegeman:2011uy} proposed such a generalization, which they coined \emph{``continuum} MERA,'' or cMERA for short. The elements of the original, discrete MERA were conceptually modified in their adaptation to the continuum. In particular, Haegeman {\it et al.}~did not propose a discretization of the field theory to more closely parallel the lattice structure and countable layers of the original MERA; instead, cMERA was formulated directly in the continuum, with the system's resolution determined by a Wilsonian UV cut-off $\Lambda_{\text{\tiny UV}}$. In lieu of a discrete layered architecture, continuous scaling transformations were used to flow the system from one scale to another. In particular, the role of the fine-graining mapping was formally implemented by infinitesimal dilation operations to increase the UV cut-off $\Lambda_{\text{\tiny UV}}$ under RG flow. Finally, the action of the unitary entangler operator was implemented in a basis of momentum-space degrees of freedom, effectively generating correlations at all scales up to the shortest distance cut-off, $1/\Lambda_{\text{\tiny UV}}$. While formally attractive and initially promising, the cMERA formulation turned out to be unwieldy and did not lead to any notable results, except in examples of free field theories  \cite{Franco-Rubio:2017tkt,zou2019magic}. Specifically, the path-ordered dilation transformations required to implement changes in scale proved to be impractical for any non-trivial (i.e., interacting) field theory. More problematic, however, was the fact that the implementation of entangling operations in momentum space obscured the locality of entanglement and failed to explicitly capture a key feature of MERA, namely, the hierarchical pattern of entanglement between {\it localized, neighboring} degrees of freedom at each scale.

The shortcomings of the cMERA formulation mentioned above were never addressed in the many attempts to implement and/or generalize it to interacting field theories  \cite{Hu:2017rsp,Cotler:2018ehb,Franco-Rubio:2019nne,Fernandez-Melgarejo:2019sjo,Fernandez-Melgarejo:2020fzw}. This hindered any meaningful progress in the study of QFTs with cMERA since it was first proposed over a decade ago. In particular, no useful/practical entangler ansatzes for interacting QFTs have been fully demonstrated to date.

In this article, we re-examine the generalization of the MERA principles for field theories, and propose an alternative to cMERA which preserves, in explicit form, key features of Entanglement Renormalization, namely, (i) locality of entanglement and (ii) the layered architecture of the original MERA. We coin this alternative MERA adaptation to QFTs ``wMERA.''
A key underlying principle of wMERA is the use of Discrete Wavelet Transforms (DWT) to discretize continuum quantum fields into countable degrees of freedom localized in position space\footnote{We are not the first to use DWTs in the context of MERA (see also \cite{Brennen:2014iqu,Evenbly:2016cly,Haegeman:2017vrx,Witteveen:2019lsk,Witteveen:2020hdp,Stottmeister:2020ezd,George:2022ysn,Sewell:2022nwi}). While there are some overlapping elements of our proposed wMERA framework with previous works in the literature, our framework is built with the explicit goal of making the generalization of MERA to interacting QFTs more straightforward. In addition, to the best of our knowledge, we are the first to use this framework to study \emph{dynamics}, i.e., to perform real-time evolution and compute correlation functions at finite timelike separations.}.
Intuitively, DWTs encode low- and high-pass filters that, upon a choice of scale, separate the degrees of freedom of a quantum field into short-range fluctuations and long-range fluctuations. This simple basis choice provides a natural framework for RG evolution, while keeping locality and separation of scales in explicit form, which makes the MERA generalization quite straightforward. In particular, the fine-graining transformations are implemented via {\it Inverse Wavelet Transforms}, which add shorter-range (wavelet) degrees of freedom at each inverse RG step to flow the system to the UV.

The outline of this article is as follows. In section \ref{cMERAintro}, we provide a brief overview of cMERA and highlight its main shortcomings. In section \ref{DaubechiesSection}, we introduce the concept and basic machinery of discrete wavelet transforms, and the specific class of Daubechies wavelets. In section \ref{wMERAFormulation}, we provide the generic formulation of wMERA using DWTs, the concrete implementation of which is then illustrated in section \ref{ExampleSection} for two concrete examples, namely, a free scalar field and a free Dirac fermion in (1+1)d. Our concluding remarks and outlook are presented in section \ref{Conclusion}.

\section{Brief overview of cMERA}
\label{cMERAintro}

In 2011, Haegeman {\it et al.}~\cite{Haegeman:2011uy} introduced cMERA as a formal adaptation of MERA to the continuum.
Briefly, in cMERA the ground state wavefunction of a system at its infrared (IR) fixed point, $|\Omega_{\text{IR}}\rangle$, is related to its microscopic counterpart at an ultraviolet (UV) scale $s_{\text{UV}}$, $|\Psi(s_{\text{UV}})\rangle$, via the following unitary transformation:
\begin{equation}
\label{wf}
|\Psi(s_{\text{UV}})\rangle ~=~ \mathcal{P}\left[e^{-i\int_{_{s_\text{\tiny IR}}}^{^{s_\text{\tiny UV}}}~ds~\left(\hat{K}(s)\,+\,\hat{L}\;\right)}\right]\;|\Omega_{\text{IR}}\rangle\,.
\end{equation}
Above, $\mathcal{P}$ denotes path-ordering in scale $s$, $\hat{L}$ is the generator of scaling transformations, and $\hat{K}(s)$ is the generator of entanglement at scale $s$. Imposing renormalization group (RG) invariance of physical observables,
\begin{equation}
\langle \Psi(s_{\text{UV}})|\,\hat{O}(s_{\text{UV}})\,|\Psi(s_{\text{UV}})\rangle ~=~
\langle \Omega_{\text{IR}}|\,\hat{O}(s_{\text{IR}})\,|\Omega_{\text{IR}}\rangle\,,
\end{equation}
they obtained the RG equation for a generic hermitian operator $\hat{O}(s)$:
\begin{equation}
\frac{\;d\hat{O}(s)}{ds} ~=~ -i\,\big[\hat{K}(s)+\hat{L}\, , \; \hat{O}(s) \big].
\end{equation}
Haegeman {\it et al.}~\cite{Haegeman:2011uy} illustrated the cMERA construction for a few examples of free field theories\footnote{These examples are shown in detail in the appendix of \emph{version 1} of their \href{https://arxiv.org/abs/1102.5524v1}{preprint} posted on the arXiv online repository.}. For instance, for a free scalar field theory in $(d+1)$ dimensions, the entangling exponent in (\ref{wf}), $\hat{K}(s)+\hat{L}$, was given by the following Gaussian generators:
\begin{eqnarray}
\hat{L}&=&\int_0^\Lambda d^dk~\frac{1}{2}\Bigg[\,\hat\pi(-\vec{k})\,\big(\vec{k}\!\cdot\!\vec{\nabla}_{k\,}\hat{\phi}(\vec{k})\big)\,+\,\big(\vec{k}\!\cdot\!\vec{\nabla}_{k\,}\hat{\phi}(\vec{k})\big)\,\hat\pi(-\vec{k})\nonumber\\
&&~~~~~~~~~~~~~~~~~+\frac{d+1}{2}\,\left(\,\hat{\phi}(\vec{k})\,\hat\pi(-\vec{k})\,+\,\hat\pi(-\vec{k})\,\hat{\phi}(\vec{k})  \,\right)  \,\Bigg]\,,
\end{eqnarray}
and
\begin{equation}
\hat{K}(s)~=~-\int_0^\Lambda d^dk~\frac{1}{4}\,\left[1+\left(\frac{\Lambda}{m_\phi}e^s\right)^2\right]^{-1}\!\left(\,\hat{\phi}(\vec{k})\,\hat\pi(-\vec{k})\,+\,\hat\pi(-\vec{k})\,\hat{\phi}(\vec{k})  \,\right),
\end{equation}
where $\hat\phi(\vec{k})$ and $\hat\pi(\vec{k})$ are the {\it momentum-space} modes of the scalar field and its canonical conjugate momentum, respectively; $m_\phi$ is the mass of the scalar field; and $\Lambda$ is an \emph{ad hoc} UV cut-off.

While formally elegant and concise, the cMERA formulation turned out to be cumbersome even for the simplest case of free field theories. As we shall see in section \ref{ExampleSection}, the \emph{ad hoc} cut-off $\Lambda$ and the  path-ordered architecture of scaling transformations in (\ref{wf}) is completely unnecessary for free field theories. More importantly, however, they are impractical for any nontrivial ({\it i.e.}, interacting) field theory, for which the entangler generator $\hat{K}(s)$ must be generalized beyond quadratic order in the canonical fields to be able to capture nontrivial $n$-point correlation functions beyond the two-point function. 

Compounded by these obstacles is the fact that all incarnations of cMERA attempted so far in the literature were, to the best of this author's knowledge, implemented in momentum space. Since interactions are not local in momentum space, neither is entanglement. More problematic is the fact that interactions entangle momentum-space degrees of freedom across vastly different scales. Hence, at the conceptual level, cMERA constructions in momentum space obscure the hierarchical pattern of entanglement between {\it localized} degrees of freedom in {\it position space} at each scale. At the practical level, momentum-space cMERA is difficult to implement numerically, since it does not share the original MERA's layered architecture, nor the quasi-locality of its entanglers.

To address these obstacles, we propose an alternative adaptation of MERA to the continuum that is directly formulated in {\it position space}, and that does away with the use of continuous path-ordered scaling transformations. In our proposed formulation, which we coin ``wMERA,'' the layers of the MERA are obtained via a {\it Multiresolution Analysis}, whereby the quantum fields are expanded in a {\it wavelet} basis. There is several practical advantages of wMERA over cMERA: (i) the wavelet discretization of quantum fields automatically regulates the theory in the UV; (ii) the locality of degrees of freedom preserves the quasi-locality of entangler ansatzes; (iii) separation of scales is explicitly built-in, enabling the layered architecture of MERA to capture the hierarchical pattern of entanglement across different scales; and (iv) with the degrees of freedom and scales automatically discretized, numerical methods developed for the original MERA can be adapted to wMERA.

Before describing wMERA in detail, we first we provide a brief introduction to Discrete Wavelet Transforms and, in particular, to Multiresolution Analysis with Daubechies wavelets, in section \ref{DaubechiesSection}.

\section{Brief overview of DWTs with Daubechies wavelets}
\label{DaubechiesSection}

Wavelet transforms offer a flexible framework to deconstruct the features of a function, image, or sequence in a hierarchy of resolutions. They are extensively used in signal processing and data compression.

In quantum field theories, on the other hand, Fourier Transforms (FTs) have been traditionally used to decompose the degrees of freedom of a quantum field into momentum modes. While such decompositions are optimal for problems dealing with scattering of plane waves, the delocalized nature of Fourier modes makes FTs highly inefficient in representing spatially localized features, and, by extension, the pattern of ground state entanglement of local, interacting field theories. A suitable alternative to FTs, which we will use in our wMERA formulation, are discrete wavelet transforms (DWTs), which will allow us to decompose the degrees of freedom of a quantum field into localized modes in a hierarchy of resolutions\footnote{The discussion in this section is heavily based on Bulut \& Polyzou \cite{Bulut:2013bg}.}.

In order to intuitively understand some of the key features of DWTs, we begin by describing Haar wavelets, which are arguably the simplest wavelet bases there are. 

\subsection{Haar wavelets}

The Haar wavelet bases are generated from discrete translations and rescalings of the Haar {\it scaling function}, which is essentially the unit box function centered around $x=1/2$:
\begin{equation}
s(x)=\begin{cases}
1~~\text{if}~&0 \leq x < 1\\
0~&\text{otherwise.}
\end{cases}
\end{equation}
Suppose one wishes to represent square integrable functions on the real line, $L^2(\mathbb{R})$, with a resolution of $\Delta x =1$. An orthonormal basis for such decomposition can be constructed from discrete translations of the Haar scaling function $s(x)$:
\begin{equation}
\{s_n(x)\}\big|_{n \in \mathbb{Z}}~~~~\text{with} ~~~~~s_{n}(x)~\equiv~{\mathcal T}^{\,n}[s(x)]~\equiv~ s(x-n).
\label{B0}
\end{equation}
Above, ${\mathcal T}$ is the unit translation operator. 

\begin{figure}
\centering\includegraphics[width=0.99\textwidth]{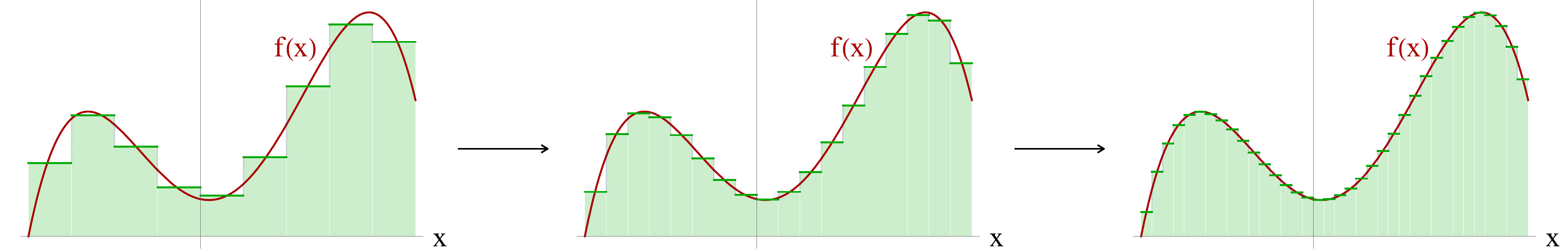}
\caption{Illustration of the decomposition of a generic function $f(x)$ in three Haar scaling function bases with different resolutions: $\Delta x=a_0$ (left), $\Delta x=a_0/2$ (middle), and $\Delta x=a_0/4$ (right).}
\label{HaarFx} 
\end{figure}

Now, suppose one wishes to improve the resolution of the decomposition basis by a factor of 2, i.e., to $\Delta x =1/2$. In this case, the appropriate Haar basis would consist of rescaled scaling functions which are narrower by a factor of 2 (to improve resolution) and taller by a factor of $\sqrt{2}$ (to preserve unit normalization):
\begin{eqnarray}
\{s_n^{1}(x)\}\big|_{n \in \mathbb{Z}}~~~~\text{with} ~~~~~
s_n^{1}(x)~\equiv~\mathcal{D}[s_n(x)]&\equiv&\sqrt{2}\;s_n(2x)\nonumber\\
&=&\sqrt{2}\;s(2x-n).\label{B1}
\end{eqnarray}
Above, ${\mathcal D}$ is the dyadic rescaling operator. An example in which a generic function is decomposed in this basis in a sequence of improved resolutions is shown in figure \ref{HaarFx}.

\begin{figure}
\includegraphics[width=0.45\textwidth]{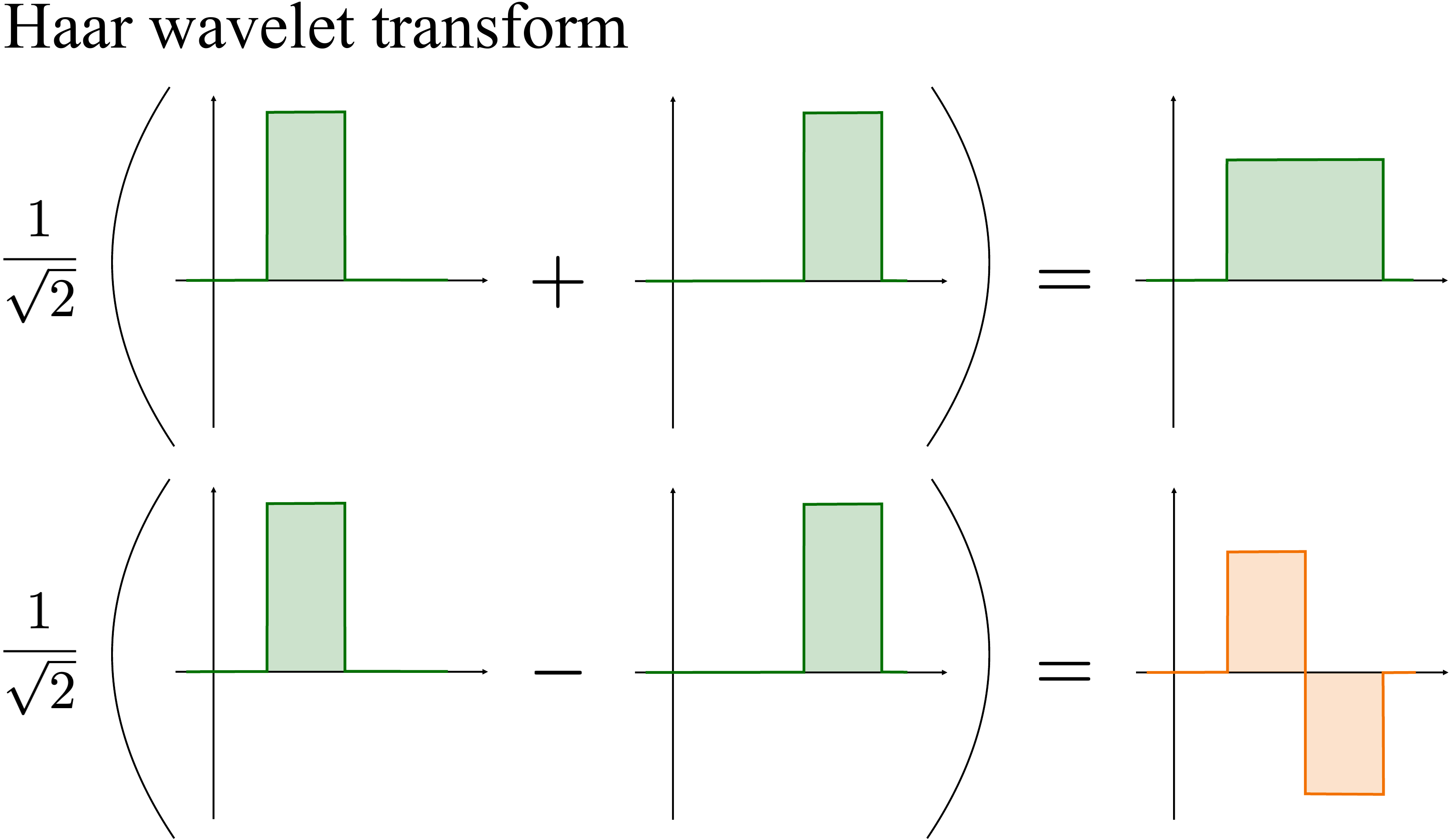}~~~~~~~~~~~~~~
\includegraphics[width=0.45\textwidth]{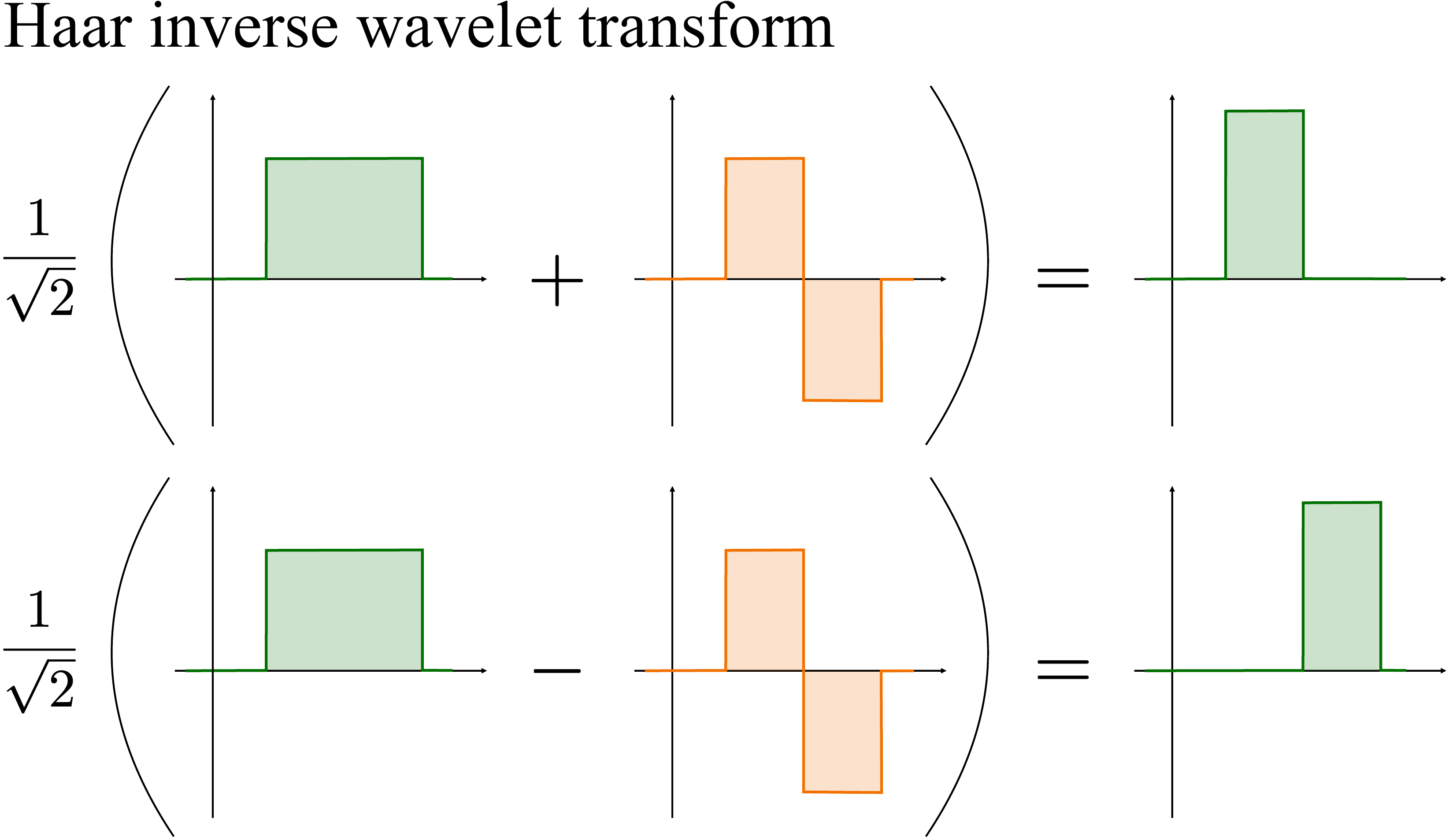}
\caption{\emph{Left panel:} a pictorial illustration of the Haar wavelet transform, which decomposes a signal into low-resolution ``collective'' features (green scaling functions) and high-resolution features (orange wavelet functions). \emph{Right panel:} the inverse Haar wavelet transform, which recombines this decomposition into a basis of finer-resolution collective features.}
\label{HaarWTs} 
\end{figure}

It is easy to see that the space $\mathcal{S}_0$ spanned by the original basis $\{s_n(x)\}$ is a subset of the space $\mathcal{S}_1$ spanned by the improved-resolution basis $\{s^1_n(x)\}$:
\begin{equation}
\mathcal{S}_{0}~\subset~\mathcal{S}_{1}.
\label{S0S1}
\end{equation}
The orthogonal complement of $\mathcal{S}_{0}$ in $\mathcal{S}_{1}$ is a nonempty set called the wavelet subspace $\mathcal{W}_0$:
\begin{equation}
\label{W0}
\mathcal{S}_{1}~=~\mathcal{S}_{0}\oplus\mathcal{W}_0~,~~~~~~~~~~\mathcal{W}_0\neq\emptyset\,.
\end{equation}
$\mathcal{W}_0$ is spanned by the Haar wavelet functions $w_n(x)$, which are generated from discrete translations of the \emph{mother wavelet} $w(x)$:
\begin{equation}
w_{n}(x)\equiv{\mathcal T}^{\,n}[w(x)]=w(x-n)~~~~~~\text{with}~~~~~~w(x)=\begin{cases}
~~\,1~~\text{if}~&0 \leq x < \frac{1}{2}\\
-1~~\text{if}~&\frac{1}{2} \leq x < 1\\
~~\,0~&\text{otherwise.}
\end{cases}
\end{equation}
In other words, the bases
\begin{equation}
\{s_n^{1}(x)\}\big|_{n \in \mathbb{Z}}~~~~~~\text{and} ~~~~~~\{s_n(x)\}\big|_{n \in \mathbb{Z}}\,\cup\;\{w_n(x)\}\big|_{n \in \mathbb{Z}}
\end{equation}
span the same subspace $\mathcal{S}_{1}$ in $L^2(\mathbb{R})$. Their elements are related by the following orthogonal transformation:
\begin{subequations}\label{HaarWT01}
\begin{alignat}{3}
&s_n(x)&=&~~\frac{1}{\sqrt{2}}\Big(s_{2n}^{1}(x)+s_{2n+1}^{1}(x)\Big)\\
&w_{n}(x)~&=&~~\frac{1}{\sqrt{2}}\Big(s_{2n}^{1}(x)-s_{2n+1}^{1}(x)\Big)
\end{alignat}
\end{subequations}
whose inverse is:
\begin{subequations}\label{HaarInvWT01}
\begin{alignat}{3}
&s^{1}_{2n}(x)&=&~~\frac{1}{\sqrt{2}}\Big(s_{n}(x)+w_{n}(x)\Big)\\
&s^{1}_{2n+1}(x)~&=&~~\frac{1}{\sqrt{2}}\Big(s_{n}(x)-w_{n}(x)\Big).
\end{alignat}
\end{subequations}
The transformation in \eqref{HaarWT01} defines the Haar \emph{wavelet transform} (WT), which can be recast in the following form:
\begin{subequations}\label{HaarWTLH}
\begin{alignat}{3}
&s_n(x)&=&~~\mathbf{L}_{nm}\,s_m^{1}(x),\\
&w_{n}(x)~&=&~~\mathbf{H}_{nm}\,s_m^{1}(x)
\end{alignat}
\end{subequations}
with
\begin{eqnarray}
&&\mathbf{L}_{nm}~\equiv~h_{m-2n}~~~~~\text{and}~~~~~h_n=\begin{cases}
~\frac{1}{\sqrt{2}}~~\text{if}~&n=0\\
~\frac{1}{\sqrt{2}}~~\text{if}~&n=1\\
~0~&\text{otherwise.}
\end{cases}\label{lowPassL}\\
&&\mathbf{H}_{nm}~\equiv~g_{m-2n}~~~~~\text{and}~~~~~g_n=\begin{cases}
~~\frac{1}{\sqrt{2}}~~\text{if}~&n=0\\
-\frac{1}{\sqrt{2}}~~\text{if}~&n=1\\
~~\,0~&\text{otherwise.}
\end{cases}\label{lowPassH}
\end{eqnarray}
The projection operators $\mathbf{L}$ and $\mathbf{H}$ in \eqref{lowPassL} and \eqref{lowPassH} are, respectively, the Haar {\it low-pass filter} and the Haar {\it high-pass filter}. They are defined in terms of the finite array of weight coefficients $\{h_n\}$ and $\{g_n\}$.
With this notation, the \emph{inverse wavelet transform} (IWT) in \eqref{HaarInvWT01} can be rewritten as:
\begin{equation}
s^{1}_{n}(x)~=~\mathbf{L}^{T}_{nm}\,s_m(x)+\mathbf{H}^{T}_{nm}\,w_m(x).
\label{HaarInvWTLH}
\end{equation}
The Haar wavelet transform and its inverse are depicted in figure \ref{HaarWTs}.

If one wishes to represent functions with resolution $\Delta x = 1/2^r$ (where the resolution index $r$ is always an integer), a Haar basis can be constructed from $r$ dyadic rescalings of $s_n(x)$, yielding basis function elements with width equal to $1/2^r$:
\begin{eqnarray}
s_n^{r}(x)~\equiv~\mathcal{D}^r[s_n(x)]&=&2^{r/2}\;s_n(2^r x)\nonumber\\
&=&2^{r/2}\;s(2^r x-n).\label{HaarSrBasis}
\end{eqnarray}
In particular, the subspace $\mathcal{S}_{r+1}$ spanned by the basis $\{s_n^{r+1}(x)\}\big|_{n \in \mathbb{Z}}$ contains all lower resolution subspaces $\mathcal{S}_{r},\, \mathcal{S}_{r-1},\, ...,\, \mathcal{S}_{0}$ in a nested form:
\begin{equation}
\label{nestedSr}
\mathcal{S}_{0}~\subset~\mathcal{S}_{1}~\subset~\mathcal{S}_{2}~\subset~...\subset~\mathcal{S}_{r}~\subset~\mathcal{S}_{r+1}~\subset~...\,.
\end{equation}
In full analogy with \eqref{W0}, the orthogonal complement of $\mathcal{S}_{r}$ in $\mathcal{S}_{r+1}$ is the wavelet subspace $\mathcal{W}_r$:
\begin{equation}
\label{Wr}
\mathcal{S}_{r+1}~=~\mathcal{S}_{r}\oplus\mathcal{W}_r,
\end{equation}
which is spanned by a basis of rescaled wavelets:
\begin{eqnarray}
w_n^{r}(x)~\equiv~\mathcal{D}^r[w_n(x)]&=&2^{r/2}\;w_n(2^r x)\nonumber\\
&=&2^{r/2}\;w(2^r x-n).
\label{HaarSWrBasis}
\end{eqnarray}
The generalization of the (inverse) wavelet transform relations between the two equivalent bases
\begin{equation}
\{s_n^{r+1}(x)\}\big|_{n \in \mathbb{Z}}~~~~~~\text{and} ~~~~~~\{s^{r}_n(x)\}\big|_{n \in \mathbb{Z}}\,\cup\;\{w^{r}_n(x)\}\big|_{n \in \mathbb{Z}}
\end{equation}
spanning $\mathcal{S}_{r+1}$ is, expectedly, similar to the  $r=0$ expressions in \eqref{HaarWTLH} and \eqref{HaarInvWTLH}:
\begin{subequations}\label{HaarWT}
\begin{alignat}{3}
&s^{r}_n(x)&=&~~\mathbf{L}_{nm}\,s_m^{r+1}(x),\\
&w^{r}_{n}(x)~&=&~~\mathbf{H}_{nm}\,s_m^{r+1}(x),
\end{alignat}
\end{subequations}
and
\begin{equation}
s^{r+1}_{n}(x)~=~\mathbf{L}^{T}_{nm}\,s^{r}_m(x)+\mathbf{H}^{T}_{nm}\,w^{r}_m(x).
\label{HaarInvWT}
\end{equation}
Combining (\ref{nestedSr}) and (\ref{Wr}), we have a multiresolution decomposition of $L^2(\mathbb{R})$:
\begin{equation}
\label{L2Rdecomposition}
L^2(\mathbb{R})~=~\mathcal{S}_{r}\oplus\mathcal{W}_{r}\oplus\mathcal{W}_{r+1}\oplus\mathcal{W}_{r+2}\oplus...\oplus\mathcal{W}_{r+j}\oplus...,
\end{equation}
i.e., for a base resolution choice $r$, the set of functions
\begin{equation}
\label{srwrjbasis}
\{s_n^r(x)\}\big|_{n \in \mathbb{Z}}\,\cup~ \{w_n^{r}(x)\}\big|_{n \in \mathbb{Z}}\,\cup~ \{w_n^{r+1}(x)\}\big|_{n \in \mathbb{Z}}\,\cup~\dots\cup \{w_n^{r+j}(x)\}\big|_{n \in \mathbb{Z}}\,\cup~\dots
\end{equation}
form an orthonormal basis for $L^2(\mathbb{R})$ with arbitrarily high resolution\footnote{Note that scaling functions $s_n^{r}(x)$ and $s_n^{r'}(x)$ with {\it different} resolutions (i.e., $r\neq r^\prime$) are {\it not orthogonal} to each other and are {\it not elements of the same basis}. Likewise,  wavelets $w_n^{\,r'}(x)$ with $r^\prime<r$ are not orthogonal to $s_n^r(x)$ and are not elements of the basis in \eqref{srwrjbasis}.}. The scaling functions $s_n^r(x)$ encode structure up to scale $1/2^r$, and the wavelets $w_n^{r+j}(x)$ capture structure at finer resolutions $1/2^{r+j+1}$.

It is easy to see that the multiresolution decomposition in \eqref{L2Rdecomposition} and its associated basis in \eqref{srwrjbasis} offer the desired architecture upon which to build the layers of a MERA for a QFT ground state. The discretized degrees of freedom of a quantum field $\phi^{r}_i$, obtained from smearing $\phi(x)$ over the scaling functions $s_i^r(x/a_0)$, heuristically describe collective field fluctuations over distances of order $\mathcal{O}(a_0/2^r)$. Shorter-distance fluctuations of the field at higher resolutions $r^\prime>r$ are captured by $\phi(x)$ projections over the wavelet functions $w_n^{r'}(x/a_0)$. Inverse wavelet transforms are a means to recombine scaling- and wavelet-degrees of freedom into shorter-distance collective modes, whereas wavelet transforms are a means to tease out UV fluctuations that need to be disentangled to enable coarse-graining.

Unfortunately, the Haar multiresolution decomposition has shortcomings that make it not completely suitable for this intended MERA application. In particular, the Haar scaling and wavelet functions are not differentiable, rendering the kinetic term in the Hamiltonian for discretized field degrees of freedom ill-defined. The naive lattice QFT discretization, which strongly parallels the Haar discretization, goes around this problem by adopting the {\it ad hoc} dictionary
\begin{equation}
\label{HaarKDictionary}
\frac{1}{2}\int dx \;(\partial_x \phi)^2~~\to~~~\frac{1}{2}\,\sum_{ij}\,\phi^{r}_i\,\mathds{K}_{ij}^{r}\,\phi^{r}_j
\end{equation}
at resolution $\Delta x = a_0/2^r$, where $a_0$ is the reference lattice spacing for $r=0$, and the discretized kinetic term coefficients are given by
\begin{equation}
\label{HaarK}
\mathds{K}_{ij}^{r}~\equiv~\frac{1}{(a_0/2^r)^2}~\big(2\delta_{ij}-\delta_{i+1,j}-\delta_{i-1,j}\big).
\end{equation}
The problem is that the ``Haar smearing'' interpretation of $\phi^{r}_i$ as
\begin{equation}
\phi^{r}_i\equiv\int dx\;\phi(x)\,s_i^r(x/a_0)
\end{equation}
is inconsistent with the scaling dependence of the discretized kinetic term in \eqref{HaarK}. Specifically, consistency requires that coarse graining $\mathds{K}^{r}$ (through low-pass filter projections) should recover the kinetic term at lower resolutions. However, that is not the case. One can verify by inspection that
\begin{equation}
\big(\mathbf{L}\,\mathds{K}^{r}\,\mathbf{L}^\mathsf{T}\big)_{ij}~=~2\times\,\mathds{K}_{ij}^{(r-1)}
\end{equation}
which is the incorrect scaling.

This problem is specific to wavelet classes whose basis elements are not differentiable---not a fatal flaw of Multiresolution Analysis in general. It can be bypassed by adopting a class of wavelets with continuous first derivatives. Indeed, Haar wavelets belong to a broader family of wavelets called {\it Daubechies wavelets,} named after their inventor Ingrid Daubechies \cite{10.1002/cpa.3160410705}. Other types of wavelets within the Daubechies family retain many of the simple and attractive properties of Haar wavelets while satisfying our  differentiability requirement, as we shall discuss shortly. In what follows, briefly review of the definition and properties of Daubechies wavelets.

\subsection{Daubechies D$N$ wavelets}
Daubechies wavelets provide a family of orthonormal bases for all square integrable functions on the real line, $L^2(\mathbb{R})$. Each type of Daubechies wavelets is characterized their number $N$ of non-vanishing weights $\{h_j\}$---hence the nomenclature Daubechies ``D$N$'' wavelets. The Haar wavelets, which have $N=2$ non-vanishing weights $h_0=h_1=1/\sqrt{2}$ (see \eqref{lowPassL}), are also called Daubechies D2 wavelets.

The Daubechies D$N$ basis elements are functions with compact support at different spatial resolutions, including basis elements which vanish outside an arbitrarily small open set. They are fractal functions, since any element of the basis can be generated from a single scaling function $s(x)$ by applying a finite number of discrete translations and rescalings to $s(x)$, and forming linear superpositions of these rescaled and translated functions. In particular, the D$N$ scaling function satisfies a linear renormalization group equation:
\begin{equation}
\label{DBs}
s(x)~=~\mathcal{D}\!\left[ \sum_{j=0}^{N-1}~h_j\,\mathcal{T}^j[s(x)] \right]\!,
\end{equation}
where the weights $\{h_j\}$ are a sequence of $N$ non-vanishing numbers (to be discussed shortly), and the unit translation ($\mathcal{T}$) and dyadic rescaling ($\mathcal{D}$) operators have been previously introduced in \eqref{B0} and \eqref{B1}. Essentially, $s(x)$ is the fixed point of an operation which takes the weighted average of a finite number of translated copies of itself and rescales it to half of its original support.

The scaling function $s(x)$ has compact support on the finite interval $[0, N-1]$. Further scaling functions with finer resolution are obtained by rescalings and translations of $s(x)$:
\begin{equation}
\label{snk}
s_n^{r}(x)~\equiv~\mathcal{D}^r\big[\mathcal{T}^n[s(x)]  \big]\,.
\end{equation}
Indeed, the subspace spanned by $\{s^r_n(x)\}$,  $\mathcal{S}_r$, has a finest resolution scale of $1/2^r$ and contains all lower resolution subspaces $\mathcal{S}_{r-1},\, \mathcal{S}_{r-2},\, ...,\, \mathcal{S}_{0}$ in the same nested form as in \eqref{nestedSr}.

The wavelet subspaces are generated from the mother wavelet $w(x)$, which also descends from the scaling function $s(x)$ via:
\begin{equation}
\label{DNw}
w(x)~=~ \sum_{j=0}^{N-1}~g_j\,\mathcal{D}\left[\mathcal{T}^j[s(x)] \right].
\end{equation}
Above, the weights $\{g_j\}$ are related to the weights $\{h_j\}$ in \eqref{DBs} by:
\begin{equation}
\label{DNg}
g_j~=~(-1)^jh_{N-1-j}.
\end{equation}
In full analogy with the case of Haar wavelets, a basis of translated and rescaled mother wavelets $\{w_n^{r}(x)\}$ will span the D$N$ wavelet subspace $\mathcal{W}_r$. A multiresolution decomposition of $L^2(\mathbb{R})$ with Daubechies D$N$ wavelets then follows with the same structure of \eqref{L2Rdecomposition} and \eqref{srwrjbasis}.

Additionally, the action of the low- and high-pass filters in (inverse) wavelet transforms for Daubechies D$N$ wavelets is formally the same as for Haar wavelets:
\begin{eqnarray}\label{DbWT}
\text{WT}~~&&\begin{cases}
\,s^{r}_n(x)=~~\mathbf{L}_{nm}\,s_m^{r+1}(x),\\
\,w^{r}_{n}(x)~=~\mathbf{H}_{nm}\,s_m^{r+1}(x),
\end{cases}\\
\nonumber\\
\text{IWT}~~&&\begin{cases}
\,s^{r+1}_{n}(x)~=~\mathbf{L}^{T}_{nm}\,s^{r}_m(x)+\mathbf{H}^{T}_{nm}\,w^{r}_m(x),
\end{cases}
\end{eqnarray}
with the difference that, above, the low-pass and high-pass filters $\mathbf{L}$ and $\mathbf{H}$ are determined by the D$N$ weights that enter in \eqref{DBs} and \eqref{DNg}:
\begin{equation}
\label{D6LH}
\mathbf{L}_{nm}~\equiv~h_{m-2n}~,~~~~~\mathbf{H}_{nm}~\equiv~g_{m-2n}.
\end{equation}
The weights $\{h_j\}$ for Daubechies D$N$ wavelets are determined by three conditions:
\begin{description}
\item[(1)]\;\;Unit area for $s(x)$:
\begin{equation}
\int dx~s(x)=1~~~\iff~~\sum_{j=0}^{N-1}h_j=\sqrt{2}\,,
\label{condition1}
\end{equation}
\item[(2)]\;\;Orthonormality of the basis $\{s_n(x)\}$:
\begin{equation}
\int dx\,s_n(x)s_m(x)=\delta_{nm} ~~~\iff~~ \sum_{j=0}^{N-1}h_j\,h_{j-2n}=\delta_{n,0}\,,
\label{condition2}
\end{equation}
\item[(3)]\;\;Ability of the basis $\{s_n(x)\}$ to locally represent $k$-degree polynomials with $0\leq k \leq (N/2-1)$. In other words, there must exist a sequence $\{c_n\}$ such that:
\begin{equation}
x^k=\sum_{n=-\infty}^{\infty}\,c_n\, s_n(x).
\label{nDegreePol}
\end{equation}
Given that wavelets and scaling functions are orthogonal to each other, the property above is equivalent to the requirement that the wavelets possess $N/2$\,$-$$1$ vanishing moments:
\begin{equation}
\int dx\; x^k\,w(x)=0~~,~~~~0\leq k \leq (N/2-1).
\end{equation}
It then follows from \eqref{DNw} and \eqref{DNg} that this last condition on the weights $\{h_j\}$ can be written as:
\begin{equation}
\sum_{j=0}^{N-1}j^n\,g_{j}~=~\sum_{j=0}^{N-1}j^n(-1)^j \,h_{N-1-n}~=~0\,.
\label{condition3}
\end{equation}
\end{description}
There are two solutions to conditions (\ref{condition1}), (\ref{condition2}), and (\ref{condition3}) above, namely, $\{h_j\}$ and $\{h_j'\}$. They are related by reverse ordering, i.e., $h_j'=h_{N-1-j}$, and their corresponding $s(x)$ and $s'(x)$ are mirror images of each other.

For our wMERA formulation, we would like to choose a Daubechies D$N$ basis with a minimum requirement of a continuous first derivative. The smallest $N$ value for which this condition is satisfied is $N=6$ \cite{doi:10.1137/0522089,doi:10.1137/0523059}. Bases of smoother functions with higher differentiability could be chosen at the expense of additional computational cost, since such bases would have a larger number of weights $\{h_j\}$. For our present purposes, we will consider Hamiltonians exhibiting at most first derivatives, and therefore, for the remainder of this paper, we will adopt Daubechies D6 wavelets for our wMERA formulation.

\begin{figure}
\centering\includegraphics[width=0.8\textwidth]{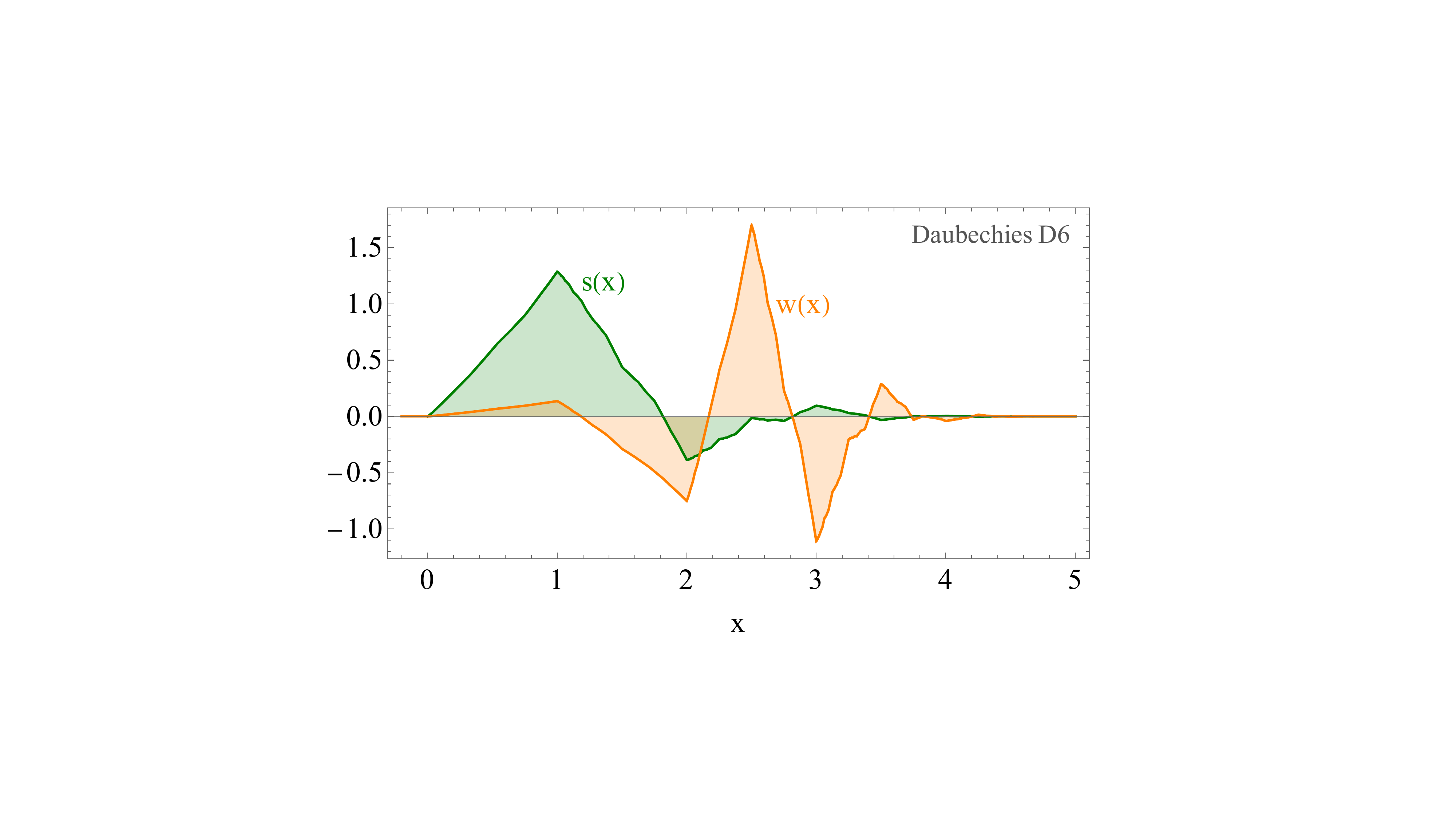}
\caption{The Daubechies D6 scaling function $s(x)$ and mother wavelet $w(x)$.}
\label{D6sw} 
\end{figure}

The weights for Daubechies D6 wavelets are given by:
\begin{subequations}\label{D6h}
\begin{alignat}{3}
&h_0&~=~&\frac{1}{16\sqrt{2}}\;\bigg(1+\sqrt{10}+\sqrt{5+2\sqrt{10}}\;\bigg)\\
&h_1&~=~&\frac{1}{16\sqrt{2}}\;\bigg(5+\sqrt{10}+3\sqrt{5+2\sqrt{10}}\;\bigg)\\
&h_2&~=~&\frac{1}{16\sqrt{2}}\;\bigg(10-2\sqrt{10}+2\sqrt{5+2\sqrt{10}}\;\bigg)\\
&h_3&~=~&\frac{1}{16\sqrt{2}}\;\bigg(10-2\sqrt{10}-2\sqrt{5+2\sqrt{10}}\;\bigg)\\
&h_4&~=~&\frac{1}{16\sqrt{2}}\;\bigg(5+\sqrt{10}-3\sqrt{5+2\sqrt{10}}\;\bigg)\\
&h_5&~=~&\frac{1}{16\sqrt{2}}\;\bigg(1+\sqrt{10}-\sqrt{5+2\sqrt{10}}\;\bigg)\\
&h_j&~=~&0~~\text{if}~~j<0~\text{or}~j>5
\end{alignat}
\end{subequations}
Figure \ref{D6sw} shows the D6 scaling function $s(x)$ and mother wavelet $w(x)$.

In anticipation of our wMERA study of free scalar and fermion theories in section \ref{ExampleSection}, we end this section with a compilation of overlap integrals of D6 scaling and wavelet functions and their first derivatives, which will determine the kinetic-term coefficients of the discretized (1+1)-d Hamiltonians we will consider. For a derivation of the numerical values of the integrals listed below, see, e.g., \cite{Polyzou:2020ifj}.

\begin{figure}
\centering\includegraphics[width=0.99\textwidth]{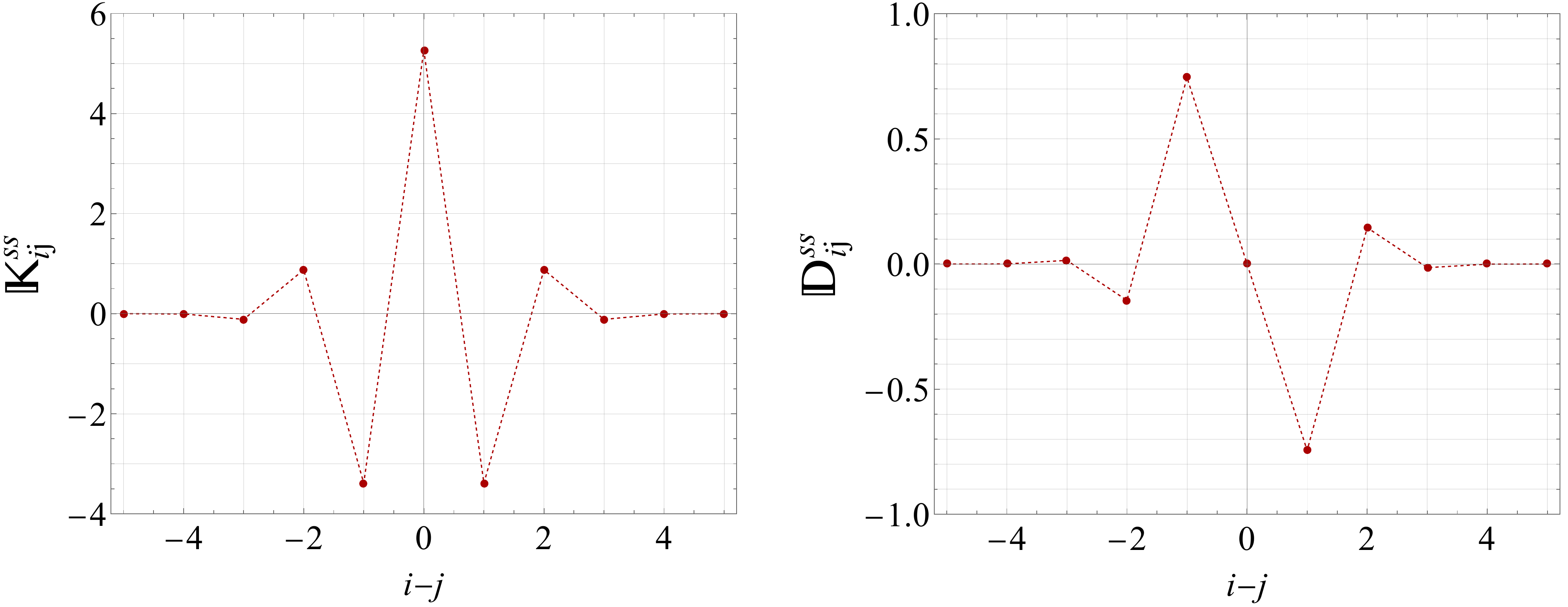}
\caption{The kinetic term coefficients for scalars (left) and fermions (right).}
\label{KssDss} 
\end{figure}

For scalar fields, there will be three types of kinetic-term coefficients, namely:
\begin{eqnarray}
\mathds{K}_{ij}^{{ss}\,(r)}~~~&\equiv&~\int\,dx~\nabla s_i^{r}(x)\,\nabla s^{r}_j(x)\\
\mathds{K}_{ij}^{{sw}\,(r)}~~&\equiv&~\int\,dx~\nabla s^{r}_i(x)\,\nabla w^{r}_j(x)\\
\mathds{K}_{ij}^{{ww}\,(r)}~&\equiv&~\int\,dx~\nabla w^{r}_i(x)\,\nabla w^{r}_j(x).
\end{eqnarray}
Discrete translational invariance implies that these kinetic coefficients depend only on the separation between lattice sites, $i-j$. In addition, the compact support of $s_n^{r}(x)$ and $w_n^{r}(x)$ on the finite interval $n/2^r\leq x \leq (n+5)/2^r$ implies that the kinetic coefficients vanish for separations $|i-j|\geq5$. Defining
\begin{equation}\label{Kss}
\mathds{K}_{ij}^{{ss}\,(r=0)}~\equiv~\mathds{K}_{ij}^{{ss}}~\equiv~K_{|i-j|},
\end{equation}
the nonzero ${K_j}$ coefficients are given by:
\begin{equation}\label{Kcoeff}
K_{0}=\frac{295}{56}~,~~~~
K_{1}=-\frac{356}{105}~,~~~~
K_{2}=\frac{92}{105}~,~~~~
K_{3}=-\frac{4}{35}~,~~~~
K_{4}=-\frac{3}{560}\,,
\end{equation}
and illustrated in figure \ref{KssDss} (left).

The other kinetic coefficients can be obtained by applications of low- and high-pass filters (defined through \eqref{D6LH}, \eqref{D6h}, and \eqref{DNg}):
\begin{eqnarray}
\mathds{K}^{{ss}\,(r)}~~&\equiv~&2^{2r}\,\mathds{K}^{{ss}}\\
\mathds{K}^{{sw}\,(r)}~&\equiv~&2^{2(r+1)}\,\mathds{K}^{{sw}}~,~~~~~~~\,\mathds{K}^{{sw}}_{ij}\!~~\equiv~\big(\mathbf{L}\,\mathds{K}^{ss}\,\mathbf{H}^\mathsf{T}\big)_{ij}\label{Ksw}\\
\mathds{K}^{{ww}\,(r)}&\equiv~&2^{2(r+1)}\,\mathds{K}^{{ww}}~,~~~~~~\,\mathds{K}^{{ww}}_{ij}~\equiv~\big(\mathbf{H}\,\mathds{K}^{ss}\,\mathbf{H}^\mathsf{T}\big)_{ij}.\label{Kww}
\end{eqnarray}
For a fermion field, the four types of kinetic coefficients will be:
\begin{eqnarray}
\mathds{D}_{ij}^{{ss}\,(r)}~~~&\equiv&~\int\,dx~s_i^{r}(x)\,\nabla s^{r}_j(x)\\
\mathds{D}_{ij}^{{sw}\,(r)}~~&\equiv&~\int\,dx~s^{r}_i(x)\,\nabla w^{r}_j(x)\\
\mathds{D}_{ij}^{{ws}\,(r)}~~&\equiv&~\int\,dx~w^{r}_i(x)\,\nabla s^{r}_j(x)~=-\mathds{D}_{ji}^{{sw}\,(r)}\\
\mathds{D}_{ij}^{{ww}\,(r)}~&\equiv&~\int\,dx~w^{r}_i(x)\,\nabla w^{r}_j(x).
\end{eqnarray}
Unlike the scalar field case with symmetric $\mathds{K}^{{ss}}$ kinetic coefficients, in the fermionic case the $\mathds{D}^{{ss}}$ kinetic coefficients are antisymmetric. Defining
\begin{equation}\label{Dss}
\mathds{D}_{ij}^{{ss}\,(r=0)}~\equiv~\mathds{D}^{{ss}}_{ij}~\equiv~\text{sign}(i-j)\,D_{|i-j|},
\end{equation}
the non-zero ${D_j}$ coefficients are given by 
\begin{equation}\label{Dcoeff}
D_{1}=-\frac{272}{365}~,~~~~
D_{2}=\frac{53}{365}~,~~~~
D_{3}=-\frac{16}{1095}~,~~~~
D_{4}=-\frac{1}{2920}\,,
\end{equation}
and illustrated in figure \ref{KssDss} (right).
The other fermionic kinetic coefficients can be obtained by applications of low- and high-pass filters:
\begin{eqnarray}
\mathds{D}^{{ss}\,(r)}~~&\equiv~&2^{r}\,\mathds{D}^{{ss}}\\
\mathds{D}^{{sw}\,(r)}~&\equiv~&2^{r+1}\,\mathds{D}^{{sw}}~,~~~~~~~\,\mathds{D}^{{sw}}_{ij}\!~~\equiv~\big(\mathbf{L}\,\mathds{D}^{ss}\,\mathbf{H}^\mathsf{T}\big)_{ij}\\
\mathds{D}^{{ww}\,(r)}&\equiv~&2^{r+1}\,\mathds{D}^{{ww}}~,~~~~~~\,\mathds{D}^{{ww}}_{ij}~\equiv~\big(\mathbf{H}\,\mathds{D}^{ss}\,\mathbf{H}^\mathsf{T}\big)_{ij}.
\end{eqnarray}

\section{Generic wMERA formulation}
\label{wMERAFormulation}

The decomposition of the Hilbert space of a QFT into localized degrees of freedom organized by a hierarchy of scales makes the connection between wMERA and its discrete counterpart, MERA, more straightforward than the cMERA approach. In this section, we outline the generic architecture of wMERA using DWTs. To make the notation concrete, we assume a scalar field theory in (1+1) spacetime dimensions, but this assumption can be easily generalized to theories with more fields, either bosonic or fermionic. The generalization to higher ($d$+1) spacetime dimensions can also be made by considering wavelet bases in $L^2(\mathbb{R}^d)$; this, however, is beyond the scope of this paper.

In the previous section, we introduced the resolution index $r$ associated with a lattice spacing $a_0/2^r$ for the scaling and wavelet bases. In particular, under our unspecified choice for unit convention, the reference lattice spacing for $r=0$ was set to $a_0=1$. Although the choice of $a_0$ is completely arbitrary, it is useful to identify it with some natural scale of the theory. For the sake of specificity, we will adopt $a_0=1/m_\phi$, where $m_\phi$ is the mass of the scalar field in our toy example. Given this choice, we have that for an arbitrary resolution $r$, the ``lattice site index'' $n$ of the scaling and wavelet functions represents a displacement from the origin by $n$ integer steps of $1/(2^r\,m_\phi)$. In addition, in order to preserve the dimensionless inner product of the scaling functions, we will also explicitly pull out a dimensionful normalization factor in their definition:
\begin{equation}\label{snrNormM}
s_n^{r}\big(x\big)~\equiv~\sqrt{2^r\,m_\phi}\;s\big(2^r m_\phi\, x\,-\,n\big)\,.
\end{equation}
The discretized degrees of freedom of a scalar field $\hat\Phi(x)$ are then obtained from smearing the field in our chosen wavelet basis:
\begin{subequations}\label{DOFphi}
\begin{alignat}{3}
&\hat\phi_n^{\,r}&~=~&\int dx~ \hat\Phi(x)\,s_n^r(x)\\
~~~~~~~~~~~~&\hat\varphi_n^{r'}&~=~&\int dx~ \hat\Phi(x)\,w_n^{r'}(x)~~~~~~(r^\prime\geq r).
\end{alignat}
\end{subequations}
The corresponding discretized degrees of freedom of the canonical conjugate momentum $\hat\Pi(x)$ are:
\begin{subequations}\label{DOFpi}
\begin{alignat}{3}
&\hat\pi_n^{r}&~=~&\int dx~ \hat\Pi(x)\,s_n^r(x)\\
~~~~~~~~~~~~&\hat\varpi_n^{r'}&~=~&\int dx~ \hat\Pi(x)\,w_n^{r'}(x)~~~~~~(r^\prime\geq r).
\end{alignat}
\end{subequations}
They obey the usual canonical commutation relations,
\begin{eqnarray}
\Big[\hat\phi_{n}^{\,r},\,\hat\pi_{m}^{r}\Big]~=~i\,\delta_{nm}~~,~~~~~~~~~~~~
\Big[\hat\varphi_{n}^{r'},\,\hat\varpi_{m}^{r''}\Big]~=~i\,\delta_{r'r''}\,\delta_{nm},
\end{eqnarray}
all other commutators vanishing.
With the definitions above, the DWT representation of the scalar field $\hat{\Phi}(x)$ at a fixed resolution ($r$+1) can be written as:
\begin{subequations}
\begin{alignat}{3}
&\hat{\Phi}^{r+1}(x)&~=&~\sum_{n}\;\hat\phi_n^{\,r+1}\,s_n^{r+1}(x)
\label{phiDecr1}\\
&&~=&~\sum_{n}\;\hat\phi_n^{\,r}\,s_n^r(x)~+~\hat\varphi_n^{r}\,w_n^{\,r}(x),
\label{phiDecSW}
\end{alignat}
\end{subequations}
and similarly for the canonical conjugate momentum $\hat{\Pi}(x)$:
\begin{subequations}
\begin{alignat}{3}
&\hat{\Pi}^{r+1}(x)&~=&~\sum_{n}\;\hat\pi_n^{r+1}\,s_n^{r+1}(x)
\label{piDecr1}\\
&&~=&~\sum_{n}\;\hat\pi_n^{r}\,s_n^r(x)~+~\hat\varpi_n^{r}\,w_n^{\,r}(x).
\label{piDecSW}
\end{alignat}
\end{subequations}
Alternatively, the discretized degrees of freedom can be recast in terms of creation and annihilation operators by defining:
\begin{subequations}\label{aadagger}
\begin{alignat}{3}
&\hat\phi_n^{\,r}&~\equiv&~\frac{1}{\sqrt{2\Delta}}\Big(\hat{\sigma}_n^{r\,\dagger}+\hat{\sigma}_n^{r}\Big)~,~~~~~~\hat\pi_n^{r}~\equiv~i\,\sqrt{\frac{\Delta}{2}}\;\Big(\hat{\sigma}_n^{r\,\dagger}-\hat{\sigma}_n^{r}\Big),\label{Saadagger}\\
&\hat\varphi_n^{r}&~\equiv&~\frac{1}{\sqrt{2\Delta}}\Big(\hat{\omega}_n^{r\,\dagger}+\hat{\omega}_n^{r}\Big)~,~~~~~~\hat\varpi_n^{r}~\equiv~i\,\sqrt{\frac{\Delta}{2}}\;\Big(\hat{\omega}_n^{r\,\dagger}-\hat{\omega}_n^{r}\Big).\label{Waadagger}
\end{alignat}
\end{subequations}
Above, $\hat{\sigma}^{r}$ and $\hat{\sigma}^{r\,\dagger}$ are the creation and annihilation operators for ``collective excitations'' at scales $1/(2^r\,m_\phi)$, whereas $\hat{\omega}^{r}$ and $\hat{\omega}^{r\,\dagger}$ are the creation and annihilation operators for the corresponding ``fluctuating'' wavelet modes. They obey the commutation relations:
\begin{eqnarray}\label{CommutatorSigma}
\big[\hat{\sigma}_n^{r},\;\hat{\sigma}_m^{r\,\dagger}\big]~=~\delta_{nm}~,~~~~~~~~~~
\big[\hat{\omega}_n^{r},\;\hat{\omega}_m^{r' \dagger}\big]~=~\delta_{rr'}\,\delta_{nm}\,,
\end{eqnarray}
with all other commutators vanishing. The width $\Delta$ appearing in (\ref{Saadagger}), (\ref{Waadagger}) is arbitrary; it will be adjusted by the entangler when the wMERA parameters are variationally optimized. An obvious initial choice is $\Delta = m_\phi$.

Note that the discretized degrees of freedom also obey (inverse) wavelet transform relations. Specifically, let $\{\hat{\text{s}},\, \hat{\text{w}}\}$ denote any one of the following pairs: $\{\hat{\phi},\, \hat{\varphi}\}$, $\{\hat{\pi},\, \hat{\varpi}\}$, $\{\hat{\sigma},\, \hat{\omega}\}$, or $\{\hat{\sigma}^{\dagger},\, \hat{\omega}^{\dagger}\}$. The (inverse) wavelet transform relations for each of these pairs are given by:
\begin{eqnarray}
\text{WT}~~&&\begin{cases}
\,\hat{\text{s}}^{r}_n=~~\mathbf{L}_{nm}\,\hat{\text{s}}_m^{r+1},\\
\,\hat{\text{w}}^{r}_{n}~=~\mathbf{H}_{nm}\,\hat{\text{s}}_m^{r+1},
\end{cases}
\label{FieldWT}\\
\nonumber\\
\text{IWT}~~&&\begin{cases}
\,\hat{\text{s}}^{r+1}_{n}~=~\mathbf{L}^\mathsf{T}_{nm}\,\hat{\text{s}}^{r}_m+\mathbf{H}^\mathsf{T}_{nm}\,\hat{\text{w}}^{r}_m.
\end{cases}
\label{FieldIWT}
\end{eqnarray}

The D6-discretization of quantum fields also leads to discretized Hamiltonians at different resolutions. Consider, for the sake of illustration, a generic Hamiltonian for our toy example of a scalar field in (1+1)d:
\begin{equation}
\label{H(x)}
\hat{H}=\int dx\,\Bigg[\,\frac{1}{2}\,\hat\Pi(x)^2~+~\frac{m_\phi^2}{2}\;\hat\Phi(x)^2~+~\frac{1}{2}\,\nabla\hat\Phi(x)\,\nabla\hat\Phi(x)~+~\mathcal{V}\big[\hat\Phi(x)\big] \Bigg].
\end{equation}
Plugging expressions \eqref{phiDecr1} and \eqref{piDecr1} into \eqref{H(x)} and integrating over $x$, we obtain the discretized system's Hamiltonian at resolution $r$ in terms of scaling modes:
\begin{equation}
\label{HrPhiPi}
\hat{H}^{r}_{\phi}~=~\frac{1}{2}\,\sum_{i}\bigg[\,\big(\hat\pi_i^{r}\big)^2~+~m_\phi^2\big(\hat\phi_i^{\,r}\big)^2~+~(2^r m_\phi)^2\,\sum_{j}\hat\phi_i^{\,r}\;\mathds{K}_{ij}^{{ss}}\;\hat\phi_j^{\,r} \bigg]   ~+~\mathcal{V}_{\phi}^{r}\big[\hat\phi^{\,r}\big],
\end{equation}
where the kinetic term coefficients $\mathds{K}^{{ss}}$ have been defined in \eqref{Kss} and \eqref{Kcoeff}.

In \eqref{H(x)}, we left the interaction potential $\mathcal{V}\big[\hat\Phi\big]$ unspecified. Assuming $\mathcal{V}\big[\hat\Phi\big]=\lambda\,\hat\Phi^n$, the corresponding discretized interaction potential in \eqref{HrPhiPi} will have the generic form:
\begin{equation}
\mathcal{V}_{\phi}^{r}\big[\hat\phi^{\,r}\big]~=~\lambda\!\!\!\!\sum_{~~i_1,i_2,\dots, i_n}\!\!\!\!I_{i_1i_2\dots i_n}\;\hat\phi^{\,r}_{i_1}\;\hat\phi^{\,r}_{i_2}\,\dots\, \hat\phi^{\,r}_{i_n}\,,
\end{equation}
where
\begin{equation}
I_{i_1i_2\dots i_n}~\equiv~\int dx~s^r_{i_1}(x)\,s^r_{i_2}(x)\,\dots\,s^r_{i_n}(x).
\end{equation}

The discretized Hamiltonian can also be written in terms of creation and annihilation operators of scaling modes by using \eqref{Saadagger}. Assuming $\Delta=m_\phi$ for convenience, we have:
\begin{eqnarray}
\hat{H}^{r}_{\sigma}&~=~&\frac{m_\phi}{2}\,\sum_{i}\left(\hat{\sigma}^{r\,\dagger}_i\hat{\sigma}^{r}_i+\hat{\sigma}^{r}_i\,\hat{\sigma}^{r\,\dagger}_i\right)\,\;+\;~m_\phi\,2^{2r-2} \,\sum_{ij}\big(\hat{\sigma}^{r\,\dagger}_i+\hat{\sigma}^{r}_i\big)\,\mathds{K}_{ij}^{{ss}}\,\big(\hat{\sigma}^{r\,\dagger}_j+\hat{\sigma}^{r}_j\big)\nonumber\\
&&+~\mathcal{V}_{\sigma}^{r}\big[\hat{\sigma}^{r(\dagger)}\big].\label{HrSaadagger}
\end{eqnarray}
Finally, through the use of IWTs defined in \eqref{FieldIWT}---or, alternatively, by using the field decompositions in \eqref{phiDecSW} and \eqref{piDecSW} combined with \eqref{aadagger}---the discretized Hamiltonian can also be cast in terms of scaling and wavelet modes. Concretely,
\begin{eqnarray}
\hat{H}_{\sigma\omega}^{(r+1)}&~=~~&\hat{H}_{\sigma}^{r}~+~\frac{m_\phi}{2}\,\sum_{i}\left(\hat{\omega}^{r\,\dagger}_i\hat{\omega}^{r}_i+\hat{\omega}^{r}_i\,\hat{\omega}^{r\,\dagger}_i\right)\nonumber\\
&&+~m_\phi\,2^{2r-2} \,\sum_{ij}\big(\hat{\omega}^{r\,\dagger}_i+\hat{\omega}^{r}_i\big)\,\mathds{K}_{ij}^{{ww}}\,\big(\hat{\omega}^{r\,\dagger}_j+\hat{\omega}^{r}_j\big)\nonumber\\
&&+~m_\phi\,2^{2r-1} \,\sum_{ij}\big(\hat{\sigma}^{r\,\dagger}_i+\hat{\sigma}^{r}_i\big)\,\mathds{K}_{ij}^{{sw}}\,\big(\hat{\omega}^{r\,\dagger}_j+\hat{\omega}^{r}_j\big)\nonumber\\
&&+~\mathcal{V}_{\sigma\omega}^{(r+1)}\big[\hat{\sigma}^{r(\dagger)}\!,\,\hat{\omega}^{r(\dagger)}\big],\label{Hr1SWaadagger}
\end{eqnarray}
where the kinetic term coefficients $\mathds{K}^{{sw}}$ and $\mathds{K}^{{ww}}$ have been defined in \eqref{Ksw} and \eqref{Kww}, respectively.

The wavelet decomposition of the field degrees of freedom also implies a decomposition of the Hilbert space of this theory into orthogonal subspaces at different scales:
\begin{equation}
\label{Hdecomposition}
\mathcal{H}~=~\mathcal{H}_{r}~\oplus~\mathfrak{h}_{r}~\oplus~\mathfrak{h}_{r+1}~\oplus...\oplus~\mathfrak{h}_{r+j}~\oplus...\,.
\end{equation}
Above, $\mathcal{H}_{r}$ is the Hilbert subspace containing the ``long-range'' scaling modes describing collective excitations of the scalar field over scales $1/(2^rm_\phi)$. It is identified with the Fock space generated by creation operators $\hat{\sigma}^{r\,\dagger}$ acting on the product state defined by:
\begin{equation}
\label{vac_k}
\big|{\mathbf{P}_{\!\sigma}\big\rangle}_{\!r}~\equiv~\prod_n\,\big|0_{\sigma_n}\big\rangle_{\!r}~,~~~\text{with}~~~~~~\hat{\sigma}_{n}^{r}\,\big|0_{\sigma_n}\big\rangle_{\!r}~=~0\,.
\end{equation}
Likewise, the finer resolution Hilbert subspaces $\mathfrak{h}_{r'}$ ($r^\prime\geq r$) contain modes describing short-range quantum fluctuations of the scalar field at scales $1/(2^{r'+1}m_\phi)$. Each of these subspaces is identified with the Fock space generated by the creation operators $\hat{\omega}^{r' \dagger}$ acting on the product state defined by:
\begin{equation}
\label{vac_k+j}
\big|{\mathbf{P}_{\!\omega}\big\rangle}_{\!r'}~=~\prod_n\,\big|0_{\omega_n}\big\rangle_{\!r'}~,~~~\text{with}~~~~~~\hat{\omega}_{n}^{r'}\,\big|0_{\omega_n}\big\rangle_{\!r'}~=~0\,.
\end{equation}

With the Hilbert space decomposition described in (\ref{Hdecomposition})-(\ref{vac_k+j}), we can sketch the main ingredients of wMERA:
\begin{description}
\item[(1)]\;\;At very low resolutions $r_{\text{\tiny IR}}\ll0$, the discretized degrees of freedom describe fluctuations at lengths much longer than the correlation length of the system (that is, $L\gg1/m_\phi$ in our toy example). At such scales, the nonlocal terms in the Hamiltonian---including the kinetic term---are suppressed, and the ground state exhibits exponentially negligible spatial entanglement. Hence, the far infrared ground state is well-approximated by a product state ansatz\footnote{If the (non-quadratic) Hamiltonian interactions are non-negligible in the far IR, the ground state might not coincide with the state in \eqref{vac_k}. Nevertheless, it should be well-approximated by a product state which can be related to the state in \eqref{vac_k} via an ultra-local entangler.},
\begin{equation}
\big|{\mathbf{0}\big\rangle}_{\!r_{_\text{IR}}}~\simeq~\big|{\mathbf{P}_{\!\sigma}\big\rangle}_{\!r_{_\text{IR}}}.
\end{equation}

At resolutions close to the correlation length of the system, (that is, $r\sim0$ and $L\sim1/m_\phi$ in our toy example), entanglement is still relatively localized (i.e., it is only non-negligible between neighboring sites, and decays exponentially for site separations $|n-m|\gg 1$). Hence, any resolution close to $r=0$ is a good starting point for building the wMERA. Choosing, for concreteness, $r=0$ as the initial resolution, the ground state can be described by the following ansatz state in the Hilbert space $\mathcal{H}_{0}$:
\begin{equation}
\big|{\mathbf{0_\sigma}\big\rangle}_{\!0}~\simeq~\hat{U}_0\big|{\mathbf{P}_{\!\sigma}\big\rangle}_{\!0}\,,
\end{equation}
where $\hat{U}_0$ is a quasi-local unitary entangler ansatz. The (\emph{a priori} unkown) parameters of $\hat{U}_0$ are obtained variationally by minimizing the expectation value of the discretized Hamiltonian at $r=0$:
\begin{equation}
{\big\langle}\mathbf{0_\sigma}\big|\,\hat{H}_\sigma^{0}\,\big|{\mathbf{0_\sigma}}\big\rangle_{\!0}~=~{\big\langle}{\mathbf{P}_{\!\sigma}}\big|\,\hat{U}_0^\dagger\, \hat{H}_\sigma^{0}\,\hat{U}_0\,\big|{{\mathbf{P}_{\!\sigma}}}\big\rangle_{\!0}.
\end{equation}
This constitutes the first wMERA layer.

\item[(2)]\;\;To build the second layer, the first step is to enlarge the Hilbert space to include the finer resolution wavelet subspace $\mathfrak{h}_{0}$ via the fine-graining mapping ${W}_{1}$:
\begin{equation}
{W}_{1}: ~~\mathcal{H}_{0} ~~\mapsto~~\mathcal{H}_{0}\,\oplus\,\mathfrak{h}_{0}\;\cong\;\mathcal{H}_{1}\,.
\end{equation}
Accordingly, the ansatz is augmented by wavelet states initialized in a product state with the rest of the system. By abuse of notation, we write:
\begin{equation}
\label{isometry}
\hat{W}_{1}\,\big|{\mathbf{0_\sigma}\big\rangle}_{\!0}  ~\equiv~ \big|{\mathbf{0_\sigma}\big\rangle}_{\!0}\otimes\big|{\mathbf{P}_{\!\omega}\big\rangle}_{\!0}\,.
\end{equation}

\item[(3)]\;\;The second step is to entangle the newly introduced wavelet states with the original, lower-resolution scaling states. This is parameterized by a quasi-local entangler $\hat{V}_1$:
\begin{equation}
\label{AnsatzSW}
\big|{\mathbf{0_{\sigma\omega}}\big\rangle}_{\!1}  ~=~ \hat{V}_{1}\,\Big[\big|{\mathbf{0_\sigma}\big\rangle}_{\!0}\otimes\big|{\mathbf{P}_{\!\omega}\big\rangle}_{\!0}\Big].
\end{equation}
Similarly to the first layer, the parameters of the entangler ansatz $\hat{V}_1$ are variationally optimized by minimizing the expectation value of the discretized Hamiltonian at $r$\,$=$\,$1$:
\begin{equation}\label{HvevVr}
{\big\langle}\mathbf{0_{\sigma\omega}}\big|\hat{H}^{1}_{\sigma\omega}\big|{\mathbf{0_{\sigma\omega}}}\big\rangle_{\!1}~=~{\big\langle}{\mathbf{P}_{\!\sigma\omega}}\big|\,\big(\hat{V}_1\,\hat{U}_0\big)^\dagger \hat{H}^{1}_{\sigma\omega}\big(\hat{V}_1\,\hat{U}_0\big)\,\big|{{\mathbf{P}_{\!\sigma\omega}}}\big\rangle_{\!1},
\end{equation}
where, above, the $r=1$ Hamiltonian $\hat{H}^{1}_{\sigma\omega}$ is expressed in terms of scaling and wavelet modes as in \eqref{Hr1SWaadagger}, and 
\begin{equation}
\big|{\mathbf{P}_{\!\sigma\omega}\big\rangle}_{\!1}~\equiv~\big|{\mathbf{P}_{\!\sigma}}\big\rangle_{\!0}\otimes\big|{\mathbf{P}_{\!\omega}}\big\rangle_{\!0}.
\end{equation}

\item[(4)]\;\;The state in \eqref{AnsatzSW} is the ansatz for the ground state at resolution $r=1$ defined in the Fock space of $r=0$ scaling and wavelet states. As we will justify shortly, at this point it is convenient to perform an {\it inverse wavelet transform} $\big(\hat{T}_1\big)^{-1}$ to re-express this ansatz as a state in the Fock space of $r=1$ scaling states: 
\begin{equation}
\label{invWTvac_k+j}
\big|{\mathbf{0_\sigma}\big\rangle}_{\!1}~=~\big(\hat{T}_1\big)^{\!-1}\,\big|{\mathbf{0_{\sigma\omega}}\big\rangle}_{\!1}\,.
\end{equation}
Note that the variational optimization of the entangler $\hat{V}_1$ can be alternatively performed \emph{after} the IWT step is applied. Noting that the $r$\,=\,1 Hamiltonians $\hat{H}^{1}_{\sigma}$ and $\hat{H}^{1}_{\sigma\omega}$ (generically defined in \eqref{HrSaadagger} and \eqref{Hr1SWaadagger}) are related by
\begin{equation}
\hat{H}^{1}_{\sigma}=\big(\hat{T}_1\big)^{\!-1}\hat{H}^{1}_{\sigma\omega}\;\hat{T}_{1},
\end{equation}
and defining
\begin{equation}\label{T1V1T1}
\hat{U}_{1}=\big(\hat{T}_1\big)^{\!-1}\big(\hat{V}_1\,\hat{U}_0\big)\,\hat{T}_{1},
\end{equation}
we can alternatively perform the variational optimization of the $r$\,=\,1 entangler parameters by minimizing the following expectation value:
\begin{equation}
{\big\langle}\mathbf{0_\sigma}\big|\,\hat{H}_\sigma^{1}\,\big|{\mathbf{0_\sigma}}\big\rangle_{\!1}~=~{\big\langle}{\mathbf{P}_{\!\sigma}}\big|\,\hat{U}_1^\dagger\, \hat{H}_\sigma^{1}\,\hat{U}_1\,\big|{{\mathbf{P}_{\!\sigma}}}\big\rangle_{\!1}.
\end{equation}

\item[(5)]\;\;Finally, these operations can be performed iteratively to construct the layers of the wMERA until one reaches either a fixed point, or the desired UV resolution $r_{\text{\tiny UV}}$:
\begin{equation}
\label{vac_UV}
\boxed{~
\big|{\mathbf{0_\sigma}\big\rangle}_{\!r_{_\text{\tiny UV}}}~=~\left[\!\!\prod_{~r~=~r_{_{\text{\tiny IR}}}}^{r_{_{\text{\tiny UV}}}}\big(\hat{T}_r\big)^{\!-1}\hat{V}_{r}\,\hat{W}_r\right]\!\!\big|{\mathbf{P}_{\!\sigma}\big\rangle}_{\!r_{_\text{\tiny IR}}}
\,}\,.
\end{equation}
This state describes the vacuum of the theory at scale $r_{\text{\tiny UV}}$ provided that the parameters of the transformations $\hat{V}_{r}$ are, at each layer, chosen so as to minimize the expectation value of the system's Hamiltonian $\hat{H}^r$. If these conditions are not sufficient to uniquely determine the ansatz parameters, one can also demand that expectation values of other physical observables are preserved under RG flow, such as $n$-point correlation functions with $n>2$.
\end{description}
We expect that the variational optimization of the wMERA parameters in (\ref{vac_UV}) can be efficiently implemented numerically with
unitary ansatzes $\hat{V}_{r}$ that are \emph{quasi-local} and \emph{sparse}.
If this is indeed the case, then computational techniques for the variational optimization of MERA tensor networks developed for discrete systems should be adaptable to wMERA for studying QFTs.

In the next section, we will show that quasi-local and sparse wMERA unitaries are indeed sufficient to describe ground state entanglement in two simple examples: a free scalar field and a free Dirac fermion in (1+1) spacetime dimensions.

\begin{figure}
\centering\includegraphics[width=0.99\textwidth]{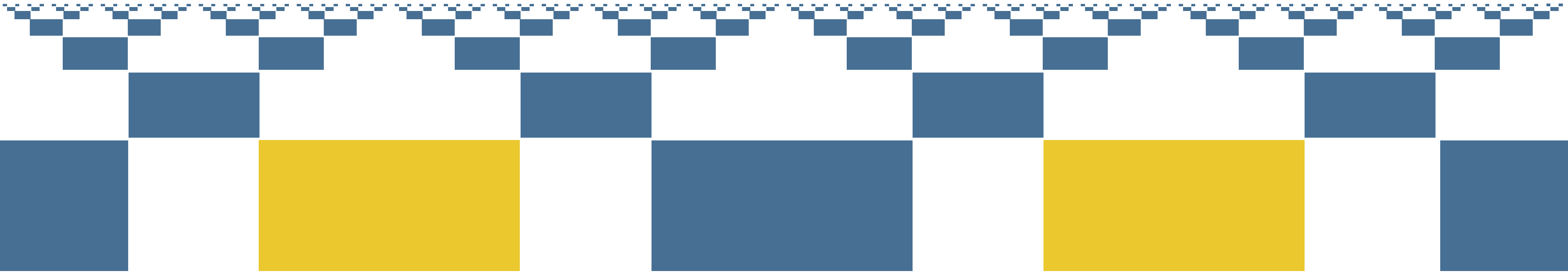}
\caption{Depiction of the lattice describing the system at a deep UV layer if inverse wavelet transforms were not incorporated in the wMERA architecture. The degrees of freedom of the first (bottom layer) consist of scaling modes (depicted in yellow) and wavelet modes (depicted in dark blue). The next layers would iteratively introduce shorter- and shorter-distance wavelet modes (depicted in dark blue) with successively denser lattices. In this lattice architecture, a scaling mode in the first layer would couple to $\mathcal{O}(n^2)$ wavelet modes, where $n$ is the number of wMERA layers that support this particular resolution.}
\label{MERAfractal} 
\end{figure}

We end this section by justifying the need for inverse wavelet transforms at every resolution layer in the wMERA formulation. Note that if these transformations were not performed, the description of the system would incur two problematic features: (1) a non-uniform lattice structure, and (2) discretized Hamiltonians $\hat{H}^r$ becoming increasingly nonlocal as the number of wMERA layers is increased. To illustrate this situation, consider a resolution layer $r=r_0+n$, where $n\gg 1$ and $r_0$ is the initial wMERA layer. In the absence of IWTs, the degrees of freedom at this layer would consist of the scaling modes, $\phi^{(r_0)}$, and $n+1$ types of wavelet modes: $\varphi^{(r_0)},\,\varphi^{(r_0+1)},\,\dots,\,\varphi^{(r_0+n)}$. These $n+2$ different types of modes would have different lattice densities, with $\phi^{(r_0)}$ and $\varphi^{(r_0)}$ having the sparsest lattices (with lattice spacing $a_{0}$), and $\varphi^{(r_0+n)}$ having the densest lattice (with lattice spacing $a_{0}/2^{n}$). In addition, any given scaling mode $\phi_j^{(r_0)}$ would couple to a total of
$\big(5n^2+23n+36\big)/2$ other modes, namely,
\begin{equation}
\begin{cases}
&\phi_{j-4}^{(r_0)}\;\dots\;\phi_{j+4}^{(r_0)}\\
&\varphi_{j-4}^{(r_0)}\;\dots\;\varphi_{j+4}^{(r_0)}\\
&\varphi_{2j-4}^{(r_0+1)}\;\dots\;\varphi_{2j+9}^{(r_0+1)}\\
&~~~\vdots\\
&\varphi_{2^nj-4}^{(r_0+n)}\;\dots\;\varphi_{2^nj+5n+4}^{(r_0+n)}.
\end{cases}\nonumber
\end{equation}
In other words, the nonlocality of the Hamiltonians $\hat{H}^{r_0+n}$ would grow quadratically with the number of wMERA layers. This situation is depicted in figure \ref{MERAfractal}. It is easy to see that not only this lattice structure becomes quickly impractical, but the ansatz would require unitary entanglers that more and more nonlocal at each layer. The IWTs address these difficulties by redefining the degrees of freedom when moving from one layer to the next so that the uniformity of the lattice structure is preserved, and the quasi-locality of the discretized Hamiltonians $\hat{H}^r$ is restored at each layer.

\section{Concrete wMERA Examples}
\label{ExampleSection}

For the remainder of this article, we work through the concrete implementation of wMERA for a couple examples of free field theories in (1+1) spacetime dimensions: a scalar field and a Dirac fermion. For each example, we begin by deriving the exact entangler directly in the continuum without invoking cMERA complications such as {\it ad hoc} UV cut-offs or path-ordered integrals over scaling transformations. We then solve these theories using our proposed variational wMERA approach: we obtain the optimized numerical results for the wMERA entanglers, which we then use to compute the two-point correlation functions for these theories for both spacelike and timelike separations.

\subsection{Free theory of a scalar field in (1+1)d}
\label{ScalarSection}

Consider a free scalar field theory in (1+1)d. Its Hamiltonian is given by
\begin{equation}
\label{HxPhiFree}
\hat{H}=\frac{1}{2}\int dx\,\Bigg[\,\hat\Pi(x)^2~+~m_\phi^2\,\hat\Phi(x)^2~+~\nabla\hat\Phi(x)\,\nabla\hat\Phi(x)\Bigg],
\end{equation}
where the scalar field $\hat\Phi(x)$ its canonical conjugate momentum $\hat\Pi(x)$ obey the equal-time commutation relations
\begin{equation}
\Big[\hat\Phi(x),\,\hat\Pi(y)\Big]~=~i\,\delta(x-y);
\end{equation}
all other equal-time commutators vanishing.

Since free theories only have Gaussian correlations, it should come as no surprise that their entanglers are Gaussian\footnote{For an excellent reference on multimode Gaussian entanglers, see \cite{PhysRevA.41.4625}.} (i.e., the entanglers' exponents are quadratic functions of $\hat\Phi$ and $\hat\Pi$). Specifically, writing the field and its canonical conjugate momentum in terms of \emph{position-space} creation and annihilation operators:
\begin{equation}\label{a(x)}
\hat\Phi(x)~\equiv~\frac{1}{\sqrt{2m_\phi}}\Big(\hat{a}(x)^{\dagger}+\hat{a}(x)\Big)~,~~~~~~\hat\Pi(x)~\equiv~i\,\sqrt{\frac{m_\phi}{2}}\;\Big(\hat{a}(x)^{\dagger}-\hat{a}(x)\Big),
\end{equation}
we would like to find the entangler $\hat{U}$ that relates the product state $\big|{\mathbf{P}\big\rangle}$ to the ground state $\big|{\mathbf{0}\big\rangle}$:
\begin{equation}
\big|{\mathbf{0}\big\rangle}~=~\hat{U}\big|{\mathbf{P}\big\rangle},
\end{equation}
where the product state $\big|{\mathbf{P}\big\rangle}$ has, by definition, no entanglement in position space and can be thought of as the generalization of the state in \eqref{vac_k} to the continuum:
\begin{equation}\label{a(x)P}
\hat{a}(x)\,\big|\mathbf{P}\big\rangle~=~0~~~\forall~~~x.
\end{equation}
In this case, the entangler ansatz is a generator of Bogoliubov transformations (a.k.a. multimode squeezing), and has the generic form:
\begin{equation}\label{UPhi(x)}
\hat{U}~=~\exp\left[-\frac{i}{2}\int dx\,dy~\zeta(x-y)~\Big(\hat\Phi(x)\hat\Pi(y)+\hat\Pi(x)\hat\Phi(y)\Big)\,\right],
\end{equation}
or, alternatively in terms of position-space creation and annihilation operators,
\begin{equation}\label{Ua(x)}
\boxed{
~\hat{U}~=~\exp\left[~\frac{1}{2}\int dx\,dy~\zeta(x-y)~\Big(\hat{a}^\dagger(x)\hat{a}^\dagger(y)-\hat{a}(x)\hat{a}(y)\Big)\,\right]\,
}.
\end{equation}
The entangler exponent $\zeta(x-y)$ defines the entangling parameter that needs to be variationally optimized. The easiest way to proceed is to go to Fourier space by defining:
\begin{subequations}
\begin{alignat}{3}
&\hat\Phi_k&\;\equiv\,\int dx~e^{-ikx}\,\hat\Phi(x)~,~~~~~~~&\hat\Pi_k\;\equiv\,\int dx~e^{-ikx}\,\hat\Pi(x)\\
&\,\hat{a}_k&\;\equiv\,\int dx~e^{-ikx}\,\hat{a}(x)~,~~~~~~~&\;\hat{a}^\dagger_k\;\equiv\,\int dx~e^{ikx}\,\hat{a}^\dagger(x)\label{akadk}
\end{alignat}
\end{subequations}
and
\begin{equation}\label{zetaFourier}
\zeta_k~\equiv~\int dx~ e^{-ikx}\,\zeta(x).
\end{equation}
Note that $\hat{a}^\dagger_k$ and $\hat{a}_k$ in \eqref{akadk} are \emph{not} the creation and annihilation operators of momentum-mode Hamiltonian eigenstates. In particular, they do \emph{not} annihilate the ground state, but, because of \eqref{a(x)P}, they do annihilate the initial product state:
\begin{equation}\label{a(k)P}
\hat{a}_k\,\big|\mathbf{P}\big\rangle~=~0~~~\forall~~~k.
\end{equation}
In Fourier space, the entangler and the Hamiltonian are given by
\begin{equation}
\hat{U}~=~\exp\left[\,\frac{1}{2}\,\int \frac{dk}{2\pi}~\zeta_k~\Big(\hat{a}^\dagger_k\,\hat{a}^\dagger_{-k}-\hat{a}_k\,\hat{a}_{-k}\Big)\,\right]
\label{Ua(k)}
\end{equation}
and
\begin{equation}
\hat{H}~=~\frac{1}{4m_\phi}\int \frac{dk}{2\pi}~~\Bigg[\big(\omega_k^2+m_\phi^2\big)\Big({\hat{a}_k}^\dagger\,\hat{a}_k+\hat{a}_k\,\hat{a}^\dagger_k\Big)+k^2\,\Big(\hat{a}^\dagger_k\,\hat{a}^\dagger_{-k}+\hat{a}_k\,\hat{a}_{-k}\Big)\Bigg],
\label{Ha(k)}
\end{equation}
where $\omega_k^2$ has the usual definition $\omega_k^2\equiv k^2+m_\phi^2$.
Recalling the commutation relations $\big[\hat a_k,\,\hat a_q^\dagger\big]=2\pi\,\delta(k-q)$, one can easily derive the action of the entangler on the creation and annihilation operators,
\begin{subequations}\label{UaU}
\begin{alignat}{3}
&\hat{U}^{\dagger\,} \hat{a}_k\, \hat{U}&~=~&\text{cosh}(\zeta_k)\,\hat{a}_k~+~\text{sinh}(\zeta_k)\,\hat{a}_{-k}^\dagger\\
&\hat{U}^{\dagger\,} \hat{a}_k^\dagger\, \hat{U}&~=~&\text{cosh}(\zeta_k)\,\hat{a}_k^\dagger~+~\text{sinh}(\zeta_k)\,\hat{a}_{-k},
\end{alignat}
\end{subequations}
and the entangler-transformed Hamiltonian,
\begin{eqnarray}
\hat{U}^\dagger \hat{H}\,\hat{U}~=~\frac{1}{4m_\phi}\int \frac{dk}{2\pi}&&\Bigg[\Big(\big(\omega_k^2+m_\phi^2\big)\,\text{cosh}(2\zeta_k)+k^2\,\text{sinh}(2\zeta_k)\Big)\,\Big({\hat{a}_k}^\dagger\,\hat{a}_k+\hat{a}_k\,\hat{a}^\dagger_k\Big)\label{UHU}\\
&&~~+~\Big(\big(\omega_k^2+m_\phi^2\big)\,\text{sinh}(2\zeta_k)+k^2\,\text{cosh}(2\zeta_k)\Big)\,\Big(\hat{a}^\dagger_k\,\hat{a}^\dagger_{-k}+\hat{a}_k\,\hat{a}_{-k}\Big)\Bigg].\nonumber
\end{eqnarray}
Obtaining the vacuum expectation value of the Hamiltonian is now straightforward. Using \eqref{UHU} and \eqref{a(k)P}, we have:
\begin{eqnarray}
\big\langle\mathbf{0}\big|\hat{H}\big|\mathbf{0}\big\rangle~&=&~{\big\langle}\mathbf{P}\big|\hat{U}^\dagger \hat{H}\,\hat{U}\big|\mathbf{P}\big\rangle\nonumber\\
&=&~\frac{1}{4m_\phi}\int dk\,\,\delta(k-k)\,\Big(\big(\omega_k^2+m_\phi^2\big)\,\text{cosh}(2\zeta_k)+k^2\,\text{sinh}(2\zeta_k)\Big).
\end{eqnarray}
The final step is to minimize the vacuum expectation value of the Hamiltonian with respect to entangler exponent $\zeta_k$:
\begin{eqnarray}
\frac{\partial}{\partial \zeta_k}\big\langle\mathbf{0}\big|\hat{H}\big|\mathbf{0}\big\rangle=0~~&\Rightarrow&~~\big(\omega_k^2+m_\phi^2\big)\,\text{sinh}(2\zeta_k)+k^2\,\text{cosh}(2\zeta_k)~=~0\nonumber\\
&\Rightarrow&~~~\boxed{~\zeta_k\,=\,\frac{1}{2}\,\text{log}\Bigg(\frac{m_\phi}{\omega_k}\Bigg)\,}\,.\label{zetak}
\end{eqnarray}
We can now use \eqref{zetaFourier} and \eqref{zetak} to obtain the \emph{exact} entangler exponent in position space,
\begin{equation}
\boxed{~
\zeta(x-y)~=~\frac{\,e^{-m_\phi|x-y|}\,}{4|x-y|}
\;}\,.
\label{zetax}
\end{equation}
It is also informative to look at the entangler-transformed Hamiltonian using the exact entangler in \eqref{zetak} and \eqref{zetax}:
\begin{subequations}
\begin{alignat}{3}
&\hat{U}^\dagger \hat{H}\,\hat{U}~~&=&~\int \frac{dk}{2\pi}\,\omega_k\,{\hat{a}_k}^\dagger\,\hat{a}_k~~+~~\text{(zero-point energy)}\\
&&=&~\int \!dx\,dy~\,\Omega(x-y)\,\hat{a}^\dagger(x)\hat{a}(y)~~+~~\text{(zero-point energy)},
\end{alignat}
\end{subequations}
where, above,
\begin{equation}
\Omega(x-y)~\equiv~-\frac{\,m_\phi^2\,}{\pi}\,\frac{K_1\big(m_\phi|x-y|\big)}{m_\phi|x-y|}~~~~\xrightarrow[~m_\phi \to \,0~]{}~~~~-\frac{1}{\pi|x-y|^2}\,.
\label{Hkernel}
\end{equation}

While the use of FTs in this exercise was convenient to obtain the exact entangler, this trick unfortunately only works well in the case of free theories for which the entangler is Gaussian (as we argued extensively in sections \ref{ERGintro} and \ref{cMERAintro}). For interacting theories, we expect that working directly in position space through the use of DWTs will make calculations much more tractable. For the remainder of this subsection, we show how this actually works by solving this theory using wMERA.

We work with the D6-discretized degrees of freedom defined in \eqref{DOFphi}, \eqref{DOFpi}, and \eqref{aadagger}, whose dynamics are governed by the discretized Hamiltonians \eqref{HrSaadagger} and \eqref{Hr1SWaadagger} with vanishing interaction potential (i.e., $\mathcal{V}=0$). To recall our notation, at any given resolution $r$ the product state $\big|\mathbf{P}_{\!\sigma}\big\rangle_{\!r}$ is related to the ground state by a unitary entangler $\hat{U}_{r}$:
\begin{equation}
\big|\mathbf{0}_\sigma\big\rangle_{\!r}~=~\hat{U}_{r}\,\big|\mathbf{P}_{\!\sigma}\big\rangle_{\!r}.
\end{equation}
In our free theory example, the ansatz for $\hat{U}_{r}$ is a generator of Bogoliubov transformations between scaling modes:
\begin{equation}\label{Urss}
\hat{U}_{r}~=~\exp\left[\,\frac{1}{2}\,\sum_{i,j}\,\zeta_{ij}^{(r)}\;\Big(\hat{\sigma}_{i}^{\,r\,\dagger}\hat{\sigma}_{j}^{\,r\,\dagger}\,-\,\hat{\sigma}_{i}^{\,r}\hat{\sigma}_{j}^{\,r}\Big)\,\right]\!.
\end{equation}
The entangling exponent ${\zeta}^{(r)}$ is a Hermitian matrix whose entries satisfy the following relation
\begin{equation}
\zeta_{ij}^{(r)}~=~\zeta_{|i-j|}^{(r)}
\end{equation}
due to discrete translational invariance.
The entangler $\hat{U}_{r}$ transforms scaling modes in the following way:
\begin{subequations}\label{UsigmaU}
\begin{alignat}{3}
&\hat{U}_r^{\dagger}\, \hat{\sigma}_i^{\,r}\, \hat{U}_r&~=~&\big(\text{Cosh}\zeta^{(r)}\big)_{ij}\,\hat{\sigma}_j^{\,r}~~+~\big(\text{Sinh}\zeta^{(r)}\big)_{ij}\,\hat{\sigma}_{j}^{\,r\,\dagger}\\
&\hat{U}_r^{\dagger}\, \hat{\sigma}_i^{\,r\,\dagger}\, \hat{U}_r&~=~&\big(\text{Cosh}\zeta^{(r)}\big)_{ij}\,\hat{\sigma}_j^{\,r\,\dagger}~\!+~\big(\text{Sinh}\zeta^{(r)}\big)_{ij}\,\hat{\sigma}_{j}^{\,r}\,.
\end{alignat}
\end{subequations}
The other type of entangler discussed in section \ref{wMERAFormulation} was the unitary operator $\hat{V}_{r}$ that entangled scaling and wavelet modes at the wMERA layer with resolution $r$. While $\hat{V}_{r}$ is still a Gaussian entangler, it has a more general parametrization than $\hat{U}_r$. Concretely,
\begin{eqnarray}
\hat{V}_{r}&=&\exp\Bigg\{\sum_{i,j}\Bigg[\,
\nu^{ww(r-1)}_{ij}\Big(\hat{\omega}_{i}^{\,r-1\,\dagger}\hat{\omega}_{j}^{\,r-1\,\dagger}\!-\hat{\omega}_{i}^{\,r-1}\hat{\omega}_{j}^{\,r-1}\Big)\!+\nu^{sw(r-1)}_{ij}\Big(\hat{\sigma}_{i}^{\,r-1\,\dagger}\hat{\omega}_{j}^{\,r-1\,\dagger}\!-\hat{\sigma}_{i}^{\,r-1}\hat{\omega}_{j}^{\,r-1}\Big)\nonumber\\
&&~~~~~~~~~~~~~~~~~+~\rho^{sw(r-1)}_{ij}\Big(\hat{\sigma}_{i}^{\,r-1\,\dagger}\hat{\omega}_{j}^{\,r-1}-\hat{\omega}_{j}^{\,r-1\,\dagger}\hat{\sigma}_{i}^{\,r-1}\Big)\,+\,\dots ~\Bigg]\,\Bigg\}.\label{VrAnsatz}
\end{eqnarray}
Above, the ellipsis denotes higher order terms in a BCH expansion which are proportional to~~$\hat{\sigma}_{i}^{\dagger}\hat{\sigma}_{j}$,~~$\hat{\omega}_{i}^{\dagger}\hat{\omega}_{j}$,~~and~~$\hat{\sigma}_{i}^{\dagger}\hat{\sigma}_{j}^{\dagger}-\hat{\sigma}_{i}\hat{\sigma}_{j}$.

As mentioned in section \ref{wMERAFormulation}, at each wMERA layer one can either variationally optimize the $\hat{V}_{r}$ ansatz parameters
in \eqref{VrAnsatz} by minimizing
\begin{equation}
{\big\langle}\mathbf{0_{\sigma\omega}}\big|\hat{H}^{r}_{\sigma\omega}\big|{\mathbf{0_{\sigma\omega}}}\big\rangle_{\!r}~=~\prescript{}{(r-1)\!}{\Big[}\big\langle{\mathbf{P}_{\!\omega}\big|}\otimes{\big\langle}\mathbf{0_\sigma}\big|\Big]\,\hat{V}_{r}^\dagger\,\hat{H}^{r}_{\sigma\omega}\,\hat{V}_{r}\,\Big[\big|{\mathbf{0_\sigma}\big\rangle}\otimes\big|{\mathbf{P}_{\!\omega}\big\rangle}\Big]_{\!(r-1)},
\end{equation}
or one can perform the optimization \emph{after} the IWT reorganizes the scaling and wavelet modes $\hat{\sigma}^{\,(r-1)(\dagger)}$ and $\hat{\omega}^{\,(r-1)(\dagger)}$ into
$\hat{\sigma}^{\,r(\dagger)}$ scaling modes. While this choice is completely up to the developer of the optimization algorithm, this author found the second option easier to implement numerically, for two main reasons. First, there is a significant amount of redundancy in the $\hat{V}_{r}$ ansatz parameters $\nu_{ij}^{ww(r)}$, $\nu_{ij}^{sw(r)}$, and $\rho_{ij}^{sw(r)}$ (they secretly depend on a smaller number of parameters, as it should became clear shortly). Second, while we have not shown explicitly how the $\hat{V}_{r}$ ansatz in \eqref{VrAnsatz} transforms the scaling and wavelet modes $\hat{\sigma}_{i}^{\,r(\dagger)}$ and $\hat{\omega}_{i}^{\,r(\dagger)}$, one can easily anticipate that these transformations will have a more complicated form compared to \eqref{UsigmaU}.

Hence, we choose to proceed with the wMERA optimization by instead minimizing the parameters of each layer \emph{after} the IWT is performed. That is, we will optimize
\begin{equation}\label{HvevUr}
{\big\langle}\mathbf{0_\sigma}\big|\,\hat{H}_\sigma^{r}\,\big|{\mathbf{0_\sigma}}\big\rangle_{\!r}~=~{\big\langle}{\mathbf{P}_{\!\sigma}}\big|\,\hat{U}_r^\dagger\, \hat{H}_\sigma^{r}\,\hat{U}_r\,\big|{{\mathbf{P}_{\!\sigma}}}\big\rangle_{\!r}.
\end{equation}
Note, however, that this does not mean we intend minimize all $\zeta_{ij}^{(r)}$ parameters of the ansatz in \eqref{Urss} at once, since this would not take any input from previous wMERA layers, and would lead to challenges with numerical efficiency and accuracy. Instead, the crucial point to note is that the ``long range'' entanglement encoded in $\zeta_{ij}^{(r)}$ has already been optimized in the previous wMERA layers. The way to take advantage of that is through the following: using WTs, we can recast $\hat{U}_{r}$ in terms of scaling and wavelet degrees of freedom\footnote{The first equality in \eqref{TUT} is essentially a generalization of the $r=1$ relation in \eqref{T1V1T1}.},
%
\begin{eqnarray}
\hat{T}_{r}\,\hat{U}_{r\,}\big(\hat{T}_{r}\big)^{\!-1}\,&=&~\hat{V}_{r}\,\hat{U}_{r-1} \label{TUT}\\
&=&~\exp\Bigg\{\,\frac{1}{2}\,\sum_{i,j}\Bigg[~\xi^{ss{(r-1)}}_{ij}\,\Big(\hat{\sigma}_{i}^{_{(r-1)}\dagger}\hat{\sigma}_{j}^{_{(r-1)}\dagger}-\hat{\sigma}_{i}^{_{(r-1)}}\hat{\sigma}_{j}^{_{(r-1)}}\Big)\,+\nonumber\\
&&~~~~~~~~~~~~~~~~~~~~\xi^{ww(r-1)}_{ij}\Big(\hat{\omega}_{i}^{_{(r-1)}\dagger}\hat{\omega}_{j}^{_{(r-1)}\dagger}-\hat{\omega}_{i}^{_{(r-1)}}\hat{\omega}_{j}^{_{(r-1)}}\Big)\,+\nonumber\\
&&~~~~~~~~~~~~~~~~~~~~\xi^{sw(r-1)}_{ij}\Big(\hat{\sigma}_{i}^{_{(r-1)}\dagger}\hat{\omega}_{j}^{_{(r-1)}\dagger}-\hat{\sigma}_{i}^{_{(r-1)}}\hat{\omega}_{j}^{_{(r-1)}}\Big)~\Bigg]\,\Bigg\}.\qquad\qquad\nonumber
\end{eqnarray}
Above, the $\xi^{(r-1)}$ exponents are the WTs of the entangling exponent $\zeta^{(r)}$,
\begin{subequations}\label{xiFromFilters}
\begin{alignat}{3}
&\xi^{ss(r-1)}_{ij}~&\equiv&~~\big(\mathbf{L}\,\zeta^{(r)}\mathbf{L}^\mathsf{T}\big)_{ij}\\
&\xi^{ww(r-1)}_{ij}~&\equiv&~~\big(\mathbf{H}\,\zeta^{(r)}\mathbf{H}^\mathsf{T}\big)_{ij}\\
&\xi^{sw(r-1)}_{ij}~&\equiv&~~2\,\big(\mathbf{L}\,\zeta^{(r)}\mathbf{H}^\mathsf{T}\big)_{ij}\,.
\end{alignat}
\end{subequations}
Note that they too have a restricted form due to discrete translational invariance:
\begin{equation}
\xi^{ss(r)}_{ij}\,=~\xi_{|i-j|}^{ss(r)},~~~~~~~~~~~\xi^{ww(r)}_{ij}\,=~\xi_{|i-j|}^{ww(r)},~~~~~~~~~~~\xi^{sw(r)}_{ij}\,=~\xi_{i-j}^{sw(r)}.
\end{equation}
The IWT relating $\zeta^{(r)}$ to $\xi^{(r-1)}$  is given by
\begin{equation}\label{IWTzetaxi}
\zeta^{(r)}_{ij}~=~\big(\mathbf{L}^\mathsf{T}\xi^{ss(r-1)}\mathbf{L}\big)_{ij}\,+\,\big(\mathbf{H}^\mathsf{T}\xi^{ww(r-1)}\mathbf{H}\big)_{ij}\,+\,\frac{1}{2}\big(\mathbf{L}^\mathsf{T}\xi^{sw(r-1)}\mathbf{H}\,+\,\mathbf{H}^\mathsf{T}\xi^{sw(r-1)\mathsf{T}}\mathbf{L}\big)_{ij}\,.
\end{equation}
This expression is key for our algorithm to minimize $\big\langle \hat{H}_\sigma^r \big\rangle$. Specifically, at a fixed wMERA layer with resolution $r$, we treat $\xi_{i-j}^{ww(r)}$ and $\xi_{i-j}^{sw(r)}$ as free parameters, and fix the condition
\begin{equation}\label{xizeta}
\xi^{ss(r-1)}_{ij}~=~\zeta^{(r-1)}_{ij}\!,
\end{equation}
that is, we use as input the numerical values for $\zeta^{(r-1)}$ obtained from the optimization of the previous layer. While in the exact solution the relation \eqref{xizeta} is only approximately true, it is quite accurate for lattice site separations $|i-j|\gtrsim 5$. For shorter lattice site separations $|i-j|\leq 4$, the optimization of the free parameters $\xi^{ww(r)}$ and $\xi^{sw(r)}$ can compensate for any inaccuracy introduced by \eqref{xizeta}.

With the parametrization of the entanglers now fully specified, we can proceed to write the vacuum expectation value to the Hamiltonian in \eqref{HvevUr} in terms of $\zeta^{(r)}_{ij}$. First, we obtain the entangler-transformed Hamiltonian in \eqref{HrSaadagger} (with $\mathcal{V}=0$) using \eqref{UsigmaU}:
\begin{eqnarray}
\hat{U}_r^\dagger \hat{H}_\sigma^r\,\hat{U}_r~=~\,\frac{m_\phi}{2}\,\sum_{i,j}&&\!\!\!\Bigg[\Big(\text{C}_{2\zeta}^{(r)}+2^{(2r-1)}\mathds{K}^{ss}\,\text{E}_{2\zeta}^{(r)}\Big)_{\!ij}\,\Big({\hat{\sigma}_i}^{\,r\,\dagger}\hat{\sigma}_j^{\,r}+\hat{\sigma}_j^{\,r}\,\hat{\sigma}^{\,r\,\dagger}_i\Big)\nonumber\\
&&~~+~\Big(\text{S}_{2\zeta}^{(r)}+2^{(2r-1)}\mathds{K}^{ss}\,\text{E}_{2\zeta}^{(r)}\Big)_{\!ij}\,\Big(\hat{\sigma}^{\,r\,\dagger}_i\hat{\sigma}^{\,r\,\dagger}_{j}+\hat{\sigma}_i^{\,r}\,\hat{\sigma}_{j}^{\,r}\Big)\Bigg],~~~~~~~~~\label{UrHrUr}
\end{eqnarray}
where, above, we have used the short-hand notation
\begin{eqnarray}
\big(\text{C}_{2\zeta}^{(r)}\big)_{ij}&\equiv&\big(\text{Cosh}\,2\zeta^{(r)}\big)_{ij}~~~~~~~~\big(\text{S}_{2\zeta}^{(r)}\big)_{ij}\equiv\big(\text{Sinh}\,2\zeta^{(r)}\big)_{ij}~~~~~~~~\big(\text{E}_{2\zeta}^{(r)}\big)_{ij}\equiv\big(\text{Exp}\,2\zeta^{(r)}\big)_{ij}~~~~~~~~~~~~\label{CoshSinhExpZeta}\\
&=&\big(\text{Cosh}\,2\zeta^{(r)}\big)_{i-j}~~~~~~~~~~~~~~~\;=\big(\text{Sinh}\,2\zeta^{(r)}\big)_{i-j}~~~~~~~~~~~~~~~~=\big(\text{Exp}\,2\zeta^{(r)}\big)_{i-j\,.}\nonumber
\end{eqnarray}
Using \eqref{UrHrUr} and \eqref{vac_k}, we then have
\begin{eqnarray}
{\big\langle}\mathbf{0_\sigma}\big|\,\hat{H}_\sigma^{r}\,\big|{\mathbf{0_\sigma}}\big\rangle_{\!r}~&=&~{\big\langle}\mathbf{P}_{\!\sigma}\big|\,\hat{U}_r^\dagger\, \hat{H}_\sigma^{r}\,\hat{U}_r\,\big|{\mathbf{P}_{\!\sigma}}\big\rangle_{\!r}\nonumber\\
&=&~\frac{m_\phi}{2}\;\text{Tr}\left[\text{C}_{2\zeta}^{(r)}+2^{(2r-1)}\mathds{K}^{ss}\,\text{E}_{2\zeta}^{(r)}\right].\label{UrHrUrZeta}
\end{eqnarray}
As it turns out, the minimization of this vacuum expectation value can be performed analytically. Differentiating \eqref{UrHrUrZeta} with respect $\zeta^{(r)}$ and setting it to zero, we obtain
\begin{equation}\label{Hvanish}
\text{S}_{2\zeta}^{(r)}+2^{(2r-1)}\mathds{K}^{ss}\,\text{E}_{2\zeta}^{(r)}~=~0,
\end{equation}
whose solution is given by
\begin{equation}\label{ExactZetar}
\zeta^{(r)}_{ij}\,=\,-\frac{1}{4}\,\Big(\text{Log}\big[2^{2r}\,\mathds{K}^{ss}+\mathds{1}\big]\Big)_{\!ij}\,.
\end{equation}
However, as previously stated, our actual goal in this demonstration of the wMERA algorithm is to minimize \eqref{UrHrUrZeta} \emph{numerically} using the parametrization in \eqref{IWTzetaxi} and the input in \eqref{xizeta}; we will only use the exact solution for $\zeta^{(r)}_{ij}$ in \eqref{ExactZetar} to validate our numerical results. We can further manipulate \eqref{UrHrUrZeta} using \eqref{CoshSinhExpZeta}, \eqref{Kss}, and \eqref{Kcoeff} to obtain:
\begin{equation}
{\big\langle}\mathbf{0_\sigma}\big|\,\hat{H}_\sigma^{r}\,\big|{\mathbf{0_\sigma}}\big\rangle_{\!r}~=~\frac{m_\phi}{2}\,\sum_{i=\infty}^{\infty}\Bigg[\big(\text{Cosh}\,2\zeta^{(r)}\big)_{\!0}+\,2^{(2r-1)}\!\sum_{j=-4}^{4} K_{|j|}\,\big(\text{Exp}\,2\zeta^{(r)}\big)_{\!j}\Bigg].\label{HrvevAlgo}
\end{equation}
The expression above has two remarkable aspects. First, even though it is a trace over a square matrix of infinite dimension, the infinite summation over one of the indices factors out (this is due to discrete translational invariance). Second, the r.h.s.~of \eqref{HrvevAlgo} depends only on a small, finite number of elements of the matrix exponential of $\zeta^{(r)}$ (this is due to the Daubechies discretization, whose scaling and wavelet functions have compact support in position space). In particular, because of the form of \eqref{HrvevAlgo}, there is no need to perform a finite volume truncation of the system in order to minimize $\big\langle \hat{H}_\sigma^{r} \big\rangle$ with respect to the variational parameters. However, a truncation is still needed in the number of variational parameters, of which there is in principle an infinite number. One may question whether any such truncation would preserve the properties of the ground state. To argue that this is the case, note that the matrix $\zeta^{(r)}_{i-j}$ is the discrete counterpart of the entangler in the continuum, $\zeta(x-y)$, whose exact solution is given in \eqref{zetax}. In particular, $\zeta(x-y)$ becomes exponentially suppressed for separations larger than the correlation length of the system, i.e., for $|x-y|\gtrsim m_\phi^{-1}$. Therefore, we expect that the entries $\zeta^{(r)}_{i-j}$ will likewise be exponentially suppressed for site separations $|i-j|\gtrsim 2^r$.

Recalling that the free parameters of $\zeta^{(r+1)}$ are given by  $\xi^{ww(r)}$ and $\xi^{sw(r)}$ as defined in \eqref{IWTzetaxi}, in practice our truncation of free parameters will amount to setting
\begin{subequations}\label{xiTruncations}
\begin{alignat}{3}
&\xi^{ww(r)}_{i}&=\;&0~~~\text{for}~~|i|>i^{(r)}_\text{max,}\\
&\xi^{sw(r)}_{j}&=\;&0~~~\text{for}~~|j|>j^{(r)}_\text{max,}
\end{alignat}
\end{subequations}
where $i^{(r)}_\text{max}$ and $j^{(r)}_\text{max}$ are finite integers which can be increased to improve numerical accuracy. Remarkably, we will see shortly that $i^{(r)}_\text{max}$ and $j^{(r)}_\text{max}$ do not need to grow with resolution $r$, and can be kept small (i.e., in the range of $\sim$ 3–5) for all wMERA layers while still retaining excellent accuracy\footnote{The parallel of this feature in the original MERA is the fact that the bond dimension of the tensor ansatzes do not need to grow with the resolution of the MERA layer \cite{Vidal:2007hda}.}. This is in contrast with the truncation of entangling parameters $\zeta^{(r)}_{i}$, which must grow with $r$ to allow the same desired degree of accuracy to be retained, as we just argued in the previous paragraph.

\begin{figure}
\centering\includegraphics[width=0.95\textwidth]{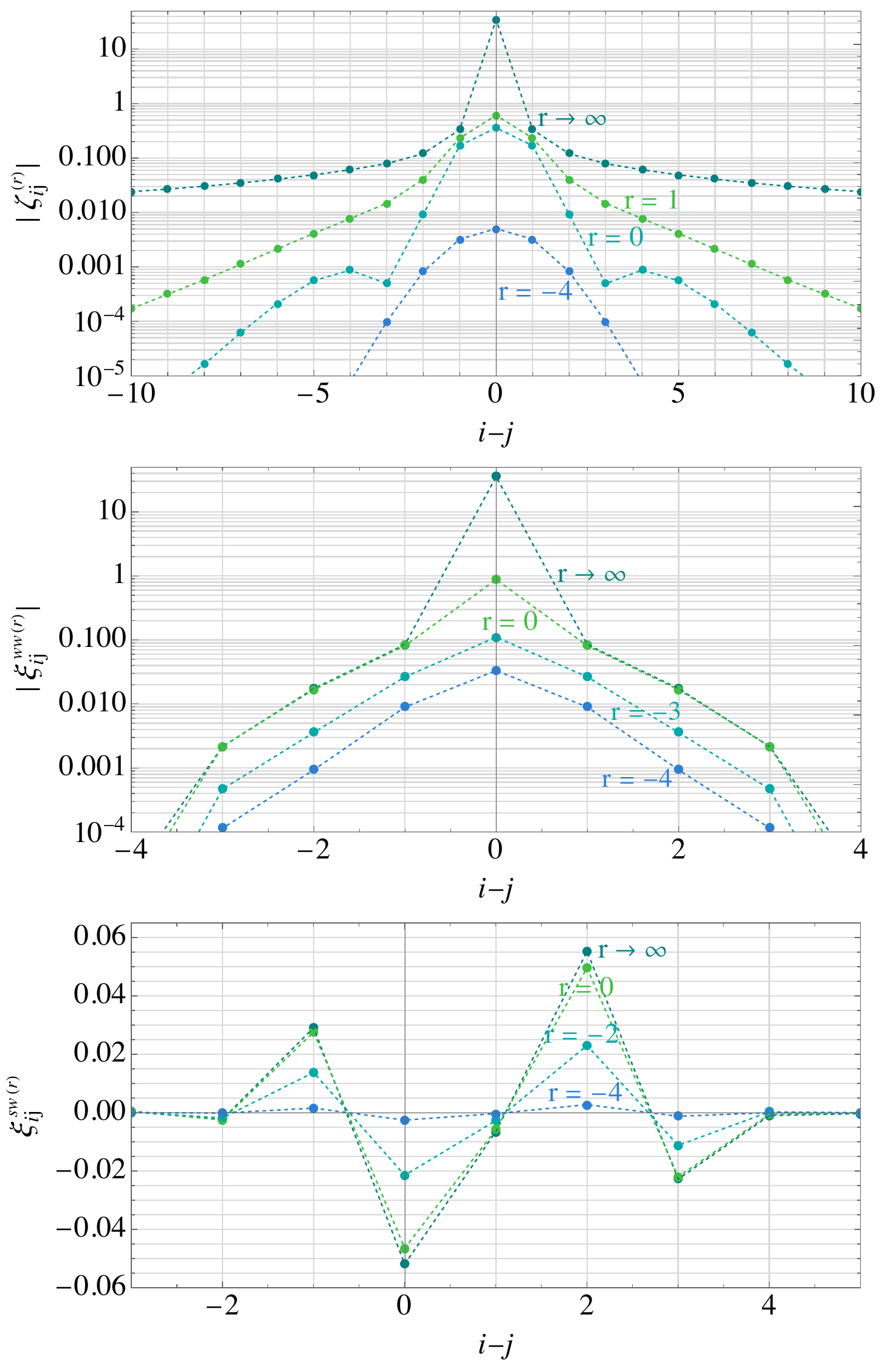}
\caption{The optimized values of the entangling parameters $\zeta^{(r)}_{ij}$,~~$\xi^{ww(r)}_{ij}$, and $\xi^{sw(r)}_{ij}$, defined in \eqref{Urss} and \eqref{xiFromFilters}, for a few illustrative choices of resolution $r$. We recall that the lattice spacing is related to the resolution index $r$ via $a_r=1/(2^r m_\phi)$.}
\label{wMERAentanglers} 
\end{figure}

In appendix~\ref{appendix1}, we discuss a method to obtain the entries of $\big(\text{Exp}\,2\zeta\big)_{\!j}$ as a function of a truncated array $\zeta_i$. This, combined with \eqref{IWTzetaxi}, can plugged into \eqref{HrvevAlgo} for the numerical variational optimization.

The exact solutions for $\zeta^{(r)}_{i-j}$,~~$\xi^{ww(r)}_{i-j}$, and $\xi^{sw(r)}_{i-j}$ are shown in figure \ref{wMERAentanglers} for a few illustrative choices of resolution $r$. The plots clearly show, as previously stated, that the range of the entangling parameters $\xi^{ww(r)}$ and $\xi^{sw(r)}$ does not grow exponentially with increasing resolution. Indeed, their optimal values not only quickly saturate at resolutions close to $r= 0$, but their range decays exponentially with increasing lattice site separations, which justifies the truncations in \eqref{xiTruncations}. This is in stark contrast to the scaling entangling parameters $\zeta^{(r)}$, whose range grows exponentially with resolution.

The results of our numerical wMERA optimization show excellent agreement with the exact solutions within the truncated range of parameters in our algorithm. Specifically, in obtaining our results, we truncated the $\xi^{ww}_{i}$, $\xi^{sw}_{j}$, and $\zeta_{k}$ arrays at $i_\text{max}=3$, $j_\text{max}=4$, and $k_\text{max}=21$, respectively, in all the layers of our calculation, which ranged from $r=-$3 to 10.

\begin{figure}
\centering\includegraphics[width=1.0\textwidth]{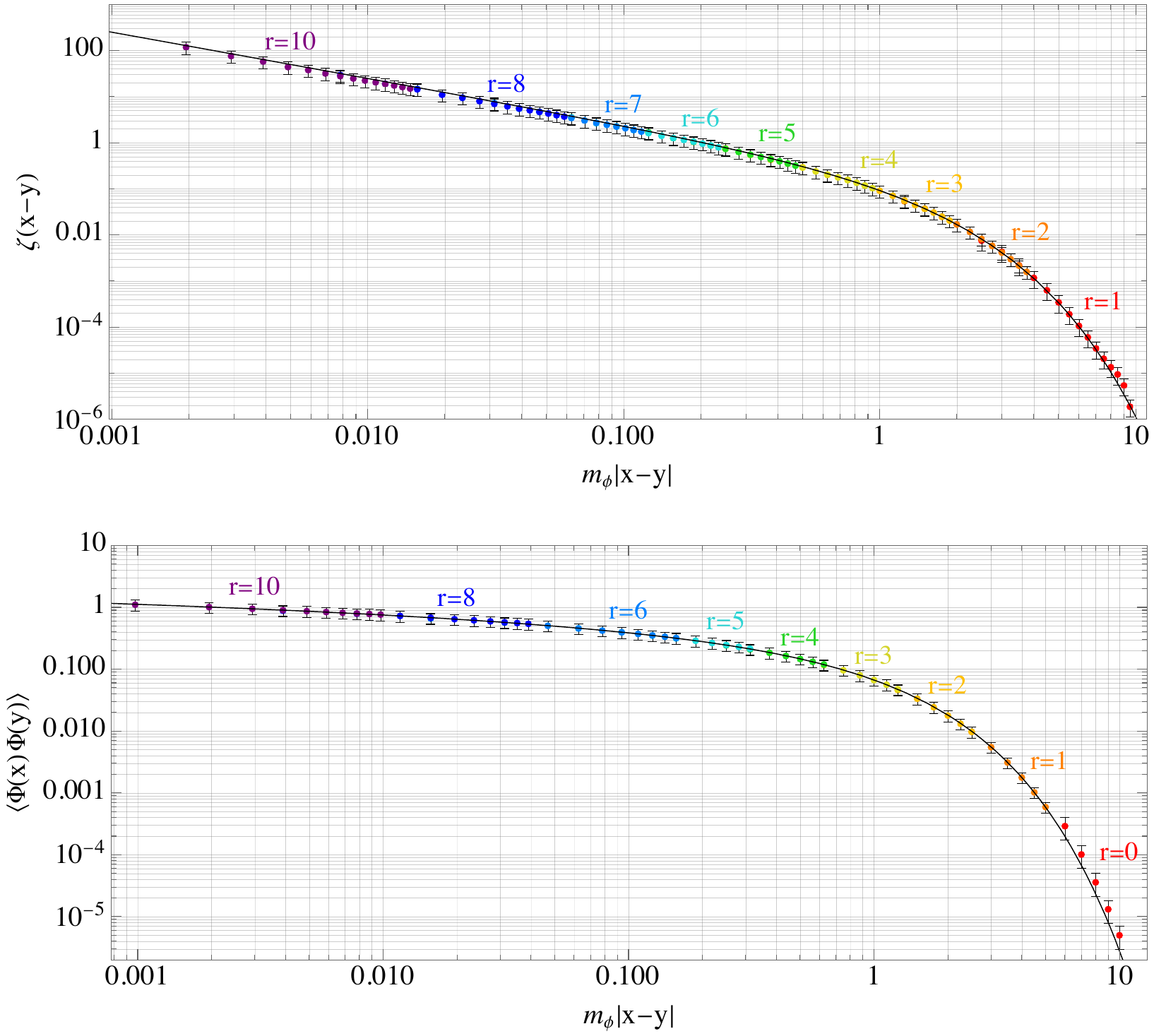}
\caption{\emph{Top:} the reconstructed continuum entangling exponent $\zeta(x-y)$ (see \eqref{Ua(x)}), calculated using the numerical output of our wMERA optimization.  The solid black curve shows the exact continuum solution for $\zeta(x-y)$ (see \eqref{zetax}). \emph{Bottom:} the reconstructed continuum scalar two-point function at equal times, $\big\langle\hat\Phi(x)\hat\Phi(y)\big\rangle$, calculated using the numerical output of our wMERA optimization. The exact continuum solution for $\big\langle\hat\Phi(x)\hat\Phi(y)\big\rangle$ is given by the solid black curve (see \eqref{phiTphiExact}). In both plots, the functions were calculated for different ranges in $m_\phi|x-y|$ using different wMERA layers, indicated by the resolution index $r$. We recall that the lattice spacing is related to the resolution index $r$ via $a_r=1/(2^r m_\phi)$.}
\label{ScalarZetaSpace2PF} 
\end{figure}

For a fixed resolution layer $r$, our numerical results for $\zeta^{(r)}_{ij}$ allow us to recover the continuum result for the entangling exponent $\zeta(x-y)$ restricted to the range $|x-y|\sim\mathcal{O}(1-10)\times(2^r m_\phi)^{-1}$:
\begin{eqnarray}
\zeta^{(r)}(x-y)~~&=&~~\sum_{i,j}\,s^r_i(x)s^r_j(y)\; \zeta^{(r)}_{ij}\\
&=&~~\sum_{i,j} \,m_\phi\;2^{r}\,s\big(2^r m_\phi\, x-i \big)\,s\big(2^r m_\phi\, y-j \big)\;\zeta^{(r)}_{i-j}\,.\nonumber
\end{eqnarray}
Piecing together the functions $\zeta^{(r)}(x-y)$ at different resolutions, we can reconstruct the continuum entangling exponent $\zeta(x-y)$ over several orders of magnitude in $|x-y|$, as shown in figure \ref{ScalarZetaSpace2PF}.

Our results for $\zeta^{(r)}_{ij}$ also allow us to reconstruct the scalar field two-point function in the continuum. For spacelike separations, it suffices to compute the equal-time two-point function; the answer for any other spacelike interval can be obtained by a Lorentz transformation. Recalling the DWT representation of the scalar field $\hat\Phi^r(x)$ at a fixed resolution $r$ in \eqref{phiDecr1}, we have:
\begin{eqnarray}\label{phiphiwMERA}
\big\langle\hat\Phi^r(x)\hat\Phi^r(y)\big\rangle~~&=&~~\sum_{i,j} \,s^r_i(x)s^r_j(y)~{\big\langle}\mathbf{0_\sigma}\big|\,\hat{\phi}_i^{r}\,\hat{\phi}_j^{r}\,\big|{\mathbf{0_\sigma}}\big\rangle_{\!r}\\
&=&~~\sum_{i,j} \,s^r_i(x)s^r_j(y)~{\big\langle}\mathbf{P_\sigma}\big|\big(\hat{U}_{r\,}^\dagger\hat{\phi}_i^{r}\,\hat{U}_r\big)\,\big(\hat{U}_{r\,}^\dagger\hat{\phi}_j^{r}\,\hat{U}_r\big)\big|{\mathbf{P_\sigma}}\big\rangle_{\!r}\nonumber\\
&=&~~\sum_{i,j} \,2^{(r-1)}\,s\big(2^r m_\phi\, x-i \big)\,s\big(2^r m_\phi\, y-j \big)\;\big(\text{Exp}\,2\zeta^{(r)}\big)_{i-j\,,}\nonumber
\end{eqnarray}
where, in the last equality, we used \eqref{snrNormM}, \eqref{Saadagger}, \eqref{vac_k}, and \eqref{UsigmaU}.

Similarly to the case of $\zeta^{(r)}(x-y)$, the two-point function in \eqref{phiphiwMERA} is restricted to the range $|x-y|\sim\mathcal{O}(1-10)\times(2^r m_\phi)^{-1}$. Piecing together $\big\langle\hat\Phi^r(x)\hat\Phi^r(y)\big\rangle$ at different resolutions, we recover the continuum result for $\big\langle\hat\Phi(x)\hat\Phi(y)\big\rangle$ over several orders of magnitude in $|x-y|$, as shown in figure \ref{ScalarZetaSpace2PF}. 

\begin{figure}
\centering\includegraphics[width=0.99\textwidth]{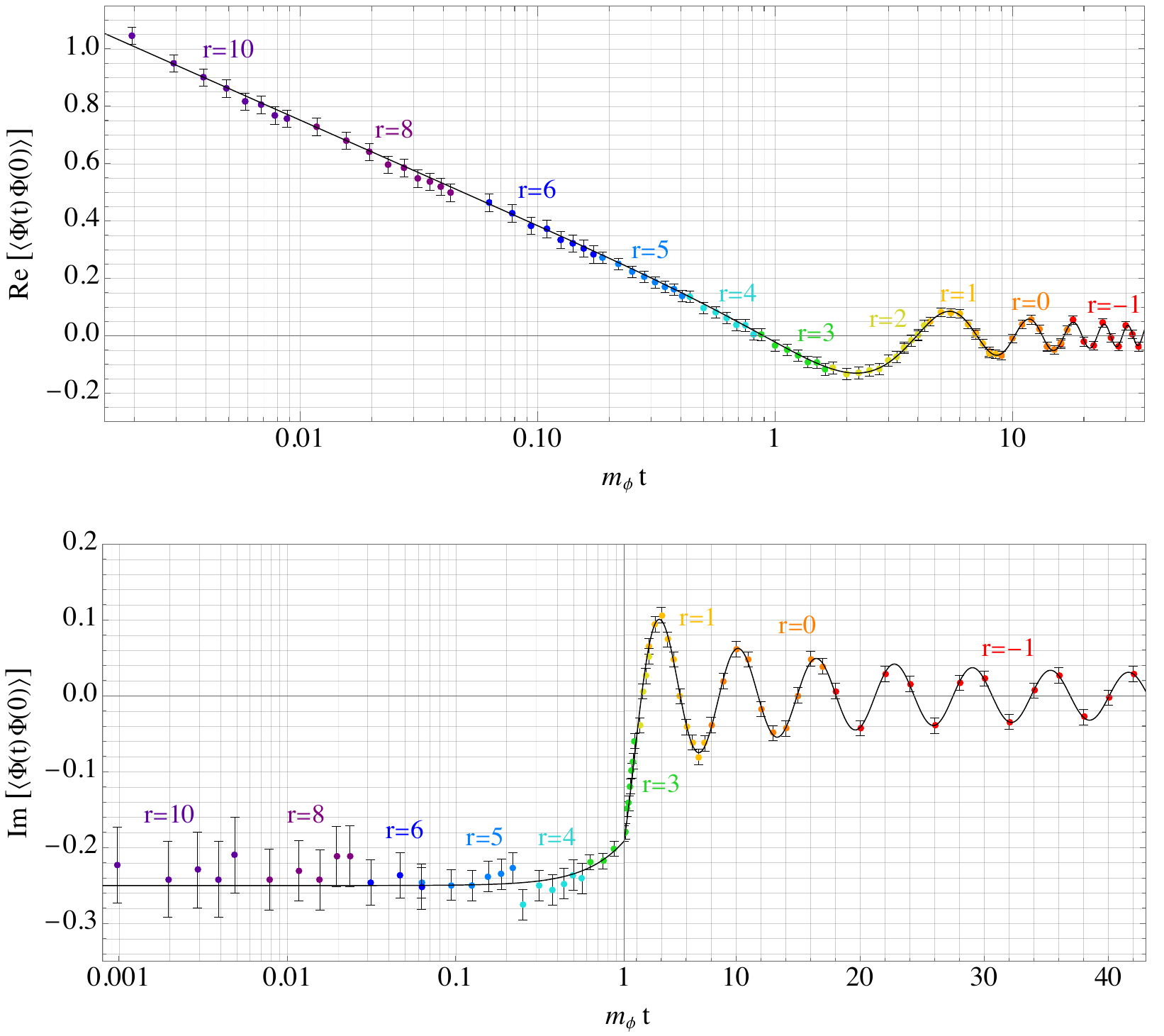}
\caption{The reconstructed continuum scalar two-point function for timelike separations, $\big\langle\hat\Phi(x,t)\hat\Phi(x,0)\big\rangle$, calculated using the numerical output of our wMERA optimization. The real and imaginary parts are shown in the top and bottom plots, respectively. The apparent derivative discontinuity at $m_{\phi\,}t=1$ in the bottom plot is an artifact of the hybrid scale of the horizontal axis: note that the scale is logarithmic for $m_{\phi\,}t<1$, and linear for $m_{\phi\,}t>1$. This choice was made so that the comparison between the numerical results and the exact continuum solution (solid black curves) is displayed with better clarity. In both plots, the functions were calculated for different ranges in $m_{\phi\,}t$ using different wMERA layers, indicated by the resolution index $r$. We recall that the lattice spacing is related to the resolution index $r$ via $a_r=1/(2^r m_\phi)$.}
\label{ScalarTime2PF} 
\end{figure}

Our final task in solving this free scalar theory with wMERA is to obtain the scalar two-point function at finite-time separations. This requires implementing real-time evolution, which is also straightforward in the wMERA framework. The first step is to obtain the entangler-transformed Hamiltonians at different resolutions using the variational solution for the entangler parameters. Plugging
\eqref{ExactZetar} into \eqref{UrHrUr}, we obtain:
\begin{equation}\label{UrHrUrOpt}
\hat{U}_r^\dagger \hat{H}_\sigma^r\,\hat{U}_r~=~m_\phi\,\sum_{i,j}\;{\hat{\sigma}_i}^{\,r\,\dagger\,}\Omega^{(r)}_{ij}\,\hat{\sigma}_j^{\,r}~+~\text{(zero-point energy)}
\end{equation}
where
\begin{equation}\label{Omegaij}
\Omega^{(r)}_{ij}~\equiv~\Big(e^{-2\zeta^{(r)}}\Big)_{ij\,.}
\end{equation}
We can also define the entangler-transformed time-evolution operator\footnote{On the r.h.s.~of \eqref{UrUtUr}, we omitted a constant phase stemming from the zero-point energy.}:
\begin{equation}\label{UrUtUr}
\hat{\mathds{U}}_r(t)~\equiv~\hat{U}_r^\dagger\Big(e^{-i\,t\,\hat{H}_\sigma^r}\Big)\hat{U}_r~=~\text{Exp}\bigg[\!\!-i\,(m_\phi\,t)\sum_{i,j}{\hat{\sigma}_i}^{\,r\,\dagger\,}\Omega^{(r)}_{ij}\,\hat{\sigma}_j^{\,r}\bigg].
\end{equation}
Its action on the scaling modes at resolution $r$ is easily derived:
\begin{subequations}\label{UtsigmaUt}
\begin{alignat}{3}
&\,\hat{\mathds{U}}_r^{\dagger}(t)\, \hat{\sigma}_i^{\,r}\, \hat{\mathds{U}}_r(t)&~=~&\big(e^{-i\,(m_\phi t)\,\Omega^{(r)}}\big)_{ij}\,\hat{\sigma}_j^{\,r}\\
&\hat{\mathds{U}}_r^{\dagger}(t)\, \hat{\sigma}_i^{\,r\dagger}\; \hat{\mathds{U}}_r(t)&~=~&\big(e^{+i\,(m_\phi t)\,\Omega^{(r)}}\big)_{ij}\,\hat{\sigma}_j^{\,r\,\dagger}.
\end{alignat}
\end{subequations}
We now have all the ingredients to obtain the scalar field two-point function at resolution $r$ for finite-time separations :
\begin{eqnarray}\label{phiTphi}
\big\langle\hat\Phi^r(x,t)\hat\Phi^r(y,0)\big\rangle~&=&~{\big\langle}\mathbf{0_\sigma}\big|\big(e^{i\,t\,\hat{H}_\sigma^r}\big)\hat\Phi^r(x)\big(e^{-i\,t\,\hat{H}_\sigma^r}\big)\hat\Phi^r(y)\big|{\mathbf{0_\sigma}}\big\rangle_{\!r}\\
\nonumber\\
&=&~\sum_{i,j} \,s^r_i(x)s^r_j(y)~{\big\langle}\mathbf{P_\sigma}\big|\hat{\mathds{U}}_r^{\dagger}(t)\,\big(\hat{U}_{r\,}^\dagger\hat{\phi}_i^{r}\,\hat{U}_r\big)\,\hat{\mathds{U}}_r(t)\,\big(\hat{U}_{r\,}^\dagger\hat{\phi}_j^{r}\,\hat{U}_r\big)\big|{\mathbf{P_\sigma}}\big\rangle_{\!r}\nonumber\\
&=&~\sum_{i,j} \,2^{(r-1)}\,s\big(2^r m_\phi\, x-i \big)\,s\big(2^r m_\phi\, y-j\big)\Big(e^{2\zeta^{(r)}-i\,(m_\phi t)\,\Omega^{(r)}}\Big)_{i-j\,.}\nonumber
\end{eqnarray}
The expression above, which generalizes \eqref{phiphiwMERA}, is restricted to the range $\sqrt{|t^2-(x-y)^2|}\sim\mathcal{O}(1-10)\times(2^r m_\phi)^{-1}$. Piecing together $\big\langle\hat\Phi^r(x,\Delta t)\hat\Phi^r(x,0)\big\rangle$ at different resolutions, we recover the continuum result for $\big\langle\hat\Phi(x,\Delta t)\hat\Phi(x,0)\big\rangle$ at timelike separations over several orders of magnitude in $\Delta t$, as shown in figure \ref{ScalarTime2PF}.

It is important to note that our wMERA results for $\big\langle\hat\Phi^r(x,t)\hat\Phi^r(y,0)\big\rangle$ should track the continuum solution over scales longer than the resolution $r$, but otherwise show some degree of unphysical structure at scales of order $|x-y| \sim (2^r m_\phi)^{-1}$. Indeed, in figures \ref{ScalarZetaSpace2PF} and \ref{ScalarTime2PF}, we chose to plot our results for a discrete array of points $\{x_i,y_i\}$ satisfying
\begin{equation}
\begin{cases}
\;2^r m_\phi\,x_i\,\equiv\,\text{floor}(2^r m_\phi\,x_i)+\epsilon\\
\;2^r m_\phi\,y_i\;\equiv\,\text{floor}(2^r m_\phi\,y_i)+\epsilon
\end{cases}\!\!\!\!,~~~~~~~~2^r m_\phi (x_i-y_i)~\equiv~ \delta_{xy} \in \mathbb{Z}.
\end{equation}
With these definitions, we can rewrite \eqref{phiTphi} as:
\begin{eqnarray}\label{phiTphiWolf}
\big\langle\hat\Phi^r(x_i,t)\hat\Phi^r(y_i,0)\big\rangle~&=&~\sum_{j,k} \,2^{(r-1)}\,s\big(\delta_{xy}-k-j+\epsilon \big)\,s\big(j+\epsilon\big)\Big(e^{2\zeta^{(r)}-i\,(m_\phi t)\,\Omega^{(r)}}\Big)_{\!k\,.}\nonumber
\end{eqnarray}
The unphysical structure of this reconstructed two-point function at scales $|x_i-y_i| \sim (2^r m_\phi)^{-1}$ can be seen by varying the parameter $\epsilon$ above over its range of $0\leq\epsilon<1$ (in other words, varying $\epsilon$ will probe the lack of smoothness of the D6 scaling function). Hence, for every plot in this paper\footnote{Namely, those in figures \ref{ScalarZetaSpace2PF}, \ref{ScalarTime2PF}, \ref{FermionSpace2PF}, \ref{FermionTimeChiChi}, and \ref{FermionTimePsiPsi}.} showing continuum functions reconstructed from numerical wMERA results, we select the value of $\epsilon$ that minimizes these unphysical fluctuations.

We can compare our numerical wMERA results for the scalar two-point function with the exact analytical expression for the Feynman propagator in position space,
given by:
\begin{equation}\label{phiTphiExact}
\big\langle\hat\Phi(x,t)\hat\Phi(0,0)\big\rangle~=~\Theta\big(s^2\big)\,\frac{(-i)}{4}\,H_0^{(2)}\big(m_\phi\sqrt{s^2}\big)~+~\Theta\big(\!\!-\!s^2\big)\,\frac{1}{2\pi}\,K_0\big(m_\phi\sqrt{-s^2}\big),
\end{equation}
where $t>0$, $\Theta$ is the Heaviside step function, $H_0^{(2)}$ is a Hankel function of the second kind, $K_0$ is a modified Bessel function of the second kind, and $s$ is the spacetime interval, i.e., $s^2\equiv t^2-x^2$.

As shown in figures \ref{ScalarZetaSpace2PF} and \ref{ScalarTime2PF}, the agreement between our numerical results and the exact continuum solution is quite satisfactory considering our aggressive truncation choices. That said, one can see that the accuracy and precision of our results slowly degrade with increasing $r$. This is not a shortcoming of wMERA itself, but an expected artifact of the approximations made in our numerical code. Specifically: (i) we kept the $\zeta^{(r)}_{k}$ array truncation fixed at $k^{(r)}_\text{max}=21$ for all layers; (ii) we restricted the order of the power expansion of the matrix exponential $\text{Exp}\,2\zeta^{(r)}$ in the $\zeta^{(r)}_{|k|}$ coefficients to 3 (see appendix~\ref{appendix1} for details); and (iii) we truncated the $\big(\text{Exp}\,2\zeta^{(r)}\big)_{l}$ array to $l^{(r)}_\text{max}=36$ for all layers.  These choices, which were not optimized, were made so we could run our code on a personal laptop with a dual Intel i5 core processor in a couple of hours (the code was written in Wolfram Mathematica \cite{Mathematica}). However, with more computational resources, one can relax the truncation choices to overcome the degradation in accuracy and precision with increasing resolution.

\subsection{Free theory of a Dirac fermion in (1+1)d}
\label{FermionSection}

Our second and final wMERA example is a free theory of a Dirac fermion in (1+1) spacetime dimensions. Many steps of this exercise (and their justifications) have a strong parallel with the scalar field theory case just considered. Therefore, in this section, we will skip over details that have already been dissected in the previous section.

We start by laying out our notation. We choose conventions such that the Dirac spinor and $\gamma$-matrices are given by:
\begin{equation}\label{spinorConvention}
\hat\Psi=\begin{pmatrix}
\!\hat\chi\\
\,\hat\psi^\dagger
\end{pmatrix}~,~~~~~~~~~~~
\gamma^0=\begin{pmatrix}
\,1 & ~\,0\,\\
\,0&-1\,
\end{pmatrix}~,~~~~~~~~~~~
\gamma^1=\begin{pmatrix}
\!~\,0 & ~1\,\\
\!-1& ~0\,
\end{pmatrix}.
\end{equation}
The system's dynamics is determined by the Hamiltonian
\begin{eqnarray}\label{FermionH(x)}
\hat{H}~&=&~\int dx\,\Big[\!-i\,\hat{\overline{\Psi}}(x)\gamma^1\partial_x\hat\Psi(x)\;+\;m_\psi\,\hat{\overline{\Psi}}(x)\hat\Psi(x)\Big]\\
&=&~\int dx\,\Big[\!-i\,\hat\chi^\dagger(x)\,\partial_x\hat\psi^\dagger(x)-i\,\hat\psi^\dagger(x)\,\partial_x\hat\chi^\dagger(x)+m_\psi\,\big(\hat\chi^\dagger(x)\hat\chi(x)-\hat\psi(x)\hat\psi^\dagger(x)\big)\Big]\nonumber
\end{eqnarray}
and equal-time anticommutation relations
\begin{equation}\label{CommutatorChiPsi}
\{\hat\chi(x),\,\hat\chi^\dagger(y)\}=\delta(x-y),~~~~~~~~~~~\{\hat\psi(x),\,\hat\psi^\dagger(y)\}=\delta(x-y);
\end{equation}
all other equal-time anticommutators vanishing.

We define the initial product state upon which the wMERA will be built as the unentangled state in position space that is annihilated by $\hat\chi$ and $\hat\psi$:
\begin{equation}\label{chipsiP}
\hat\chi(x)\,\big|\mathbf{P}\big\rangle\,=\;\hat\psi(x)\,\big|\mathbf{P}\big\rangle~=~0~~~~\forall~~~x.
\end{equation}
Our goal is to find the entangler $\hat{U}$ that relates the product state $\big|{\mathbf{P}\big\rangle}$ to the Hamiltonian ground state $\big|{\mathbf{0}\big\rangle}$:
\begin{equation}
\big|{\mathbf{0}\big\rangle}~=~\hat{U}\big|{\mathbf{P}\big\rangle}.
\end{equation}
The ansatz for $\hat{U}$ is a Gaussian operator (serving the analogous role for fermions that the generator of Bogoliubov transformations does for scalars \cite{Khanna_2007}). Its parametric form in generic ($d$+1)-spacetime dimensions (with $d$ odd) is given by:
\begin{equation}\label{UPsi(x)}
\hat{U}~=~\exp\left[-i\int dx^d\,dy^d~\,\zeta(|\vec{r}_{\,}|)~\,\hat{\Psi}^{\dagger\!}(\vec{x}_{\,})\,(\hat{r}\cdot\vec{\gamma})\,\hat\Psi(\vec{y}_{\,})\,\right]\!,
\end{equation}
where $\vec{r}\equiv\vec{x}-\vec{y}$.
For $d$\,=\,1 and in terms of the Dirac spinor components in \eqref{spinorConvention}, \eqref{UPsi(x)} becomes
\begin{equation}\label{Uchipsi(x)}
\boxed{
~\hat{U}~=~\exp\left[-i\int dx\,dy~\zeta(x-y)~\Big(\hat{\chi}^\dagger(x)\hat{\psi}^\dagger(y)+\hat{\psi}(y)\hat{\chi}(x)\Big)\,\right]\,
}.
\end{equation}
A key difference compared to the scalar case is that, for fermions, the entangling exponent in \eqref{Uchipsi(x)} is anti-symmetric, that is, $\zeta(y-x)=-\zeta(x-y)$.

To proceed with the variational optimization of the entangler, we go to Fourier space. Defining
\begin{subequations}\label{chikpsik}
\begin{alignat}{3}
&\hat\chi_k&\;\equiv\,\int dx~e^{-ikx}\,\hat\chi(x)~,~~~~~~~&\hat\psi_k\;\equiv\,\int dx~e^{-ikx}\,\hat\psi(x)\\
&\hat\chi^\dagger_k&\;\equiv\,\int dx~e^{ikx}\,\hat\chi^\dagger(x)~,~~~~~~~&\hat\psi^\dagger_k\;\equiv\,\int dx~e^{ikx}\,\hat\psi^\dagger(x)
\end{alignat}
\end{subequations}
and
\begin{equation}\label{zetaFourierFermion}
\zeta_k~\equiv~i\!\int dx~ e^{-ikx}\,\zeta(x),
\end{equation}
we can write the Hamiltonian as:
\begin{equation}
\hat{H}~=~\int \frac{dk}{2\pi}~~\Big[k\,\big(\hat\chi^\dagger_k\,\hat\psi^\dagger_{-k}+\hat\psi_{-k}\,\hat\chi_k\big)\;+\;m_\psi\,\big(\hat\chi^\dagger_k\hat\chi_k\,-\,\hat\psi_k\hat\psi^\dagger_k\big)\Big].
\label{Hchipsi(k)}
\end{equation}
The entangler in Fourier space is given by:
\begin{equation}
\hat{U}~=~\exp\left[-\,\int \frac{dk}{2\pi}~\zeta_k~\Big(\hat{\chi}^\dagger_k\,\hat{\psi}^\dagger_{-k}-\hat{\psi}_{-k}\,\hat{\chi}_{k}\Big)\,\right],
\label{Uchipsi(k)}
\end{equation}
where, above, we used the fact that $\zeta_{-k}=-\,\zeta_{\,k}$.
The action of the entangler on the spinor components is easily derived\footnote{Recalling that, in Fourier space, the anticommutation relations are given by $\big\{\hat \chi_k,\,\hat \chi_q^\dagger\big\}=2\pi\,\delta(k-q)$ and $\big\{\hat \psi_k,\,\hat \psi_q^\dagger\big\}=2\pi\,\delta(k-q)$.}:
\begin{subequations}\label{UchipsiU}
\begin{alignat}{3}
&\,~\hat{U}^{\dagger\,} \hat{\chi}_k\, \hat{U}&~=~&\text{cos}(\zeta_k)\,\hat{\chi}_k~-~\text{sin}(\zeta_k)\,\hat{\psi}_{-k}^\dagger\\
&\hat{U}^{\dagger\,} \hat{\psi}_{-k}^\dagger\, \hat{U}&~=~&\text{sin}(\zeta_k)\,\hat{\chi}_{k}~+~\text{cos}(\zeta_k)\,\hat{\psi}_{-k}^\dagger.
\end{alignat}
\end{subequations}
With \eqref{Hchipsi(k)} and  \eqref{UchipsiU}, we obtain the entangler-transformed Hamiltonian,
\begin{eqnarray}
\hat{U}^\dagger \hat{H}\,\hat{U}~=~\int \frac{dk}{2\pi}&&\Bigg[\Big(k\,\text{cos}(2\zeta_k)-m_\psi\,\text{sin}(2\zeta_k)\Big)\,\Big({\hat{\chi}_k}^\dagger\,\hat{\psi}^\dagger_{-k}+\hat{\psi}_{-k}\,\hat{\chi}_k\Big)\label{UHUfermion}\\
&&~~+~\Big(k\,\text{sin}(2\zeta_k)+m_\psi\,\text{cos}(2\zeta_k)\Big)\,\Big(\hat{\chi}^\dagger_k\,\hat{\chi}_{k}-\hat{\psi}_{k}\,\hat{\psi}_{k}^\dagger\Big)\Bigg].\nonumber
\end{eqnarray}
Using \eqref{UHUfermion} and \eqref{chipsiP}, we then obtain the vacuum expectation value of the Hamiltonian,
\begin{eqnarray}
\big\langle\mathbf{0}\big|\hat{H}\big|\mathbf{0}\big\rangle~&=&~{\big\langle}\mathbf{P}\big|\hat{U}^\dagger \hat{H}\,\hat{U}\big|\mathbf{P}\big\rangle\nonumber\\
&=&\,-\int dk\,\,\delta(k-k)\,\Big(k\,\text{sin}(2\zeta_k)+m_\psi\,\text{cos}(2\zeta_k)\Big).
\end{eqnarray}
Minimizing $\langle\hat{H}\rangle$ with respect to entangler exponent $\zeta_k$, we get:
\begin{eqnarray}
\frac{\partial}{\partial \zeta_k}\big\langle\mathbf{0}\big|\hat{H}\big|\mathbf{0}\big\rangle=0~~~~&\Rightarrow&~~~~k\,\text{cos}(2\zeta_k)-m_\psi\,\text{sin}(2\zeta_k)~=~0\nonumber\\
&\Rightarrow&~~~~~~~\boxed{~\zeta_k\,=\,\frac{1}{2}\,\text{arctan}\bigg(\frac{k}{m_\psi}\bigg)\,}\,.\label{zetakFermion}
\end{eqnarray}
Using \eqref{zetaFourierFermion}, we finally obtain the fermionic entangler exponent in position space,
\begin{equation}
\boxed{~
\zeta(x-y)~=~\frac{\,e^{-m_\psi|x-y|}\,}{4(x-y)}
\;}\,.
\label{zetaxFermion}
\end{equation}

Finally, we can plug in the exact entangler expression into \eqref{UHUfermion} to write the entangler-transformed Hamiltonian as
\begin{subequations}
\begin{alignat}{3}
&\hat{U}^\dagger \hat{H}\,\hat{U}~~&=&~\int \frac{dk}{2\pi}\,\omega_k\,\Big(\hat{\chi}^\dagger_k\,\hat{\chi}_{k}-\hat{\psi}_{k}\,\hat{\psi}_{k}^\dagger\Big),\\
&&=&~\int dx\,dy~\,\Omega(x-y)\,\Big(\hat{\chi}^\dagger(x)\,\hat{\chi}(y)-\hat{\psi}(x)\,\hat{\psi}^\dagger(y)\Big)\\
&&=&~\int dx\,dy~\,\hat{\overline\Psi}(x)\Omega(x-y)\,\hat{\Psi}(y).
\end{alignat}
\end{subequations}
where $\Omega(x-y)$ has already been defined in \eqref{Hkernel}.

We now move on to the discretized theory to build the wMERA. Using the Daubechies D6 wavelet basis\footnote{In this section, we will assume that the reference lattice spacing at $r=0$ is given by $a_0=1/m_\psi$. Therefore, our definition of scaling functions $s^r_i(x)$ in \eqref{snrNormM} should be modified by the replacement $m_\phi\to m_\psi$.} to smear the fermion fields, we define:
\begin{equation}
\hat{\Psi}_i^r~=~\begin{pmatrix}
\!\hat\chi_i^r\\
\,\hat\psi_i^{r\,\dagger}
\end{pmatrix}~\equiv~\int dx~s_i^r(x)
\begin{pmatrix}
\hat\chi(x)\\
\hat\psi^\dagger(x)
\end{pmatrix}\!,
\end{equation}
so that the fermionic field representation at resolution $r$ is given by\footnote{Fermionic \emph{wavelet} modes can be analogously defined. However, similarly to the scalar field example, we will variationally optimize each wMERA after the IWT is performed, and therefore will only be working with fermionic \emph{scaling} modes.}:
\begin{equation}\label{chipsi_i}
\hat{\Psi}^r(x)~=~\begin{pmatrix}
~\hat\chi^r(x)\\
\,\hat\psi^{r\,\dagger}(x)
\end{pmatrix}~\equiv~\sum_{i}~s_i^r(x)
\begin{pmatrix}
\!\hat\chi^r_i\\
\,\hat\psi^{r\,\dagger}_i
\end{pmatrix}\!.
\end{equation}
Plugging \eqref{chipsi_i} into \eqref{FermionH(x)}, we obtain the discretized Hamiltonian at resolution $r$:
\begin{equation}\label{HiFermion}
\hat{H}^r~=~m_\psi\sum_{i}\bigg[\hat\chi^{r\,\dagger}_i\hat\chi^{r}_i\,-\,\hat\psi^{r\,}_{i}\hat\psi^{r\,\dagger}_i-\,i\,2^r\sum_j \Big(\hat\chi^{r\,\dagger}_i\,\mathds{D}^{ss}_{ij}\,\hat\psi^{r\,\dagger}_j\,+\,\hat\psi^{r}_i\,\mathds{D}^{ss}_{ij}\,\hat\chi^{r}_j\Big)\bigg],
\end{equation}
where the kinetic term coefficients $\mathds{D}^{ss}_{ij}$ have been defined in \eqref{Dss} and \eqref{Dcoeff}.

The ansatz for the unitary entangler that transforms the initial product state into the ground state of the discretized Hamiltonian in \eqref{HiFermion} can be written as:
\begin{equation}\label{UrF}
\hat{U}_{r}~=~\hat{R}_r(\theta)\;\exp\!\left[-\,i~\sum_{i,j}\,\zeta_{ij}^{(r)}\;\Big(\hat{\chi}_{i}^{\,r\,\dagger}\hat{\psi}_{j}^{\,r\,\dagger}+\,\hat{\psi}_{j}^{\,r}\hat{\chi}_{i}^{\,r}\Big)\,\right],
\end{equation}
where entangling exponent ${\zeta}^{(r)}$ is an \emph{anti}-Hermitian matrix, and the operator $\hat{R}_r(\theta)$, with a single additional variational parameter $\theta$, is given by:
\begin{equation}\label{Rr}
\hat{R}_{r}(\theta)~=~\exp\!\left[-\,\theta~\sum_{i}\,\Big(\hat{\chi}_{i}^{\,r\,\dagger}\hat{\psi}_{i}^{\,r\,\dagger}-\,\hat{\psi}_{i}^{\,r}\hat{\chi}_{i}^{\,r}\Big)\,\right].
\end{equation}
This operator mixes the two components of the Dirac spinor \emph{locally}. It turns out that, since we are working with the Dirac representation of $\gamma$-matrices, $\theta=0$. Had we chosen to work with the Weyl representation, we would find that $\theta=\pi/4$. For simplicity, in our wMERA exercise we will set $\theta=0$ from the start, although including $\theta$ as a ansatz parameter in our wMERA optimization would have been straightforward and would not have affected the final results.

Setting $\hat{R}_r(\theta)=\mathds{1}$ in \eqref{UrF}, the action of the entangler $\hat{U}_{r}$ on the fermionic scaling modes is given by:
\begin{subequations}\label{UsigmaUFermion}
\begin{alignat}{3}
&~\hat{U}_r^{\dagger}\, \hat{\chi}_i^{\,r}\, \hat{U}_r&~=~&\big(\text{Cosh}\zeta^{(r)}\big)_{ij}~\hat{\chi}_j^{\,r}~\,-~i\,\big(\text{Sinh}\zeta^{(r)}\big)_{ij}\,\hat{\psi}_{j}^{\,r\,\dagger}\\
&\hat{U}_r^{\dagger}\, \hat{\psi}_i^{\,r\,\dagger}\, \hat{U}_r&~=~&\big(\text{Cosh}\zeta^{(r)}\big)_{ij}\;\hat{\psi}_j^{\,r\,\dagger}+~i\,\big(\text{Sinh}\zeta^{(r)}\big)_{ij}\;\hat{\chi}_{j}^{\,r}.
\end{alignat}
\end{subequations}
As with the scalar example in subsection \ref{ScalarSection}, given a wMERA layer $r$, our free variational parameters $\xi^{ww(r-1)}$ and $\xi^{sw(r-1)}$ will be embedded in $\zeta^{(r)}$ via an IWT: 
\begin{equation}\label{IWTzetaxiFermion}
\zeta^{(r)}_{ij}~=~\big(\mathbf{L}^\mathsf{T}\xi^{ss(r-1)}\mathbf{L}\big)_{ij}\,+\,\big(\mathbf{H}^\mathsf{T}\xi^{ww(r-1)}\mathbf{H}\big)_{ij}\,+\,\frac{1}{2}\big(\mathbf{L}^\mathsf{T}\xi^{sw(r-1)}\mathbf{H}\,-\,\mathbf{H}^\mathsf{T}\xi^{sw(r-1)\mathsf{T}}\mathbf{L}\big)_{ij},
\end{equation}
while $\xi^{ss(r-1)}$ will be fixed by the result of the optimization of the previous layer through
\begin{equation}
\xi^{ss(r-1)}_{ij}~=~\zeta^{(r-1)}_{ij}.
\end{equation}
Similarly to the scalar example, the entangling matrices $\zeta^{(r)}$, $\xi^{ww(r)}$ and $\xi^{sw(r)}$ have their forms restricted due to discrete translational invariance. However, unlike the scalar example, $\zeta^{(r)}$ and $\xi^{ww(r)}$ are \emph{anti}-Hermitian:
\begin{equation}
\zeta_{ij}^{(r)}=-\zeta_{ji}^{(r)}=\,\zeta_{i-j}^{(r)},~~~~~~~~~\xi^{ww(r)}_{ij}=-\xi_{ji}^{ww(r)}=\,\xi_{i-j}^{ww(r)},~~~~~~~~~\xi^{sw(r)}_{ij}=\;\xi_{i-j}^{sw(r)}.
\end{equation}
We now proceed to obtain the entangler-transformed Hamiltonian using \eqref{UsigmaUFermion}:
\begin{eqnarray}
\hat{U}_r^\dagger \,\hat{H}^r\,\hat{U}_r~=~m_\psi\,\sum_{i,j}&&\!\!\!\Bigg[\Big(\text{C}_{2\zeta}^{(r)}+2^{r}\,\mathds{D}^{ss}\,\text{S}_{2\zeta}^{(r)}\Big)_{\!ij}\,\Big({\hat{\chi}_i}^{\,r\,\dagger}\hat{\chi}_j^{\,r}-\hat{\psi}_i^{\,r}\,\hat{\psi}^{\,r\,\dagger}_j\Big)\nonumber\\
&&~-i~\Big(\text{S}_{2\zeta}^{(r)}+2^{r}\,\mathds{D}^{ss}\,\text{C}_{2\zeta}^{(r)}\Big)_{\!ij}\,\Big({\hat{\chi}_i}^{\,r\,\dagger}\hat{\psi}_j^{\,r\,\dagger}+\hat{\psi}_i^{\,r}\,\hat{\chi}^{\,r}_j\Big)
\Bigg],~~~~~~~~~\label{UrHrUrFermion}
\end{eqnarray}
where we have used the short-hand notation defined in \eqref{CoshSinhExpZeta}. The vacuum expectation value of the Hamiltonian is then given by:
\begin{eqnarray}
{\big\langle}\mathbf{0}\big|\hat{H}^{r}\big|{\mathbf{0}}\big\rangle_{\!r}~&=&~{\big\langle}\mathbf{P}\big|\,\hat{U}_r^\dagger\, \hat{H}^{r}\,\hat{U}_r\,\big|{\mathbf{P}}\big\rangle_{\!r}\nonumber\\
&=&-\,m_\psi\;\text{Tr}\left[\text{C}_{2\zeta}^{(r)}+2^{r}\,\mathds{D}^{ss}\,\text{S}_{2\zeta}^{(r)}\right].\label{UrHrUrZetaFermion}
\end{eqnarray}
Differentiating \eqref{UrHrUrZetaFermion} with respect $\zeta^{(r)}$ and setting it to zero, we obtain
\begin{equation}\label{HvanishFermion}
\text{S}_{2\zeta}^{(r)}+2^{r}\,\mathds{D}^{ss}\,\text{C}_{2\zeta}^{(r)}~=~0,
\end{equation}
which admits a formal solution given by
\begin{equation}\label{ExactZetarFermion}
\zeta^{(r)}_{ij}\,=\,-\frac{1}{2}\,\big(\text{Tanh}^{-1}2^{r}\,\mathds{D}^{ss}\big)_{\!ij}\,.
\end{equation}
This solution is well-defined when the argument of the ArcTanh matrix function falls within its domain of convergence. This, however, is not of our concern here, since we will minimize \eqref{UrHrUrZetaFermion} \emph{numerically} using the parametrization in \eqref{IWTzetaxiFermion}.

Further manipulating \eqref{UrHrUrZetaFermion} using \eqref{Dss} and \eqref{Dcoeff}, we obtain:
\begin{equation}
{\big\langle}\mathbf{0}\big|\hat{H}^{r}\big|{\mathbf{0}}\big\rangle_{\!r}~=~m_\psi\,\sum_{i=\infty}^{\infty}\Bigg[\!-\big(\text{Cosh}\,2\zeta^{(r)}\big)_{\!0}+\,2^{(r+1)}\sum_{j=1}^{4} D_{j}\,\big(\text{Sinh}\,2\zeta^{(r)}\big)_{\!j}\Bigg].\label{HrvevAlgoFermion}
\end{equation}
We use this expression in our optimization algorithm, supplemented by the truncations $i_\text{max}=7$, $j_\text{max}=9$, and $k_\text{max}=21$ for the $\xi^{ww}_{i}$, $\xi^{sw}_{j}$, and $\zeta_{k}$ arrays, respectively, in all the layers of our calculation, which ranged from $r=$\;0 to 7.

With the numerical results for the entangling exponents, we can now reconstruct the fermionic two-point function. Before proceeding, we first write explicitly the exact continuum solution for the Feynman propagator in position space:
\begin{eqnarray}\label{FeynmanPropExact}
\big\langle\hat\Psi(x,t)\hat{\overline\Psi}(0,0)\big\rangle\,&=&\,
\begin{pmatrix}
\big\langle\hat\chi\,\hat{\chi}^\dagger\big\rangle &~ \,-\big\langle\hat\chi\,\hat{\psi}\big\rangle \\
\big\langle\hat\psi^\dagger\hat{\chi}^\dagger\big\rangle &~\, -\big\langle\hat{\psi}^\dagger\hat\psi\big\rangle
\end{pmatrix}\\
&=&\,m_\psi\,\mathds{1}_{2\times2\!}\left[\Theta\big(s^2\big)\,\frac{(-i)}{4}\,H_0^{(2)}\big(m_\phi\sqrt{s^2}\big)+\Theta\big(\!\!-\!s^2\big)\,\frac{1}{2\pi}\,K_0\big(m_\phi\sqrt{-s^2}\big)\right]\nonumber\\
&&~+m_\psi^2\,x_{\mu}\gamma^{\mu\!} \left[\Theta\big(s^2\big)\,\frac{(-1)}{4}\,\frac{H_1^{(2)}\big(m_\phi\sqrt{s^2}\big)}{m_\psi\sqrt{s^2}}+\Theta\big(\!\!-\!s^2\big)\,\frac{i}{2\pi}\,\frac{K_1\big(m_\phi\sqrt{-s^2}\big)}{m_\phi\sqrt{-s^2}}\right]\!,\nonumber
\end{eqnarray}
where $t\geq 0$ and $s^2\equiv t^2-x^2$. From \eqref{FeynmanPropExact} we see that we need to compute three expectation values (namely, $\big\langle\hat\chi\,\hat{\chi}^\dagger\big\rangle$, $\big\langle\hat{\psi}^\dagger\hat\psi\big\rangle$, and $\big\langle\hat\chi\,\hat{\psi}\big\rangle$) in order to compare our wMERA results with the continuum solution.

Plugging \eqref{HvanishFermion} into \eqref{UrHrUrFermion}, we obtain the entangler-transformed Hamiltonian in terms of the optimized entangler parameters:
\begin{equation}\label{UrHrUrOptFermion}
\hat{U}_r^\dagger \,\hat{H}^r\,\hat{U}_r~=~m_\psi\,\sum_{i,j}\,\bigg(\,{\hat{\chi}_i}^{\,r\,\dagger}\,\Omega^{(r)}_{ij}\,\hat{\chi}_j^{\,r}-\hat{\psi}_i^{\,r}\,\Omega^{(r)}_{ij}\,\hat{\psi}^{\,r\,\dagger}_j\bigg),
\end{equation}
with
\begin{equation}\label{OmegaijFermion}
\Omega^{(r)}_{ij}~\equiv~\Big(\text{C}_{2\zeta}^{(r)}+2^{r}\,\mathds{D}^{ss}\,\text{S}_{2\zeta}^{(r)}\Big)_{\!ij\,.}
\end{equation}
The entangler-transformed time-evolution operator is then given by:
\begin{eqnarray}\label{UrUtUrFermion}
\hat{\mathds{U}}_r(t)~&\equiv&~\hat{U}_r^\dagger\Big(e^{-i\,t\,\hat{H}^r}\Big)\hat{U}_r \nonumber\\
&=&~\text{Exp}\bigg[\!\!-i\,(m_\psi\,t)\sum_{i,j}\bigg(\,{\hat{\chi}_i}^{\,r\,\dagger}\,\Omega^{(r)}_{ij}\,\hat{\chi}_j^{\,r}-\hat{\psi}_i^{\,r}\,\Omega^{(r)}_{ij}\,\hat{\psi}^{\,r\,\dagger}_j\bigg)\bigg].
\end{eqnarray}
It transforms the fermionic scaling modes as :
\begin{subequations}\label{UtsigmaUtFermion}
\begin{alignat}{3}
&\hat{\mathds{U}}_r^{\dagger}(t)\; \hat{\chi}_i^{\,r}\; \hat{\mathds{U}}_r(t)&~=~&\big(e^{-i\,(m_\phi t)\,\Omega^{(r)}}\big)_{ij}\,\hat{\chi}_j^{\,r}\\
&\hat{\mathds{U}}_r^{\dagger}(t)\; \hat{\psi}_i^{\,r}\; \hat{\mathds{U}}_r(t)&~=~&\big(e^{-i\,(m_\phi t)\,\Omega^{(r)}}\big)_{ij}\,\hat{\psi}_j^{\,r}.
\end{alignat}
\end{subequations}

\begin{figure}
\centering\includegraphics[width=0.99\textwidth]{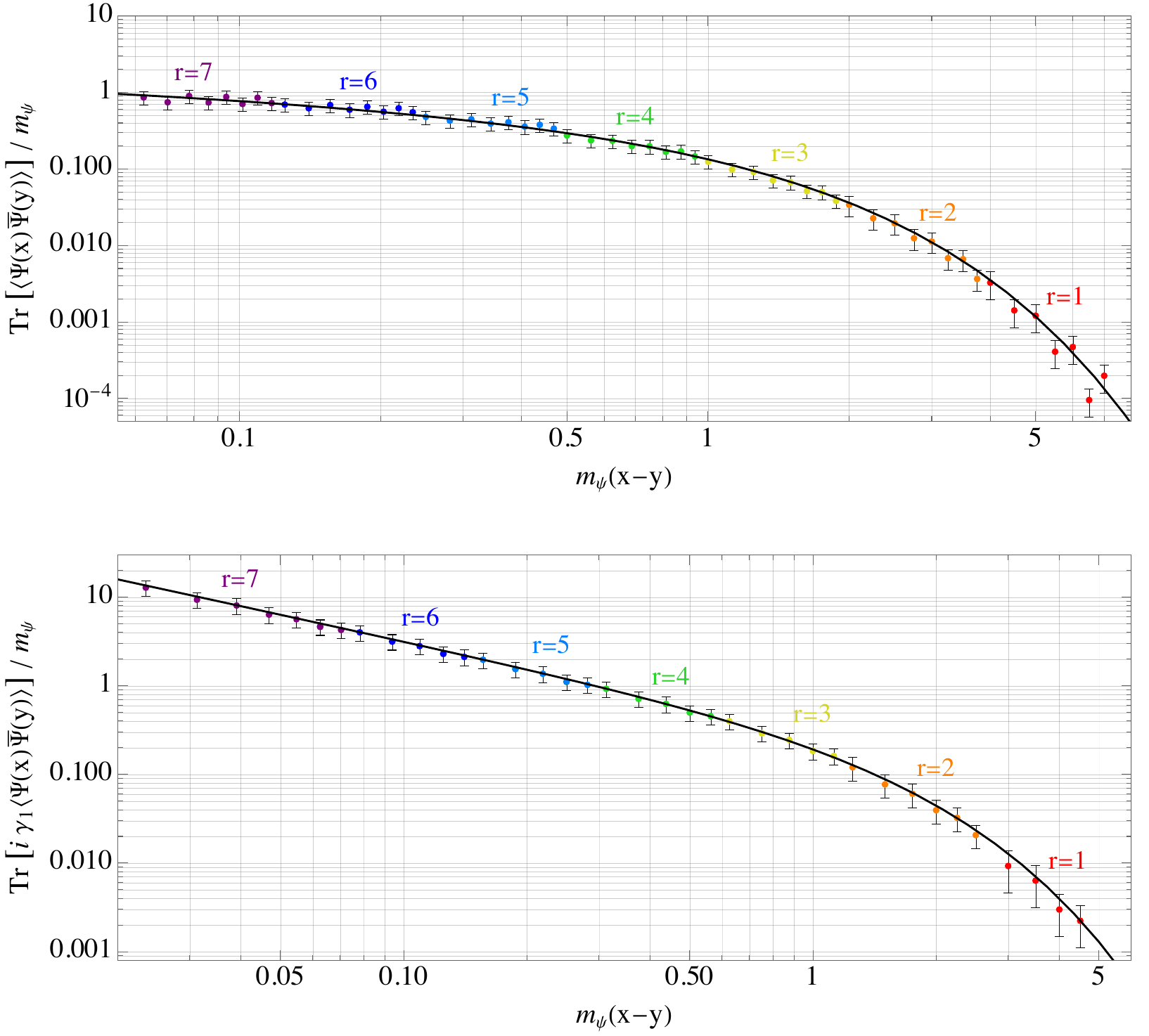}
\caption{The reconstructed continuum fermionic two-point function at equal times calculated using the numerical output of our wMERA optimization. The top and bottom plots show, respectively, $\text{Tr}\big[\big\langle\hat\Psi(x)\hat{\overline\Psi}(y)\big\rangle\big]=\big\langle\hat\chi(x)\hat{\chi}^\dagger(y)\big\rangle-\big\langle\hat\psi^\dagger(x)\hat{\psi}(y)\big\rangle$ and $\text{Tr}\big[i\,\gamma^1\big\langle\hat\Psi(x)\hat{\overline\Psi}(y)\big\rangle\big]=i\,\big\langle\hat\chi(x)\hat{\psi}(y)\big\rangle+i\,\big\langle\hat\psi^\dagger(x)\hat{\chi}^\dagger(y)\big\rangle$. The exact continuum solution is given by the solid black curves (see \eqref{FeynmanPropExact}). In both plots, the functions were calculated for different ranges in $m_{\psi}(x-y)$ using different wMERA layers, indicated by the resolution index $r$. We recall that the lattice spacing is related to the resolution index $r$ via $a_r=1/(2^r m_\psi)$.}
\label{FermionSpace2PF} 
\end{figure}

Using \eqref{UsigmaUFermion} and \eqref{UtsigmaUtFermion}, we can obtain the fermionic two-point correlation functions at any resolution $r$. For instance,
\begin{eqnarray}\label{chiTchi}
\big\langle\hat\chi^r(x,t)\hat\chi^{r\,\dagger\!}(y,0)\big\rangle&=&{\big\langle}\mathbf{0}\big|\big(e^{i\,t\,\hat{H}_\sigma^r}\big)\hat\chi^r(x)\big(e^{-i\,t\,\hat{H}_\sigma^r}\big)\hat\chi^{r\,\dagger\!}(y)\big|{\mathbf{0}}\big\rangle_{\!r}\\
\nonumber\\
&=&\sum_{i,j} \,s^r_i(x)s^r_j(y)~{\big\langle}\mathbf{P}\big|\hat{\mathds{U}}_r^{\dagger}(t)\,\big(\hat{U}_{r\,}^\dagger\,\hat{\chi}_i^{r}\,\hat{U}_r\big)\,\hat{\mathds{U}}_r(t)\,\big(\hat{U}_{r\,}^\dagger\,\hat{\chi}_j^{r\,\dagger}\hat{U}_r\big)\big|{\mathbf{P}}\big\rangle_{\!r}\nonumber\\
&=&m_{\psi}\sum_{i,j} 2^{r\,}s\big(2^r m_\psi\, x-i \big)\,s\big(2^r m_\psi\, y-j\big)\Big(\text{Cosh}^2\zeta^{(r)}e^{-i\,(m_\phi t)\,\Omega^{(r)}}\Big)_{\!i-j\,.}\nonumber
\end{eqnarray}
Analogous steps can be followed to obtain
\begin{eqnarray}\label{psiTpsi}
\big\langle\hat\psi^{r\,\dagger\!}(x,t)\hat\psi^r(y,0)\big\rangle
&=&m_{\psi}\sum_{i,j} 2^{r\,}s\big(2^r m_\psi\, x-i \big)\,s\big(2^r m_\psi\, y-j\big)\Big(\!\!-\!\text{Sinh}^2\zeta^{(r)}e^{-i\,(m_\phi t)\,\Omega^{(r)}}\Big)_{\!i-j}~~~~~~~~~
\end{eqnarray}
and
\begin{eqnarray}\label{chiTpsi}
\big\langle\hat\psi^{r\,\dagger\!}(x,t)\hat\chi^{r\,\dagger\!}(y,0)\big\rangle&=&-\,\big\langle\hat\chi^r(x,t)\hat\psi^r(y,0)\big\rangle\\
&=&~\frac{i}{2}\,m_{\psi}\sum_{i,j} 2^{r\,}s\big(2^r m_\psi\, x-i \big)\,s\big(2^r m_\psi\, y-j\big)\Big(\text{Sinh}2\zeta^{(r)}e^{-i\,(m_\phi t)\,\Omega^{(r)}}\Big)_{\!i-j\,.}\nonumber
\end{eqnarray}
As usual, the expressions above are restricted to their regimes of validity, $\sqrt{|t^2-(x-y)^2|}\sim\mathcal{O}(1-10)\times(2^r m_\phi)^{-1}$.

\begin{figure}
\centering\includegraphics[width=0.99\textwidth]{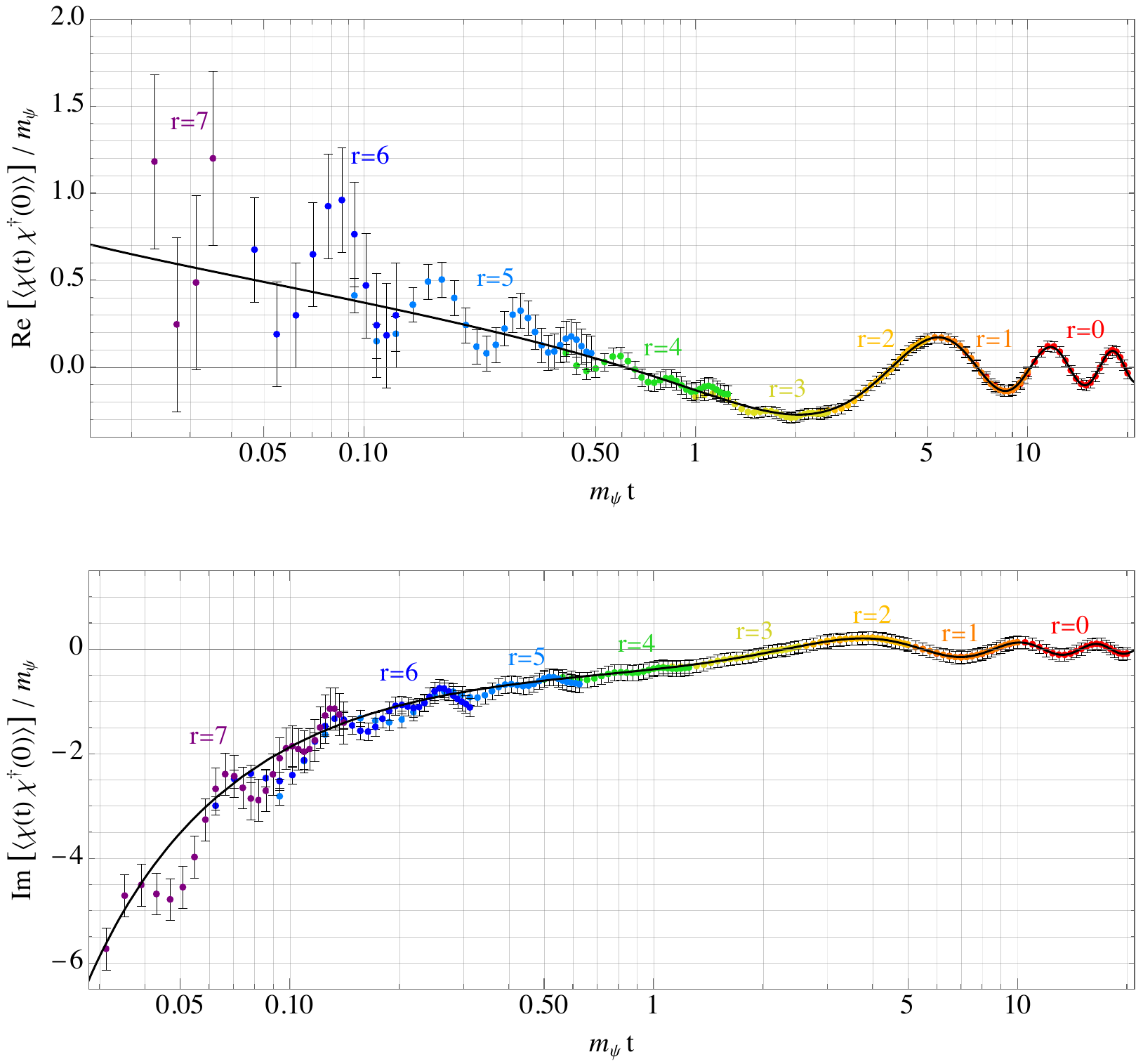}
\vspace{-5pt}
\caption{The reconstructed continuum fermionic two-point function $\big\langle\hat\chi(x,t)\hat\chi^\dagger(x,0)\big\rangle$ for timelike separations calculated using the numerical output of our wMERA optimization. The real and imaginary parts are shown in the top and bottom plots, respectively. The oscillatory behavior of the wMERA results around the exact continuum solution, given by solid black curves, stems from presence of fermion doublers in the spectrum of the discretized theory---this effect becomes significant at short time scales $m_{\psi\,}t <1$.  In both plots, the functions were calculated for different ranges in $m_{\psi\,}t$ using different wMERA layers, indicated by the resolution index $r$. We recall that the lattice spacing is related to the resolution index $r$ via $a_r=1/(2^r m_\psi)$.}
\label{FermionTimeChiChi} 
\end{figure}

\begin{figure}
\centering\includegraphics[width=0.99\textwidth]{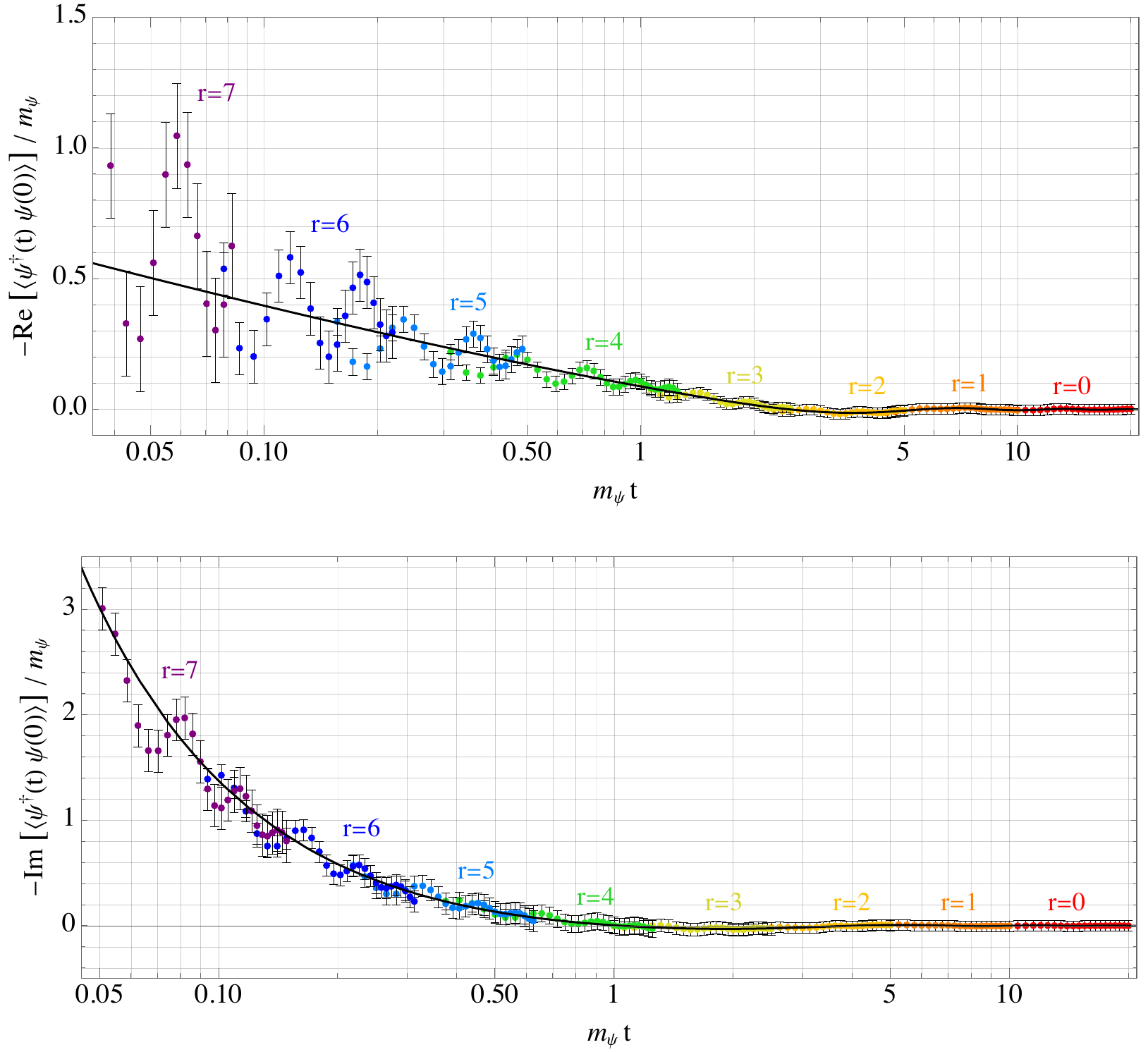}
\caption{Similar to figure \ref{FermionTimeChiChi}, but for $-\big\langle\hat\psi^\dagger(x,t)\hat\psi(x,0)\big\rangle$.}
\label{FermionTimePsiPsi} 
\end{figure}

Piecing together the reconstructed two-point functions \eqref{chiTchi}-\eqref{chiTpsi} at different resolutions, we can compare them with the exact continuum solution for $\big\langle\hat\Psi(\Delta x,\Delta t)\hat{\overline\Psi}(0,0)\big\rangle$ in \eqref{FeynmanPropExact} at both spacelike and timelike separations over several orders of magnitude in $\Delta x$ and $\Delta t$, respectively. Our results are shown in figures \ref{FermionSpace2PF}, \ref{FermionTimeChiChi}, and \ref{FermionTimePsiPsi}.
We find very good agreement between the wMERA and exact solutions for spacelike separations (see figure \ref{FermionSpace2PF}).
For timelike separations, there is good agreement at a resolutions comparable to or lower than the system's correlation length, $m_\psi^{-1}$. However, at shorter time scales, there are two competing effects worth mentioning: the first is a loss of accuracy and precision stemming from the truncation choices in our numerical algorithm. We saw this effect in the scalar case as well, and it can be easily mitigated with modified truncations choices.
The second effect is an oscillatory behavior of the wMERA two-point function around the exact solution (see figures \ref{FermionTimeChiChi} and \ref{FermionTimePsiPsi}). This effect is actually \emph{physical}, and it reflects the fact that the continuum and discretized theories have a different spectrum of states---an effect commonly know as the ``fermion doubler problem'' in discretized field theories with fermions. Indeed, our Daubechies D6 discretization is subject to the Nielsen-Ninomiya theorem \cite{Nielsen:1981hk,Friedan:1982nk}.
This means that, in the massless limit, our D6-discretized theory of Dirac fermions has indeed twice the number of zero-modes compared with the continuum theory \cite{Fries:2018pcd}. This explains the presence of oscillatory features in our wMERA results: since the Hamiltonian mass term becomes less important as the system flows to the UV, the system's correlation function should approach the massless solution at very high resolutions; therefore, it should not be surprising that the effects of the doubler states become more and more pronounced at shorter and shorter time scales.

It is notable, however, that modulo these UV oscillations due to doublers, the two-point function for D6-discretized theory \emph{actually tracks} the shape of the continuum solution, a feature that is not shared by the naive lattice field theory discretization based on Haar wavelets.
Indeed, for the sake of comparison, we also obtained the wMERA solution for the Haar-discretized fermionic theory, and found that (i) the resulting two-point function had a completely different shape than the continuum solution---this also follows from solving the Haar-discretized theory using other methods, such as discrete Fourier transforms; and (ii) the real-time Hamiltonian evolution of the Haar-discretized theory suffered from much higher noise than the real-time evolution of the D6-discretized theory. This noise was present at \emph{all scales} and was not mitigated by increasing the truncations orders in the numerical algorithm.  This seems to indicate that the D6 discretization might be a better choice than the traditional, Haar discretization of fermionic theories. It is not clear to this author, however, whether this preliminary conclusion will hold once these theories are properly regularized to remove the doubler states.

Given the ubiquity of the fermion doubler problem in lattice field theory, many methods have been developed, in various contexts, to rid the spectrum of fermionic doubler states (see, e.g.,  \cite{Rothe:1992nt}). At least some of these methods should be adaptable to the Daubechies-based discretization. Exploration of these possibilities, however, is deferred to future work.

\section{Conclusions and Outlook}
\label{Conclusion}

Traditionally, the concept of entanglement has been completely obscured in the analytical and numerical treatments of QFTs. More recently, entanglement has emerged as an important element in the context of the holographic principle, black holes, and gravity  \cite{Hayden:2007cs,Maldacena:2013xja,Swingle:2014uza,Pastawski:2015qua,Brown:2015lvg}, and as a useful measure to quantify complexity and decoherence in quantum field theories \cite{Berges:2017hne,Carney:2017jut,Agon:2014uxa,Boyanovsky:2018fxl,Balasubramanian:2011wt,Calabrese:2004eu}. In this article, we hope to have conveyed another important role that entanglement can play, namely, as a \emph{computational tool}\footnote{This insight has emerged in condensed matter theory and has led to the emerging field of tensor networks.} to develop new numerical methods for quantum field theories, which could potentially lead to novel insights.

In this study we introduced wMERA, an adaptation of entanglement renormalization for quantum field theories that preserves several key features of the original MERA, namely, (i) discretized degrees of freedom localized in position space; (ii) a discrete, layered architecture of the ansatz paralleling an inverse RG flow; and (iii) the quasi-locality of entanglers at each layer. We considered two simple free field theory examples for which we demonstrated the concrete implementation of the wMERA algorithm in detail. In particular, we showed that wMERA allowed us to fully extract the two-point correlators for spacetime separations spanning several orders of magnitude. Notably, we were able to efficiently perform real-time evolution over many decades in $\Delta t$ by working in the appropriate resolution for the ansatz and for the system's Hamiltonian. In other words, the use of the RG flow played a key role in the efficiency of these computations.

In these specific wMERA examples, however, we did depart from the original MERA implementation in one respect: instead of having the entanglers acting on the Hilbert space (\emph{\`a la} Schr\"{o}dinger picture), we chose to have the entanglers transform the field operators (\emph{\`a la} Heisenberg picture). This, combined with translational invariance, allowed us to preserve the systems' infinite length (that is, we were able to bypass finite ``volume'' artifacts). In addition, we were able to preserve the infinite dimension of the local Hilbert spaces for bosonic degrees of freedom.
Whether this ``Heisenberg picture'' approach will be easily generalizable to wMERA implementations for interacting field theories remains to be seen; either way, efficient numerical MERA algorithms in the ``Schr\"{o}dinger picture'' have been developed for condensed matter systems, and there is no reason to believe that they could not be successfully adapted for (suitably regularized and discretized) relativistic field theories.

Our motivation for proposing wMERA is obviously not to have an alternative method for solving free field theories. The real goal is to use this approach to explore new variational methods for solving challenging problems in quantum field theory (such as strongly coupled dynamics) as well as the incorporation of RG flow in quantum algorithms (to, e.g., improve their efficiency and accuracy, including in the preparation of initial states, in the simulation of real-time evolution, and in the readout and interpretation of the simulation's output.)

The next important step towards this goal is explore non-Gaussian entanglers for {\it interacting} QFTs, and to understand their properties in different dynamical regimes. Possible avenues would be to explore hybrid entangling layers composed of the intercalation of {\it ultra-local} non-Guassian entanglers with quasi-local Bogoliubov transformations, which parallels the perturbative calculation approach; and, for systems that exhibit dualities or confinement, to explore the introduction of \emph{``ancillary''} degrees of freedom to represent the dual or confined states, and devise entanglers whose action is to map the original states to the dual or confined states. Another important next step would be to generalize the wMERA formulation to (3+1)d QFTs by exploring optimal wavelet classes in 3 spatial dimensions. In this exploratory phase, the wMERA results would have to be validated against standard perturbative calculations in the weakly coupled regime, or against exact solutions, when available.

\acknowledgments
I thank Dan Carney, Lukasz Cincio, Burak Sahinoglu, and Varun Vaidya for collaboration in very early stages of this project. I am also grateful to Tanmoy Bhattacharya, Soonwon Choi, Rolando Somma, and Claudia Frugiuele for helpful discussions. This work was supported by the program for Quantum Information Science Enabled Discovery for High Energy Physics (QuantISED for HEP), contract number KA2401032, under the U.S. Department of Energy Office of Science. I also acknowledge support from the Quantum Science Center, one of the National Quantum Information Science Research Centers supported by the U.S. Department of Energy Office of Science. Los Alamos National Laboratory is operated by Triad National Security, LLC, for the National Nuclear Security Administration of the U.S. Department of Energy under contract number 89233218CNA000001. I thank the hospitality of the Aspen Center for Physics (ACP) where this work was partially completed---the ACP is supported by the National Science Foundation under grant PHY-1607611.


\appendix
\section{Matrix exponentials of entangling parameters}
\label{appendix1}
In this appendix, we describe the approximations made in our numerical wMERA algorithm in order to represent the exponential matrix function of the entangling parameters $\zeta_{ij}$.

We begin by recalling that $\zeta_{ij}$ is a square matrix associated with an infinite lattice system (that is, the integer indices $i,\,j$ run from $-\infty$ to $+\infty$). However, due to discrete translational invariance, $\zeta_{ij}$ obeys the restricted form:
\begin{equation}\label{zetaIsArray}
\zeta_{ij}=\zeta_{i-j}.
\end{equation}
This property enables the matrix $\zeta$ to be represented by an infinite array of parameters $\zeta_k$ through the following decomposition:
\begin{eqnarray}
\zeta&=&\zeta_{\,0}
\begin{bmatrix}
\ddots&0&~~~~&~~~~&~~~~\\
0&1&0& & \\
 &0&1&0& \\
 & &0&1&0\vspace{-3pt}\\
~~~~&~~~~&~~~~&0&\ddots\end{bmatrix}
+\zeta_{\,1}\!
\begin{bmatrix}
\ddots&1&~~~~&~~~~&~~~~\\
0&0&1& & \\
 &0&0&1& \\
 & &0&0&1\vspace{-3pt}\\
~~~~&~~~~&~~~~&0&\ddots\end{bmatrix}
+\zeta_{-1}\!
\begin{bmatrix}
\ddots&0&~~~~&~~~~&~~~~\\
1&0&0& & \\
 &1&0&0& \\
 & &1&0&0\vspace{-3pt}\\
~~~~&~~~~&~~~~&1&\ddots\end{bmatrix}\!+\,\dots\nonumber\\
\nonumber\\
\nonumber\\
&\,=&~\sum_{k=-\infty}^{+\infty}\zeta_k\,\mathbb{M}^{(k)}\,,~~~~~\text{with}~~~~~\big(\mathbb{M}^{(k)}\big)_{ij}\,\equiv\,
\begin{cases}
\,1~~~\text{if}~~~j-i=k\\
\,0~~~\text{otherwise.}\label{zetaExpansion}
\end{cases}
\end{eqnarray}
%
Importantly, the matrices $\mathbb{M}^{(k)}$ satisfy the following relations:
\begin{equation}
\big[\mathbb{M}^{(i)},\mathbb{M}^{(j)}\big]\,=\,0~~\forall~~i,\,j~~~~~~~~~\text{and}~~~~~~~~~~\mathbb{M}^{(i)}\mathbb{M}^{(j)}\,=\,\mathbb{M}^{(i+j)}.
\end{equation}
These relations allow us to trade the basis of matrices $\mathbb{M}^{(k)}$ for powers of a commuting variable (a $c$-number) denoted by $m$. That is, in \eqref{zetaExpansion}, we make the replacement
\begin{equation}\label{mtok}
\mathbb{M}^{(k)}~~\longrightarrow~~~m^k.
\end{equation}
With \eqref{zetaExpansion} and \eqref{mtok}, we then write the exponential matrix function of $\zeta$ as:
\begin{eqnarray}
e^{\,\zeta}~=~\text{Exp}\left[\!\!\sum_{~~k=-\infty}^{+\infty}\!\!\!\zeta_k\,m^k\right]~&=&~\prod_{k=-\infty}^{+\infty} e^{\,\zeta_k\,m^k}\nonumber\\
&=&~\left(\;\prod_{k=1}^{\infty} e^{\,\zeta_{-k} m^{-k}}\right)\times~e^{\,\zeta_0}~\times \left(\;\prod_{k=1}^{\infty} e^{\,\zeta_{k}\,m^k}\right).~~~~\label{ExpZeta1}
\end{eqnarray}
Defining $\vec{\zeta}\equiv(\zeta_1,\,\zeta_2,\,\zeta_3,\,\dots,\zeta_n,\,\dots)$, and recalling that the entangling matrix $\zeta_{ij}$ is symmetric for scalars and anti-symmetric for fermions, we can then expand the product of exponentials in \eqref{ExpZeta1} as:
\begin{eqnarray}
\left(\;\prod_{k=1}^{\infty} e^{\,\zeta_{k}\,m^k}\right)~&\equiv&~~\sum_{j=0}^{\infty}\,c_j(\vec{\zeta}\,)\,m^j\label{prodMplus}\\
\nonumber\\
\left(\;\prod_{k=1}^{\infty} e^{\,\zeta_{-k}\,m^{-k}}\right)&=&~~\sum_{j=0}^{\infty}\,c_j(\pm\vec{\zeta}\,)\,m^{-j}~~~~\text{with}~~~~
\begin{cases}
\,c_k(+\vec{\zeta}\,)~~\text{for scalars,}\\
\,c_k(-\vec{\zeta}\,)~~\text{for fermions.}
\end{cases}\label{prodMminus}
\end{eqnarray}
Note that expression \eqref{prodMplus} \emph{defines} the functions $c_j(\vec{\zeta}\,)$. That is, we can obtain $c_j(\vec{\zeta}\,)$ by expanding the left-hand side of \eqref{prodMplus} as a Taylor series in $m$ and selecting the coefficient multiplying $m^j$.

Because of \eqref{zetaIsArray}, the exponential matrix function of $\zeta$ also obeys discrete translational invariance,
\begin{equation}\label{ExpZetaIsArray}
\big(e^{\,\zeta\,}\big)_{ij}=\big(e^{\,\zeta\,}\big)_{i-j},
\end{equation}
and it can be expanded in powers of $m$,
\begin{equation}\label{ExpZetaIsArray}
e^{\,\zeta\,}~=\sum_{~k=-\infty}^{+\infty}\big(e^{\,\zeta\,}\big)_{k}\,m^k\,.\\
\end{equation}
Using \eqref{ExpZeta1}, \eqref{prodMplus}, and \eqref{prodMminus}, we can obtain the components of $e^{\,\zeta\,}$. In particular, for a \emph{scalar} entangler:
\begin{equation}\label{ExpZetaS}
\big(e^{\,\zeta\,}\big)_{k}~=~\,e^{\,\zeta_0}~\sum_{j=0}^{\infty}~c_j(\vec{\zeta}\,)\,c_{j+|k|}(\vec{\zeta}\,)~~~~~~\text{(for~~$\zeta_{-i}=\zeta_{\,i}$)}.
\end{equation}
Similarly, for a \emph{fermionic} entangler:
\begin{equation}\label{ExpZetaF}
\big(e^{\,\zeta\,}\big)_{k}~=~\,e^{\,\zeta_0}\times\begin{cases}
\;\sum_{j=0}^{\infty}~c_{k+j}(\vec{\zeta}\,)\,c_{j}(-\vec{\zeta}\,)~~~~\text{if}~~~~k\geq 0\\
\vspace{-10pt}\\
\;\sum_{j=0}^{\infty}~c_j(\vec{\zeta}\,)\,c_{j-k}(-\vec{\zeta}\,)~~~~\text{if}~~~~k<0
\end{cases}~~~~\text{(for~~$\zeta_{-i}=-\zeta_{\,i}$)}.~~~~
\end{equation}
The definition of $c_j(\vec{\zeta}\,)$ in \eqref{prodMplus} as well as the expansions in \eqref{ExpZetaS} and \eqref{ExpZetaF} are exact. In order to use these expansions in our numerical algorithm, we are forced to make a few truncations:
\begin{itemize}
\item
As mentioned in subsection \ref{ScalarSection}, at each wMERA layer we work with a finite number of entangling parameters. That is, we replace the infinite array $\vec{\zeta}$ by a finite one,
\begin{equation}
\vec{\zeta}~=~(\zeta_1,\,\zeta_2,\,\zeta_3,\,\dots,\zeta_n\,).
\end{equation}
\item
We also set a maximum power $p$ for the power series expansion in any of the $\zeta_i$ coefficients. That is, we make the approximation
\begin{equation}
e^{\zeta_i m^i}~\approx~1+\zeta_{i\,} m^i+\frac{1}{2!}\big(\zeta_{i\,} m^i\big)^2+\dots+\frac{1}{p!}\big(\zeta_{i\,} m^i\big)^p.
\end{equation}
\item
Finally, we truncate the number of $c_j(\vec{\zeta}\,)$ coefficients in the expansion in \eqref{prodMplus}. That is, we set
\begin{eqnarray}
c_j(\vec{\zeta}\,)\,=\,0~~\text{for}~~j>j_{max}~~~~\Rightarrow~~~~\big(e^{\,\zeta\,}\big)_{j}\,=\,0~~\text{for}~~|j|>j_{max}.
\end{eqnarray}
\end{itemize}
This set of truncations was not optimized in the numerical wMERA algorithm we used in the examples discussed in section \ref{ExampleSection}. Mapping out the optimal truncations choices and how they affect the precision and accuracy of the calculation is deferred to a future study.

\bibliographystyle{JHEP} 
\bibliography{ERG_refs}

\end{document}